\documentclass[english,latin9]{aa}
\usepackage[LGR,T1]{fontenc}
\usepackage{float}
\usepackage{textcomp}
\usepackage{amstext}
\usepackage{graphicx}

\makeatletter

\newcommand*{\LyXFourPerEmSpace}{\hskip0.25em\relax}
\newcommand*{\LyXThreePerEmSpace}{\hskip0.3333em\relax}
\DeclareRobustCommand{\greektext}{%
  \fontencoding{LGR}\selectfont\def\encodingdefault{LGR}}
\DeclareRobustCommand{\textgreek}[1]{\leavevmode{\greektext #1}}
\ProvideTextCommand{\~}{LGR}[1]{\char126#1}

\newcommand{\lyxmathsym}[1]{\ifmmode\begingroup\def\b@ld{bold}
  \text{\ifx\math@version\b@ld\bfseries\fi#1}\endgroup\else#1\fi}

\providecommand{\tabularnewline}{\\}

\titlerunning { Wide-field GMRT imaging of X-shaped Radio-Galaxies} 
\subtitle{Spectral properties of 4C32.25 and 4C61.23}
\usepackage{rotfloat}

\makeatother

\usepackage{babel}
\begin{document}
\title{}
\title{Wide-Field uGMRT Band-3 Imaging of the Fields Around X-Shaped Radio-Galaxies}
\author{E. Retana-Montenegro\inst{1} }
\offprints{E. Retana-Montenegro}
\institute{Astrophysics Research Centre, School of Mathematics, Statistics and
Computer Science, University of KwaZulu-Natal, Durban 4041, South
Africa\and Wits Centre for Astrophysics, School of Physics, University of the Witwatersrand, Private Bag 3, Johannesburg 2050, South Africa\\
\email{edwinretana@gmail.com}\\
}
\date{Received June xx, xxxx; accepted March xx, xxxx}
\abstract{We present wide-field upgraded Giant Metrewave Radio Telescope (uGMRT)
images of the fields around the X-shaped radio-galaxies (XRGs) 4C32.25,
4C61.23, and MRC 2011-298 obtained at 400 MHz. The observations are
calibrated using the extreme peeling method to account for direction-dependent
effects across the field of view, as previously applied to Low-frequency
array (LOFAR) data. Our 400 MHz images capture in fine detail the
radio-morphology of the XRGs, as well as other serendipitous radio-sources
located in these fields. We use these images along with archival low-frequency
and high-frequency radio data to investigate the spectral properties
of the XRGs 4C32.25 and 4C61.23. Under the assumption of conditions
corresponding to the maximum radio-source age, we estimate the spectral
ages of both the primary lobes and the wings. These ages indicate
that the wings are the oldest component of the XRGs and are a product
of past radio activity. Moreover, we have used the radio images available
to derive high-resolution spectral index maps for these two XRGs.
We find that the spectral index steepens from the primary lobes towards
the wings, consistent with our spectral age estimates. These results
suggest that precessional and backflow models explain the X-shaped
radio-morphology of 4C32.25 and 4C61.23, respectively. Finally, taking
advantage of our wide-area images, we identify several serendipitous
diffuse radio-sources located in our XRG fields and cross-reference
them with previous surveys.}
\maketitle

\section{Introduction \label{sec:intro}}

The observed scaling relations between supermassive black holes (SMBH)
masses and their host galaxies suggest a clear link between the growth
of both components (e.g. \citealt{1998AJ....115.2285M,2000ApJ...539L...9F,2013ARA&A..51..511K}).
The growth of SMBH occurs via mergers and episodes of gas accretion.
During the accretion or active phase, known as active galactic nuclei
(AGN), the SMBH releases vast amounts of energy in the form of electromagnetic
radiation, relativistic plasma jets, and high-speed gas outflows \citep{2025NewAR.10101733A}.
With modern radio-interferometers it is possible to observe with detail
the extended radio morphologies of low-z radio-galaxies hosting SMBH
\citep{2022AJ....163..280M,2024ApJ...964...98X,2025MNRAS.541.3452C}.
These morphologies are usually divided into two categories: Fanaroff-Riley
type I (FRI) sources, which are characterized by dimmed edges exhibiting
bending jets and radio emission that peaks near the core, and edge-brightened
Fanaroff-Riley type II (FRII) sources, which feature well-collimated
jets, an expansive pair of radio lobes, and prominent hotspots \citep{1974MNRAS.167P..31F}.
A sub-class of radio-galaxies is characterized by the presence of
two pairs of radio-lobes associated with a single host galaxy. These
galaxies are called X-shaped radio-galaxies (XRGs) due to the two
radio-lobe pairs having high average alignment angles of $\sim70^{\lyxmathsym{\textdegree}}$
(e.g., \citealt{2002A&A...394...39C,2019A&A...631A.173B,2019ApJ...887..266J}),
which give rise to their distinctive X-shaped morphology. The primary
lobe pair (i.e., active) usually shows hot spots, whereas the secondary
lobe pair (i.e., wings) has lengths shorter or comparable to those
of the primary pair \citep{10.1093/mnras/210.4.929,2002A&A...394...39C,2019ApJ...887..266J},
and is often devoid of hot spots. The majority of XRGs can be classified
as FRII, while the rest are either FRI or present a hybrid FRI/FRII
morphology \citep{2002Sci...297.1310M}. 

Several theoretical models have been proposed in the literature to
explain the origins of XRGs (see \citealt{2012RAA....12..127G} or
\citealt{2024FrASS..1171101G} for a review). The main formation models
differ significantly: in the backflow model the wings are formed as
a result of the deflection of plasma coming from the terminal hotspots
due to buoyancy forces upon impinging on the host thermal halo \citep{10.1093/mnras/210.4.929}
or due to a pressure gradient between the host axes \citep{2002A&A...394...39C};
while in the re-orientation models the wings are fossil emission from
previous jets which undergo a slow axis precession \citep{1978Natur.275..516R,1985A&AS...59..511P},
or may experience an abrupt change in their direction (``spin-flip'')
due to a minor merger event \citep{2002MNRAS.330..609D,2002Sci...297.1310M}.
Other models from the literature include the jet-shell model that
predicts that a merger with a gas-rich disk galaxy activates the SMBH
and creates a series of stellar shells \citep{1983ApJ...266..713O,2012RAA....12..127G}.
The wings are the result of the jet disruption by the stellar shells.
Another model is the dual-AGN system, in which the radio lobes and
wings are fueled by an unresolved dual-AGN system at the core of the
host galaxy \citep{2007MNRAS.374.1085L,2019AJ....157..195L}. In this
model, both AGNs are active at the same, thus, there should not be
systematic differences in the spectral indices of the radio lobes
and wings. This will be against the expectations of the other models,
as the wings are expected to be filled with older radio emission than
the primary lobes and hence the wings should not present a radio spectrum
statistically indistinguishable from that of the primary lobes. Evidence
against the dual-AGN models was found by \citet{2023MNRAS.524.3270P},
who using a wide frequency range (150-1400 MHz) found that their XRGs
sample presents a rich variety of spectral features, in contrast with
the results of the XRGs sample by \citet{2019AJ....157..195L}, in
which the lobes and wings show similar spectral properties using a
narrower frequency range (240-610 MHz). However, \citet{2022ApJ...933...98Y}
provided evidence for a XRG associated with a dual-AGN system using
5 GHz VLBI observations. 

Over the past two decades, the advent of large-area radio-sky surveys
such as FIRST \citep{1995ApJ...450..559B}, NVSS \citep{1998AJ....115.1693C},
TGSS \citep{2017A&A...598A..78I}, and LoTSS \citep{2017A&A...598A.104S,2022A&A...659A...1S}
has increased significantly the number of known XRGs since their designation
as a new radio source class by \citet{1992ersf.meet..307L}. For instance,
\citet{2007AJ....133.2097C} compiled a total of 100 XRG candidates
in the FIRST survey doubling the number of XRGs at the time. Later,
\citet{2011ApJS..194...31P} using an automated pattern recognition
method identified 134 new XRG candidates using FIRST data. \citet{2019ApJS..245...17Y}
presented a new catalog of 265 XRG candidates by combining FIRST and
TGSS data. More recently, \citet{2022MNRAS.512.4308B} and \citet{2022ApJS..260....7B}
discovered about 40 and 14 new XRG candidates using TGSS and LoTSS
DR1 radio maps, respectively.

Low-radio frequency observations (< 1 GHz) provide unique data which
may help to resolve some questions related to the link between AGNs
and their radio-morphologies (e.g., \citealt{2019AJ....157..195L,2020A&A...638A..29B,2022ApJS..260....7B}),
and the mechanisms triggering SMBH activity (e.g., \citealt{2019A&A...622A..17S,2020AA,2022A&A...663A.153R}).
However, such observations have been limited by direction-dependent
effects (DDEs) such as: the ionospheric corruption of the visibility
data, imperfect knowledge of the antenna beams, and strong radio-frequency
interference. Therefore, it is of paramount importance to account
for DDEs to obtain deep high-fidelity images at low radio frequencies.
Different calibration strategies have been used to correct DDEs for
different radio-telescopes: GMRT \citep{2009A&A...501.1185I,2017A&A...598A..78I},
LOFAR \citep{2016ApJS..223....2V,2016MNRAS.460.2385W,2018A&A...620A..74R,2021A&A...648A...1T},
and MeerKAT \citep{2014arXiv1410.8706T,2018A&A...611A..87T}. In this
work, we use a modified implementation of the extreme-peeling technique
used by \citet{2018A&A...620A..74R} (hereafter, RM18) in the NDWFS-Bo\"otes
field to calibrate upgraded GMRT (uGMRT, \citealt{2017CSci..113..707G})
250-500 MHz Band-3 observations of three XRGs fields (4C32.25, 4C61.23,
and MRC 2011-298). These observations provide a testbed for our uGMRT
direction-dependent calibration strategy, which allows us to obtain
deep high-fidelity images.

The main purpose of studying XRGs fields is threefold. Firstly, to
understand the physical mechanisms that lead to their distinct X-shaped
morphology. The radio-galaxies 4C32.25 \citep{1985A&AS...59..511P,1992A&AS...94...13G,1994A&AS..103..157M,1995A&A...303..427K,1997A&AS..124..259R,1998AJ....115.1693C,2007AJ....134.1245C}
and 4C61.23 \citep{1998AJ....115.1693C,2001A&A...370..409L,2007AJ....134.1245C}
offer an excellent opportunity to probe the formation mechanisms of
XRGs. The higher sensitivity and angular resolution of uGMRT Band-3
allows us study in great detail the spectral properties of these two
XRGs at low-frequencies. Secondly, to investigate in detail the entire
fields surrounding these XRGs, searching for diffuse sources or AGNs
serendipitously located within them. Previously, analyses of the same
data focused only on the central XRGs \citep{2024MNRAS.530.4902S,2024A&A...690A.160B},
while we calibrate, image, and analyze the entire fields. Thirdly,
to demonstrate the robustness of our uGMRT direction-dependent calibration
strategy.

This paper is organized as follows. In Section \ref{sec:targets},
we describe the XRGs analized in this work. A summary of the datasets
used is presented in Section \ref{sec:data}. In Section \ref{sec:data_processing},
we introduce our direction-dependent calibration method based on the
extreme peeling technique. Section \ref{sec:images_catalogs} discuss
our mosaics and source catalogs. Section \ref{sec:spectral_properties}
presents our results. These include our wide-field uGMRT images, integrated
synchrotron spectra, spectral ages, spectral-index maps of 4C32.25
and 4C61.23. In Section \ref{sec:diffuse_sources}, we present a sample
of serendipitous radio-sources located in our XRGs fields. Finally,
we provide our summary and conclusions in Section \ref{subsec:conclusions}.
Throughout this paper, we use a $\Lambda$ cosmology with the matter
density $\Omega_{m}=0.30$, and the is the energy density associated
with a cosmological constant $\Omega_{\Lambda}=0.70$, the Hubble
constant $H_{0}=70\,\textrm{km}\,\textrm{s}^{-1}\,\textrm{Mpc}^{-1}$.
We assume a definition of the form $S_{\nu}\propto\nu^{\alpha}$,
where $S_{\nu}$ is the source flux density, $\nu$ the observing
frequency, and $\alpha$ the spectral index. 

\section{Targets \label{sec:targets}}

\subsection{Radio-galaxy 4C32.25 (B2 0828+32) }

The radio-galaxy 4C32.25 (also known as B2 0828+32) presents prominent
wings which are over a factor of two more extended than its radio
lobes (e.g., \citealt{1997A&AS..124..259R,1998AJ....115.1693C}).
The luminosity profiles of the host galaxy show weak signatures of
a recent merger event \citep{1996A&A...313..750U}. The radio-galaxy
has a total flux density of $S_{\textrm{1.4GHz}}=1.8\;\textrm{Jy}$
\citep{1998AJ....115.1693C}, and a redshift of $z_{\textrm{spec}}=0.0507$
\citep{2010MNRAS.408.1103L}. While its radio lobes indicate an FRII
type morphology, the radio-luminosity, $L_{178MHz}=2.60\times10^{24}\:\textrm{W}\,\textrm{Hz}^{-1}$,
is approximately one order of magnitude below the FRI/FRII threshold
($1\times10^{25}\:\textrm{W}\,\textrm{Hz}^{-1}$, at 178 MHz; \citealt{1974MNRAS.167P..31F}).
The source presents high fractional levels of polarization in the
lobes and wings \citep{1985A&AS...59..511P,1992A&AS...94...13G,1994A&AS..103..157M,2001PhDT.......173R}.
Previous spectral studies found that at frequencies higher than 1.4
GHz there is a progressive steepening of the spectral index along
the wings, with values of $\alpha\sim-0.6$ in the radio lobes to
steeper values of $\alpha\sim-1.2$ in the wings \citep{1995A&A...303..427K,2001PhDT.......173R}.
In contrast, low-resolution observations at 325-609 MHz show that
the spectral index is relatively flat along the wings with values
of $\alpha\sim-0.6$ \citep{2001PhDT.......173R}. 

\subsection{Radio-galaxy 4C61.23 }

4C61.23 is associated with a spheroidal galaxy with strong emission
lines \citep{2001A&A...378..826L}. The radio-galaxy has a total flux
density of $S_{\textrm{1.4GHz}}=1.18\;\textrm{Jy}$, and a redshift
of $z_{\textrm{spec}}=0.111$ \citep{2001A&A...370..409L}. Morphologically,
4C61.23 is a symmetric FRII type source, whereas the radio-luminosity,
$L_{178MHz}=8.55\times10^{24}\:\textrm{W}\,\textrm{Hz}^{-1}$. This
radio-luminosity is just below the FRI/FRII division. Additionally,
there are strong indications that suggest a backflow emanating from
both radio lobes, which is redirected in opposite directions perpendicular
to the jet axes \citep{1998AJ....115.1693C,2001A&A...370..409L}.
There is evidence from high-resolution VLBA images that show that
4C61.23 may be a binary black hole candidate \citep{2018ApJ...854..169L,2024MNRAS.530.4902S}.
Finally, the spectral study by \citet{2024MNRAS.530.4902S} found
that the jets exhibit steeper spectra than the wings. 

\subsection{Radio-galaxy MRC 2011-298}

MRC 2011-298 ($z_{\textrm{spec}}=0.1366$, \citealt{2009MNRAS.399..683J})
is the brightest cluster galaxy (BCG) of the galaxy cluster A3670,
a richness-class 2 galaxy cluster \citep{2009AJ....137.4795C}. This
BCG is a giant elliptical galaxy with an ellipticity of $\epsilon=0.14$
\citep{2014A&A...570A..13M} and is classified as a dumbbell galaxy
due to the presence of two bright optical cores surrounded by a stellar
halo \citep{1992A&A...266..127A,1992A&AS...95....1G}. MRC 2011-298
was first classified as an XRG candidate by \citet{1994A&AS..106....1G},
and confirmed later by \citet{2019A&A...631A.173B}. The XRG presents
a pair of north-south lobes with a pair of weak wings oriented almost
perpendicularly to the lobes. Recently, \citet{2024A&A...690A.160B}
found that the S-shaped morphology of the jets makes precession a
strong explanation for the jet bending. They also show that spin-flip
is the likely formation process for the overall X-shaped structure.
These authors also found that the wings are fainter than the lobes
at all frequencies. The radio-galaxy has a flux density of $S_{\textrm{1.4GHz}}=407.3\;\textrm{mJy}$
\citep{1998AJ....115.1693C}, and a radio-luminosity of $L_{178MHz}=2.60\times10^{24}\:\textrm{W}\,\textrm{Hz}^{-1}$. 

\section{Data \label{sec:data}}

In this section, we present the observations of the XRGs and describe
the archival radio-data employed in this work. The details of the
radio data are summarized in Table \ref{fig:summary_data_table}. 

\begin{table*}
\noindent \begin{centering}
\caption{Summary of the uGMRT Band-3 observations}
\begin{tabular}{cccccccc}
\hline 
Name & Rigth Ascension & Declination & ID & Date & Freq. & Calibrator & Int. time\tabularnewline
\hline 
 & (J2000) & (J2000) &  &  & MHz &  & Hours\tabularnewline
4C32.25 & 08h31m27.49s & +32d19m26.84s & 33\_083 & 2017-11-03 & 400 & 3C147 & 2.8\tabularnewline
4C61.23 & 11h37m21.36s & +61d20m01.16s & 33\_083 & 2017-11-03 & 400 & 3C147 & 3.0\tabularnewline
MRC 2011-298 & 20h14m18.86s & -29d42m36.02s & 41\_020 & 2022-03-18 & 400 & 3C48 & 8.0\tabularnewline
\hline 
\end{tabular}
\par\end{centering}
\begin{centering}
\centering
\par\end{centering}
$\;$

Notes: The sky coordinates refer to the pointing phase center. \label{fig:summary_data_table}
\end{table*}

\subsection{uGMRT Band-3 data \label{sec:ugmrt_data}}

The XRGs 4C32.25, 4C61.23, and MRC 2011-298 were observed with the
uGMRT wide-band receiver (GMRT Wideband Backend, GWB, \citealt{2017CSci..113..707G})
in band 3 (250\textendash 500 MHz) for 2.8, 3, and 8 hours, respectively.
The observations include 10-minute pointings on 3C147 and 3C48, which
were used as flux density calibrators. The calibration process is
described in Section \ref{sec:ugmrt_processing}.

\subsection{LoTSS data }

The LOw-Frequency ARray (LOFAR) Two-metre Sky Survey (LoTSS; \citealt{2017A&A...598A.104S,2022A&A...659A...1S})
is an ongoing LOFAR survey of the northern sky at 144 MHz. The LoTSS
images have a median noise level of $83\,{\textstyle \mu}$Jy/beam
with a resolution of $\sim6^{\prime\prime}$. The fields of 4C32.25
and 4C61.23 are both covered by LoTSS. We use the high-resolution
LoTSS images available from the LoTSS archive\footnote{https://lofar-surveys.org/dr2\_release.html}.

\subsection{VLA data }

First, we browse the NRAO VLA Archive Survey (NVAS\footnote{http://www.vla.nrao.edu/astro/nvas/};
\citealt{2008SPIE.7016E..0OC}) looking for archival images of our
XRGs. The NVAS provides access to VLA images from 1976 to 2006 processed
using the AIPS pipeline \citet{2003ASSL..285..109G}. We find that
4C32.25 and 4C61.23 have images available at 1.5 GHz and 4.89 GHz,
respectively. The angular resolutions and noise levels of these images
are $13^{\prime\prime}$, $5^{\prime\prime}$ and $0.483\,$mJy/beam,
$0.167\,$mJy/beam, respectively. We complement these data with images
from the NVSS \citep{1998AJ....115.1693C}, FIRST \citep{1995ApJ...450..559B},
and VLASS \citep{2020PASP..132c5001L} surveys at 1.4 and 3.0 GHz,
respectively. The angular resolutions and typical noise levels of
these surveys are $5^{\prime\prime}$, $45^{\prime\prime}$, $2.5^{\prime\prime}$;
and $0.15\,$mJy/beam, $0.45\,$mJy/beam, $120\,{\textstyle \mu}$Jy/beam,
respectively. Finally, for 4C32.25, we also analyze archival S-Band
data (ID: VLA/22B-144) from the NRAO archive\footnote{https://data.nrao.edu/}.
The observations were performed on January 1st, 2023. The target was
observed in a single scan for a total integration time of 0.56 hours.
Observations were carried out with the VLA in C configuration. The
source 3C147 was used as flux density calibrator. The sampling time
was set to 5 s and four polarization products (RR, LL, RL, and LR)
were obtained. The total bandwidth, equal to 2 GHz in the range 2\textendash 4
GHz, was divided into 16 sub-bands of 128 MHz with 64 frequency channels.
We used the Common Astronomy Software Applications (CASA, \citealt{2007ASPC..376..127M})
to perform a standard calibration\footnote{https://casaguides.nrao.edu/index.php/VLA\_Self-calibration\_Tutorial-CASA5.7.0}
of the VLA dataset. The final image of the target is $0.64^{\circ}\times0.64^{\circ}$
in size and was made using WSClean \citep{2014MNRAS.444..606O}, where
we impose an inner uv-limit of $50\,{\textstyle \lambda}$, and \citet{1995PhDT.......238B}
Brigg\textquoteright s weighting with a robustness parameter of 0.
The image has a resolution of $\sim5{}^{\prime\prime}$ and central
noise of $14\,{\textstyle \mu}$Jy/beam.

\subsection{RACS data}

The Rapid ASKAP Continuum Survey (RACS) is a large-area survey with
the full 36-dish ASKAP radio telescope. RACS aims to image the entire
sky (south of Dec $51^{\circ}$) at 887.5 MHz (RACS-low), 1296 MHz
(RACS-mid), and 1656 MHz (RACS-high), with a bandwidth of 288 MHz
\citep{2020PASA...37...48M,2021PASA...38...58H,2023PASA...40...34D,2025PASA...42...38D}.
RACS-low has an angular resolution of $15^{\prime\prime}$ with an
average noise level of $0.25\,$mJy/beam, RACS-mid has a resolution
of $10.2^{\prime\prime}$ and median noise of $182\,{\textstyle \mu}$Jy/beam,
and RACS-High maps have a resolution of $9.78^{\prime\prime}$ and
median noise of $195\,{\textstyle \mu}$Jy/beam. The fields of 4C32.25
and MRC 2011-298 are covered by RACS-low and RACS-mid. For these two
fields, we retrieve the RACS images from the CASDA archive\footnote{https://research.csiro.au/casda/}. 

\section{uGMRT data processing \label{sec:data_processing}}

\label{sec:ugmrt_processing}

To process the uGMRT data, we modify the extreme-peeling technique
used by RM18 to calibrate LOFAR data of the NDWFS-Bo\"otes field.
The direction-independent and direction-dependent stages are described
in the following sections.

\subsection{Direction-independent calibration \label{sec:ugmrt_di}}

First, we retrieve the raw data from the uGMRT Online archive\footnote{https://naps.ncra.tifr.res.in/}.
The raw visibilities in UVFITS format are converted to the Measurement
Set (MS) format\footnote{http://casa.nrao.edu/} using the CASA routine
importgmrt\footnote{https://casadocs.readthedocs.io/}. The MS files
are later converted to a LOFAR-compatible format \citep{2013A&A...549A..11O}.
These MS files are regularly shaped, meaning that all the time slots
have the same baselines and channels contained within a single spectral
window. The primary flux density calibrator and target scans are separated
from the original MS file with the CASA task split. The unaveraged
scans are flagged to remove radio frequency interference (RFI) using
a custom LUA AOflagger script \citep{2010MNRAS.405..155O,2012A&A...539A..95O}.
The flagging includes the bandpass edges, and the high-RFI frequency
ranges 360-379.6 MHz and 486-500 MHz, which generally present poor
signal\textendash to-noise (S/N). After flagging, the data are averaged
to a resolution of 512 channels per scan from the original resolution
of 2048 channels. The data is not averaged in the time axis to capture
the rapid changes in the ionospheric conditions. 

\begin{figure}[tp]
\begin{centering}
\includegraphics[clip,scale=0.55]{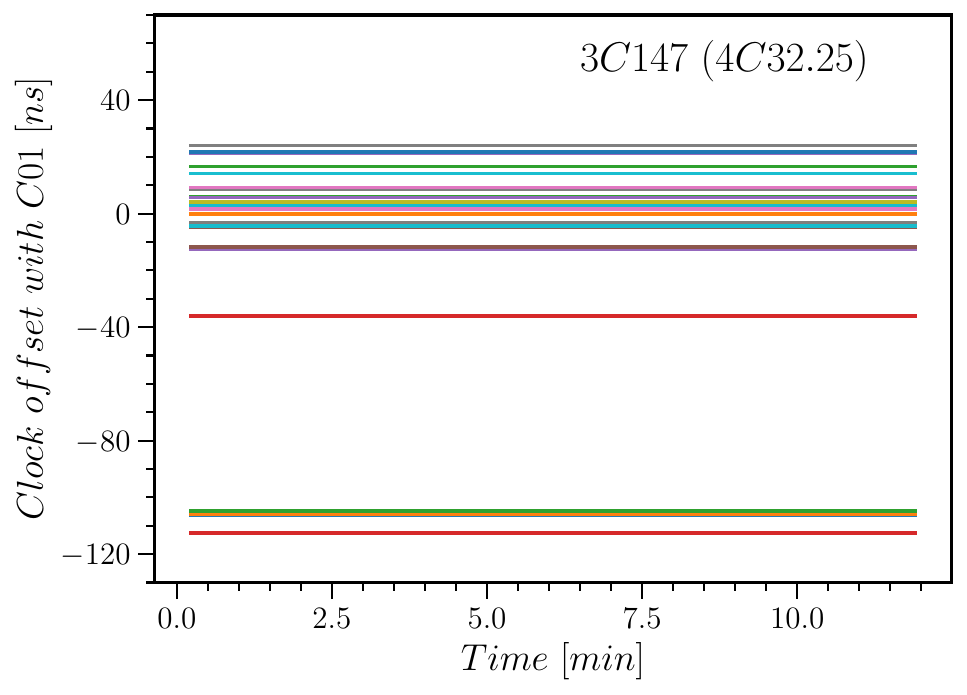}
\par\end{centering}
\centering{}\centering\caption{\label{fig:clockoffsets} Fitted clock differences using the gain
solutions from 3C147. The values show a bimodal distribution between
core and arm antennas (mainly from the Southern arm). The colors represent
different antennas.}
\end{figure}

The first stage involves using the primary flux density scan in three
steps. The first step is obtaining the XX and YY gain solution toward
the flux density calibrator using the LOFAR skymodel. The primary
flux density calibrator of each field is indicated in Table \ref{fig:summary_data_table}.
This skymodel is normalized to the flux density scale of \citet{2012MNRAS.423L..30S}
(hereafter SH12). The gains are computed using the open-source solver
DP3 \citep{2018ascl.soft04003V}. The second step is to determine
the clock offsets between the core and the (Southern, Western, and
Eastern) arms uGMRT antennas using the primary flux density calibrator
phases solutions. The clock offsets for the 4C32.25 field are shown
in Figure \ref{fig:clockoffsets}. The largest negative between core
and arms is close to 120 ns. The other two fields have similar bimodal
distributions in their clock offset values: for 4C61.23, the maximum
offset is about 100 ns, while the offsets for MRC 2011-298 are smaller,
with the largest offset being 30 ns. The third step is to measure
the XX and YY phase offsets for the calibrator. 

\begin{figure}[tp]
\centering{}\includegraphics[clip,scale=0.4]{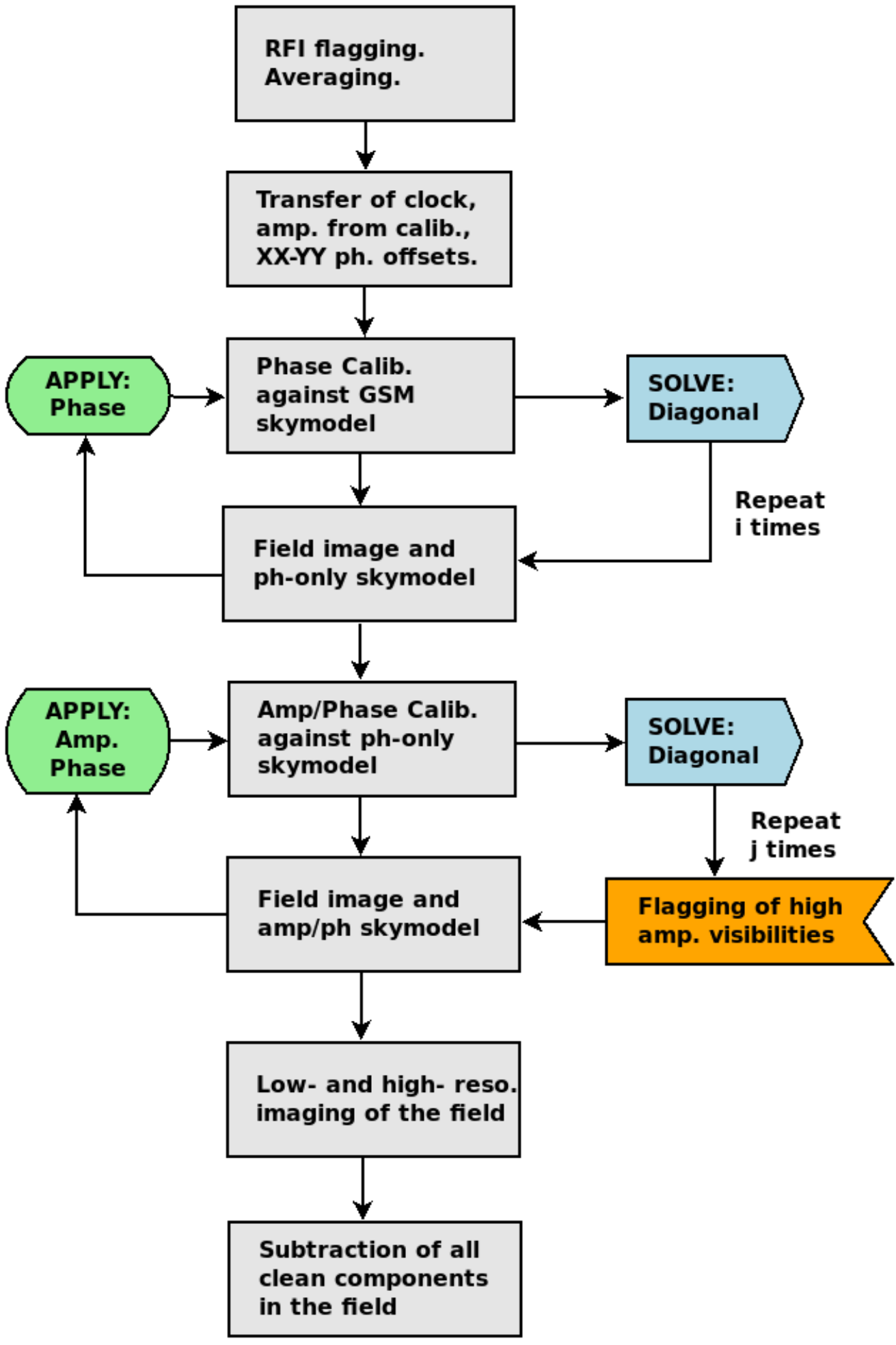}\centering\caption{\label{fig:gmrt_di_steps} Schematic view of the direction-independent
calibration steps for uGMRT Band 3 data.}
\end{figure}

For the second stage, a series of steps involving the target scan
are performed to obtain the DI images of the target field. The flow
diagram of the direction-independent calibration part is shown in
Figure \ref{fig:gmrt_di_steps}. Firstly, the target scans are created,
averaged, and flagged using the same method as described for the calibrator
scans. Secondly, DP3 is used to the transfer of amplitudes, clock
values and XX-YY phase offsets from the flux density calibrator to
the target. The gain amplitudes are filtered and smoothed to eliminate
significant outliers. This flags visibility data related to the eliminated
gain outliers. Thirdly, with DP3 we perform three phase-only calibration
cycles combining all the target scans using a beam-attenuated TGSS
skymodel \citep{2017A&A...598A..78I}. We used the polynomial beam
model described in the uGMRT users manual\footnote{https://naps.ncra.tifr.res.in/}
to attenuate the skymodel. Imaging is carried out using WSClean \citep{2014MNRAS.444..606O},
for which we impose an inner uv-limit of 50 \textgreek{l}, and a \citet{1995PhDT.......238B}
robust weighting parameter of 0. This value provides a compromise
in weighting between the core and edge baselines. WSClean is run with
multi-frequency deconvolution and multiscale modes enabled, in order
to address the wide bandwidth of the data and to better recover the
diffuse emission in the uGMRT maps. The total size of our images is
$6500\times6500$ pixels with a pixel size of $2^{\prime\prime}$.
Fourthly, three additional amplitude-phase calibration cycles are
performed using the skymodel obtained in the last step. 

Finally the skymodel from the last calibration cycle is used to subtract
the high-resolution clean components from the visibilities. We image
the new residual visibilities at low-resolution using a gaussian taper
of $10^{''}$ to account for any diffuse emission component resolved
out at high-resolution. This low-resolution skymodel is subtracted
from the high-resolution subtracted visibilities to obtain new residual
DI visibilites with both low- and high-resolution CLEAN components
subtracted to obtain DI residual visibilities. This subtraction will
be improved iteratively during the directional self-calibration process.
Finally, the low-resolution model is combined with the high-resolution
one to obtain the complete DI skymodel of the target field. 

\subsection{Direction-dependent calibration \label{sec:ugmrt_dd}}

\begin{figure*}[tp]
\centering{}\includegraphics[clip,scale=0.4]{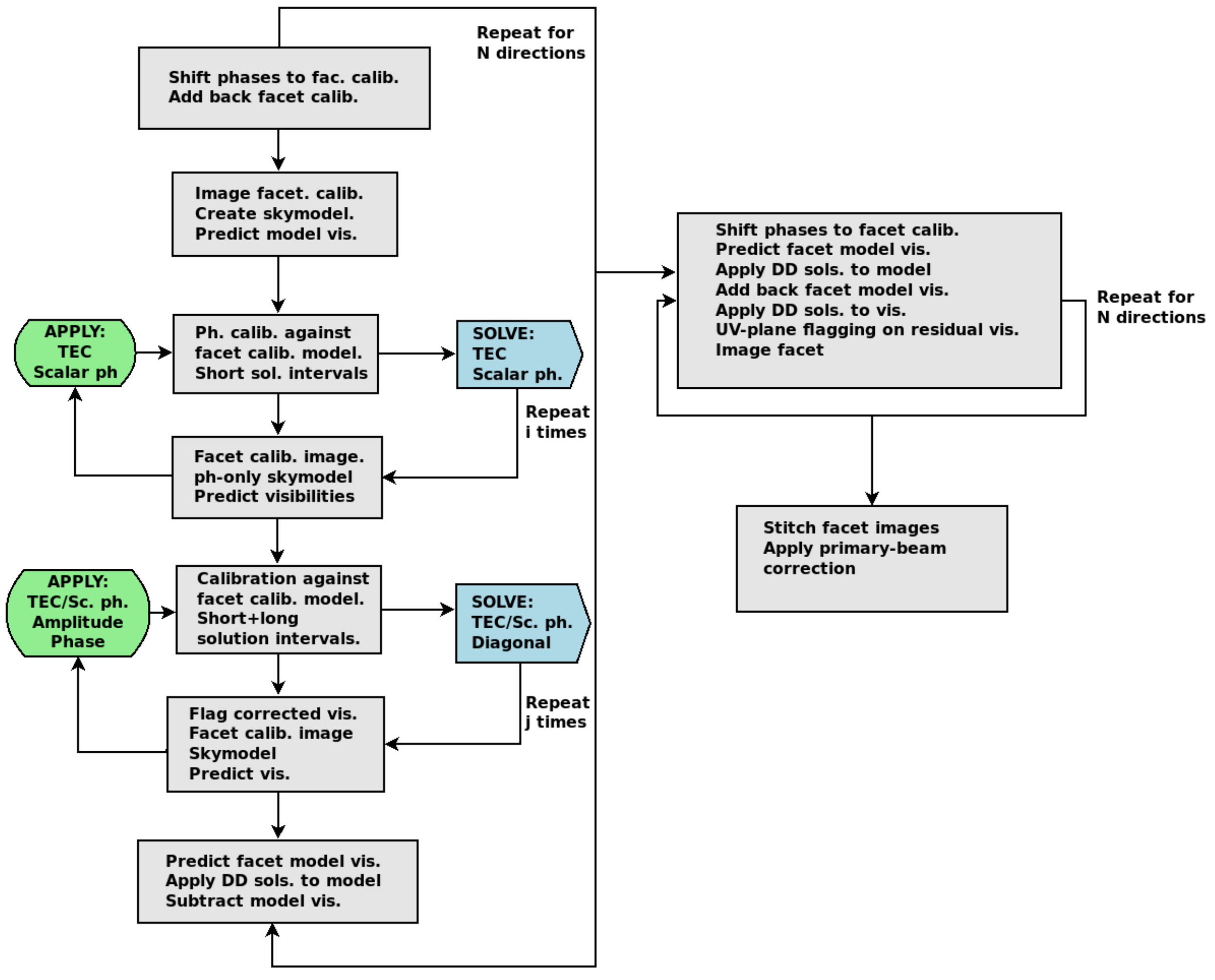}\centering\caption{\label{fig:gmrt_dd_steps} Schematic view of the direction-dependent
calibration steps for uGMRT Band 3 data.}
\end{figure*}

\begin{figure*}[tp]
\centering{}\includegraphics[clip,scale=0.26]{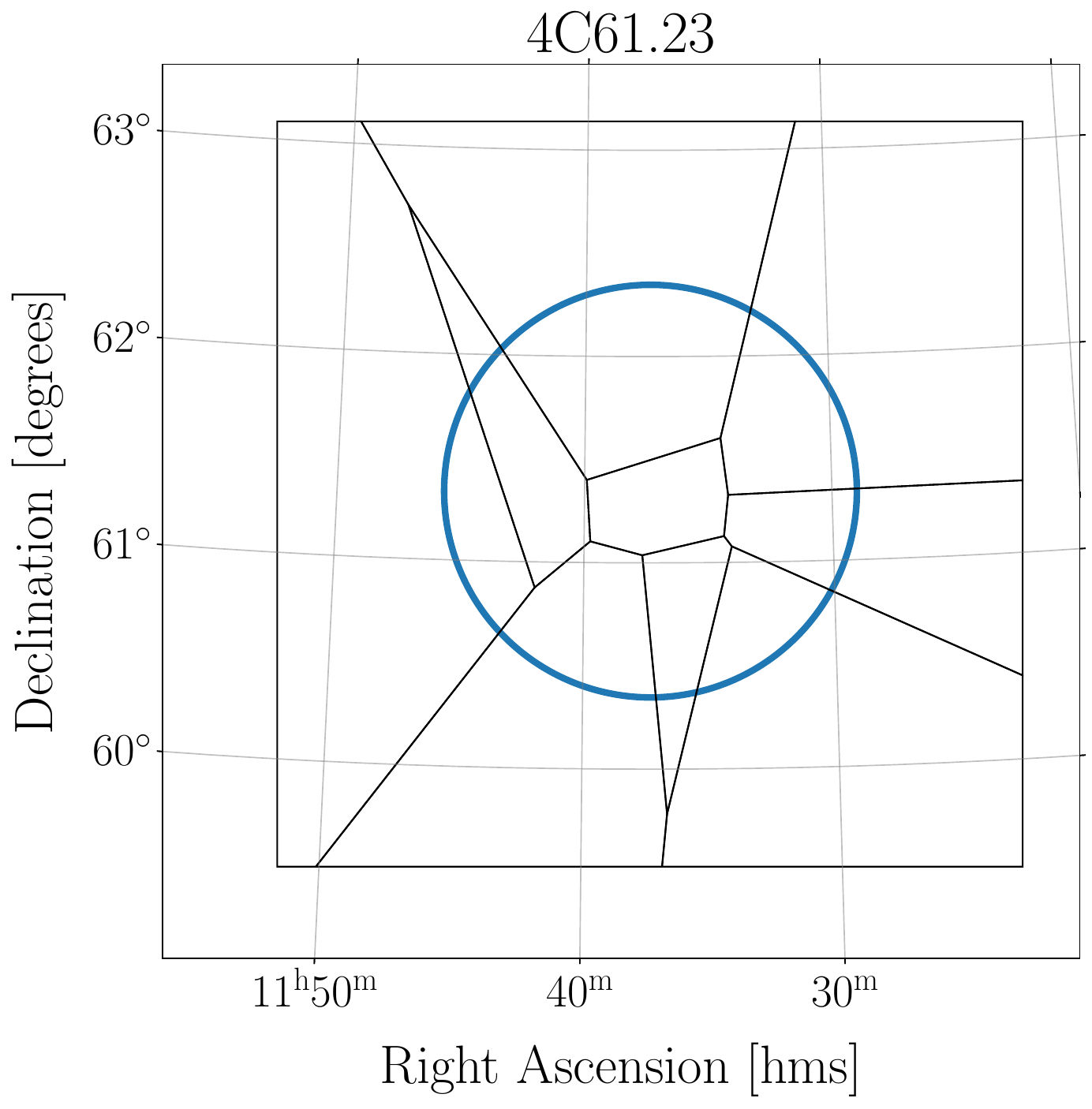}\includegraphics[clip,scale=0.26]{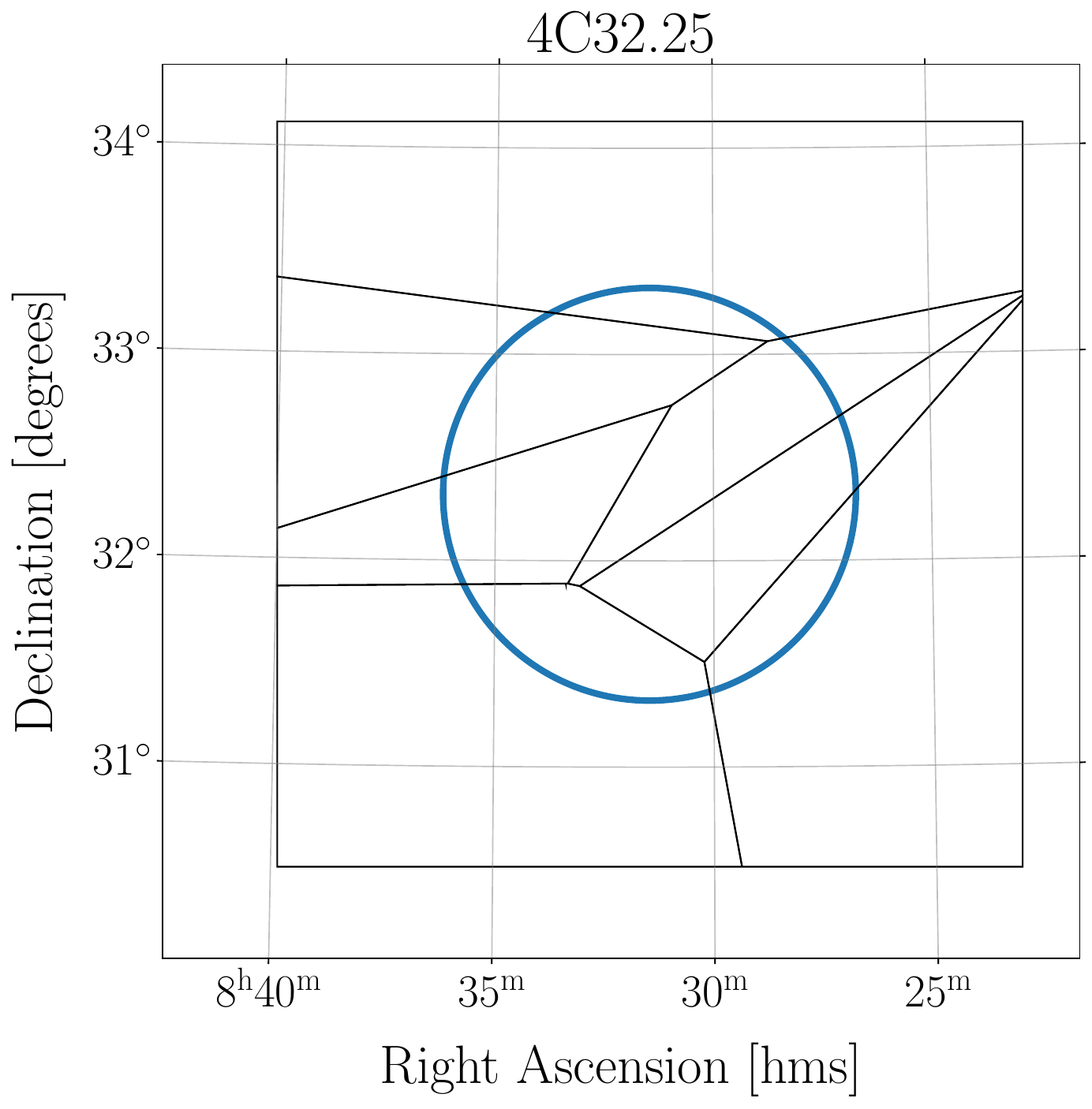}\includegraphics[clip,scale=0.26]{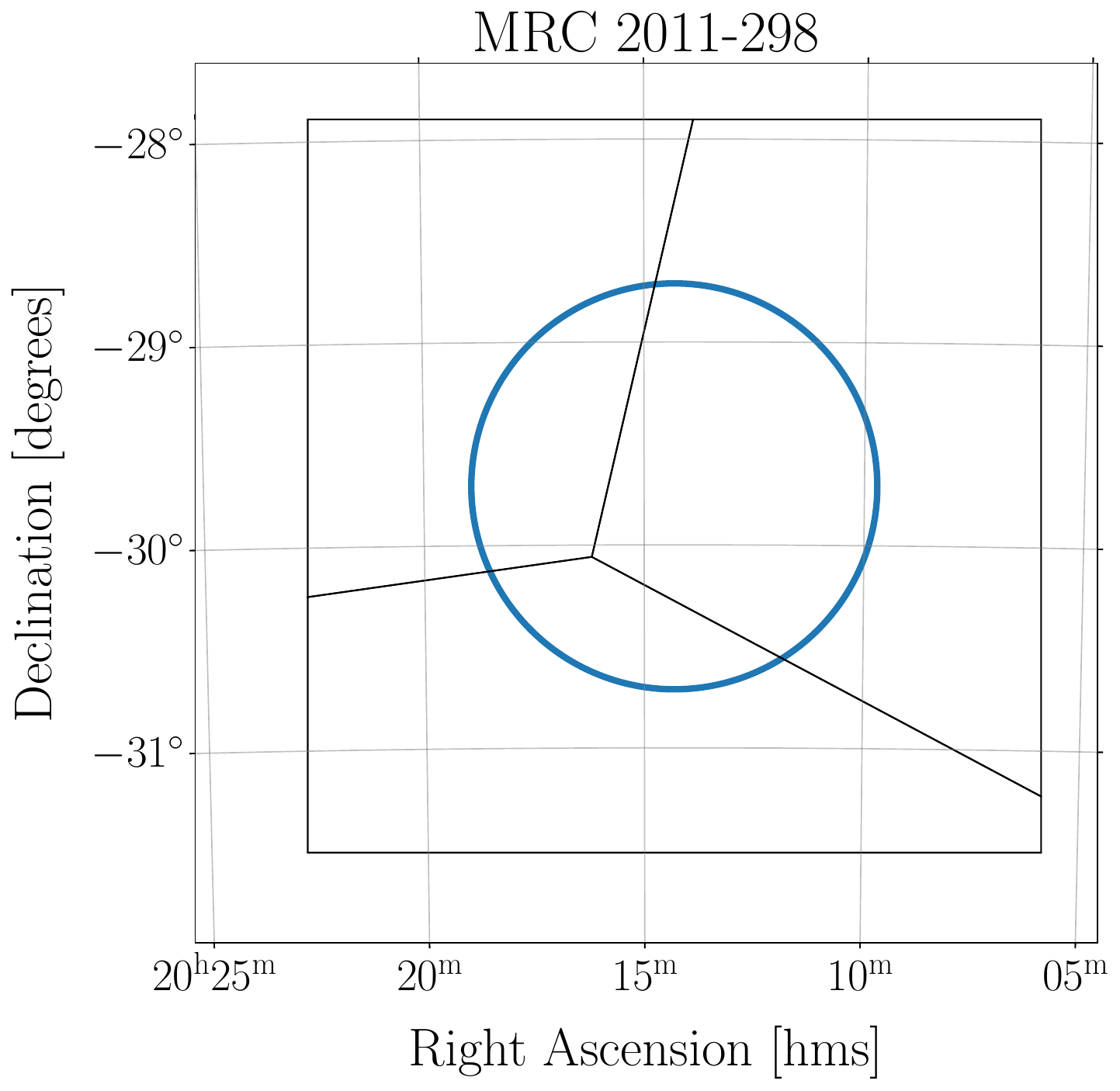}\centering\caption{\label{fig:facets} The spatial distribution of facets in the 4C61.23,
4C25.25, and MRC 2011-298 fields. The blue circle denotes angular
distance where the uGMRT beam is 50 percent of that at the pointing
center, that is about $0.74$ degrees.}
\end{figure*}

Artifacts caused by direction-dependent (DD) effects, as antenna-antenna
beam variations and ionospheric distortions are still present in our
DI images. \citet{1984AJ.....89.1076S} suggested that as long DD
effects vary slowly accross the field of view (FOV) it is possible
to discretize it into smaller regions or facets. For each facet, a
self-calibration process can be done to correct the visibilities for
DD effects. This can be repeated until the facets in the FOV are corrected.
A critical part in the FOV discretization, is that each facet must
have a bright source or a group of closely spaced bright sources in
order for the self-calibration process to converge. The resulting
facet images are used to improve the DI residual visibilities. Based
on our experience, sources brighter than $0.1\:\textrm{Jy}$ are good
facet calibrators. The facet distribution for our three fields are
shown in Figure \ref{fig:facets}. The facet calibrators vary from
field to field, with nine facets for the 4C61.23 field, while the
MRC 2011-298 field requires only three facets. 

\begin{figure}[tp]
\begin{centering}
\includegraphics[clip,scale=0.25]{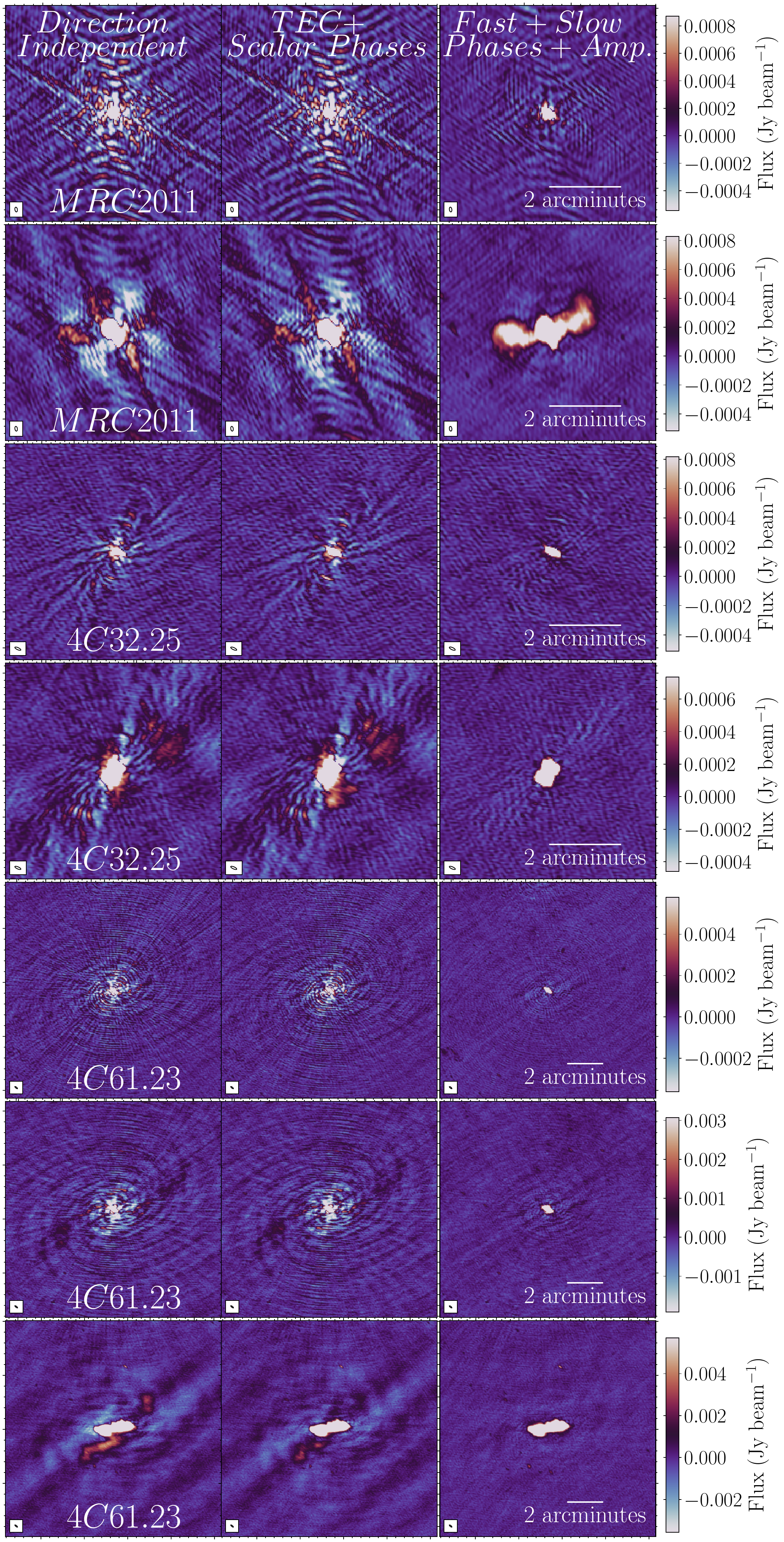}
\par\end{centering}
\centering{}\centering\caption{\label{fig:dd_selfcal} Images showing a direction-dependent self-calibration
sequence for facet calibrators from the different fields. The corresponding
field name is indicated in the images of the first column. The first
column displays the images made with only the DI self-calibration
solutions. The central column displays the improvements after two
iterations of fast phase (TEC and phase-offset)-only DDE calibration.
The right-hand column shows the improvement after two further iterations
of fast phase (TEC+phase offset) and slow phase and amplitude (XX
and YY gain) DDE calibration. The scale is the same for all images
of the self-calibration sequence.}
\end{figure}

The flow diagram of the direction-independent (DD) calibration part
is shown in Figure \ref{fig:gmrt_dd_steps}. The first step is to
shift the DI residual visibilities to the direction of the facet calibrator,
adding it back to the UV data. In the first and second cycles, we
solve for scalar Jones phase offsets and total ionospheric electron
content (TEC) terms in short time scales (fast gains), which cause
frequency-dependent ionospheric distortion on the phases. In subsequent
cycles, we begin by solving for scalar phase+TEC only followed by
the diagonal Jones phase+amplitude solutions over longer timescales
to capture the slower variations in the beam (slow gains). The solution
intervals used for fast gains are 4 channels and 1 time step; while
for slow gains, they are 30 channels and 100 time steps. The self-calibration
cycle can be repeated until artifacts around bright facet calibrators
are reduced significantly and convergence is achieved. When the DD
self-calibration cycle is completed, the DI residual visibilities
are shifted to the direction of the facet center. The facet skymodel
subtracted at the end of DI calibration is added back and the DD gain
solutions are applied. The new facet skymodel is subtracted from the
DI residual visibilities. By doing this, we ensure that the effect
of the presence of bright calibrators is reduced and improves the
subsequent subtraction of fainter facets. This process is done in
descending order according to the facet calibrator flux density. Figure
\ref{fig:dd_selfcal} displays the DD self-calibration sequence for
various facet calibrators from the three different fields. 

A relevant aspect to consider in uGMRT wide-field imaging is the mitigation
of artifacts. For instance, amplitude artifacts usually manifests
as ripples across the image background. We employ the UV-plane outlier
flagging approach by \citet{2018AJ....156....9S} to eliminate ripples
in our images. The principle behind this approach is that the ripples
can be pinpointed as localized high-amplitude features in the UV-plane
using the residual (model - corrected) visibilities. These features
are identified in annular regions in the UV-plane. To prevent over-flagging
due to the larger visibility density towards the center of the UV-plane,
we calculate a threshold for each annulus that is used to flag all
outliers within the respective annulus. This method is used for the
final facet imaging, and it is optional in each DD self-calibration
cycle if it is needed to eliminate persistent ripples. 

The final facets are imaged using the improved residual visibilities
at full resolution, and are stitched together using casacore\footnote{https://github.com/casacore}.
The primary-beam correction is applied using the same model used in
Section \ref{sec:ugmrt_di}. A radial-cutoff is imposed, where the
uGMRT beam is 50 percent of that at the pointing center, which corresponds
to a radius of $\sim0.74\;\textrm{degrees}.$ 

\section{Images and source catalog \label{sec:images_catalogs}}

\begin{figure*}[tp]
\begin{centering}
\includegraphics[clip,scale=0.33]{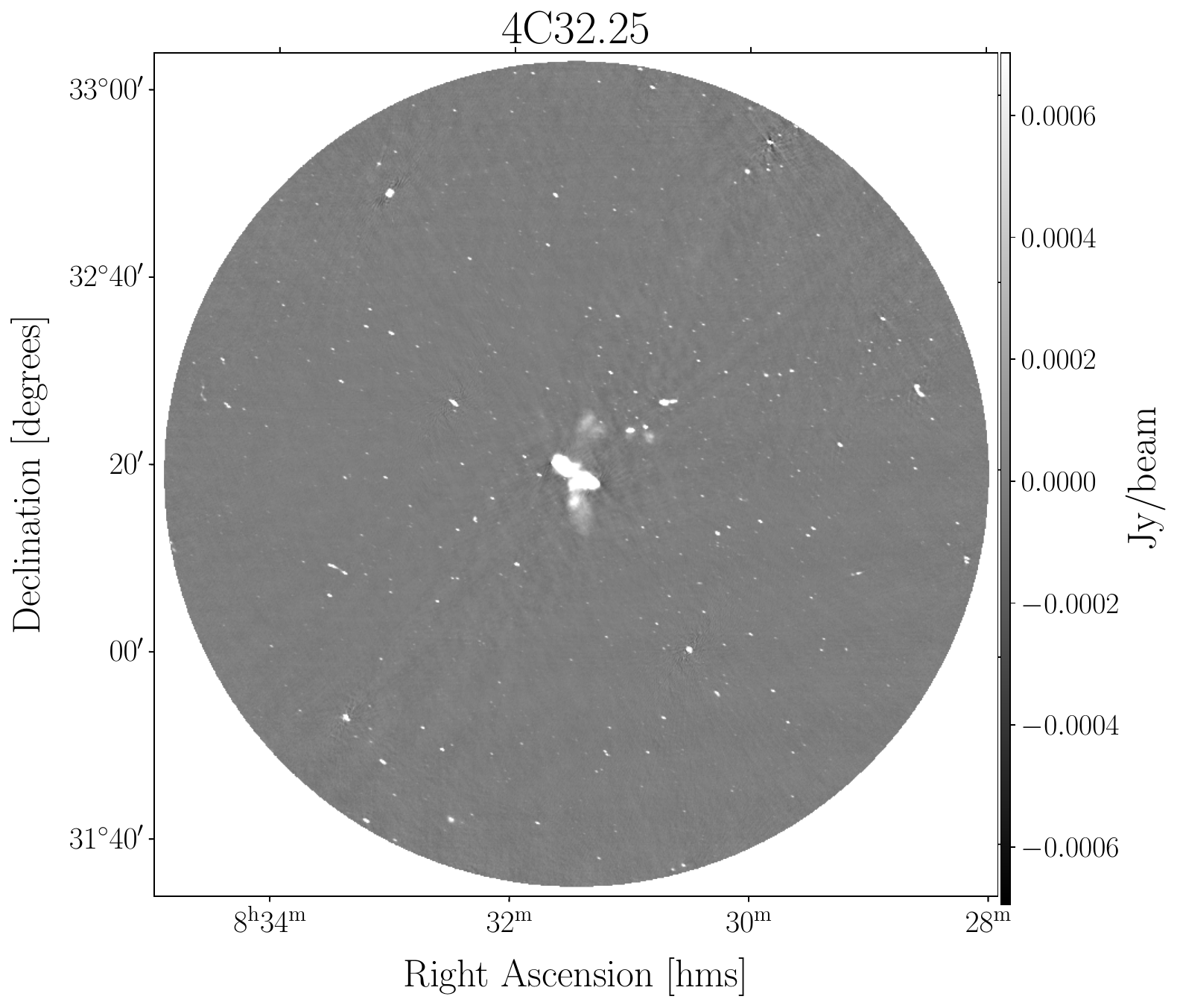}\includegraphics[clip,scale=0.38]{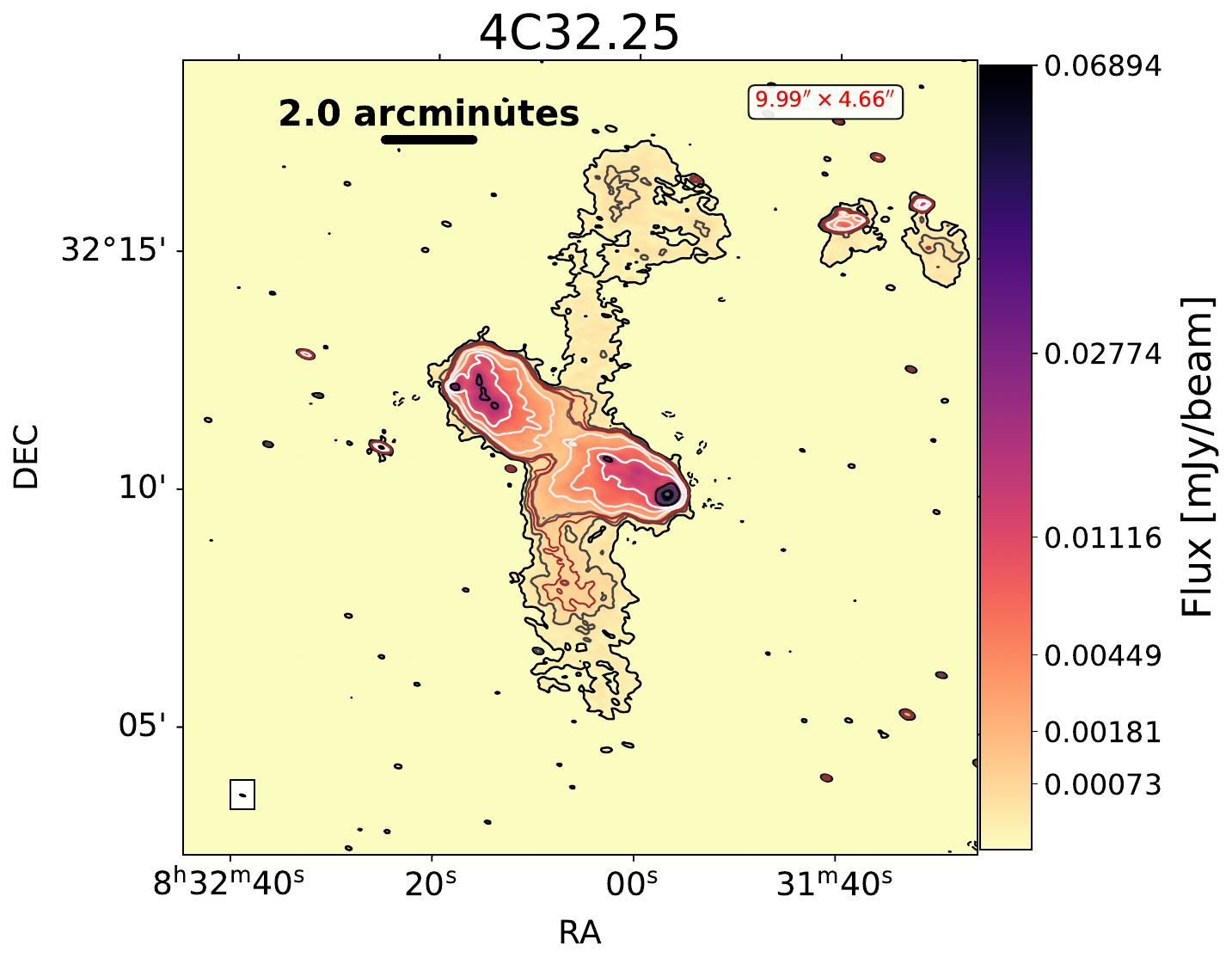}
\par\end{centering}
\begin{centering}
\includegraphics[clip,scale=0.33]{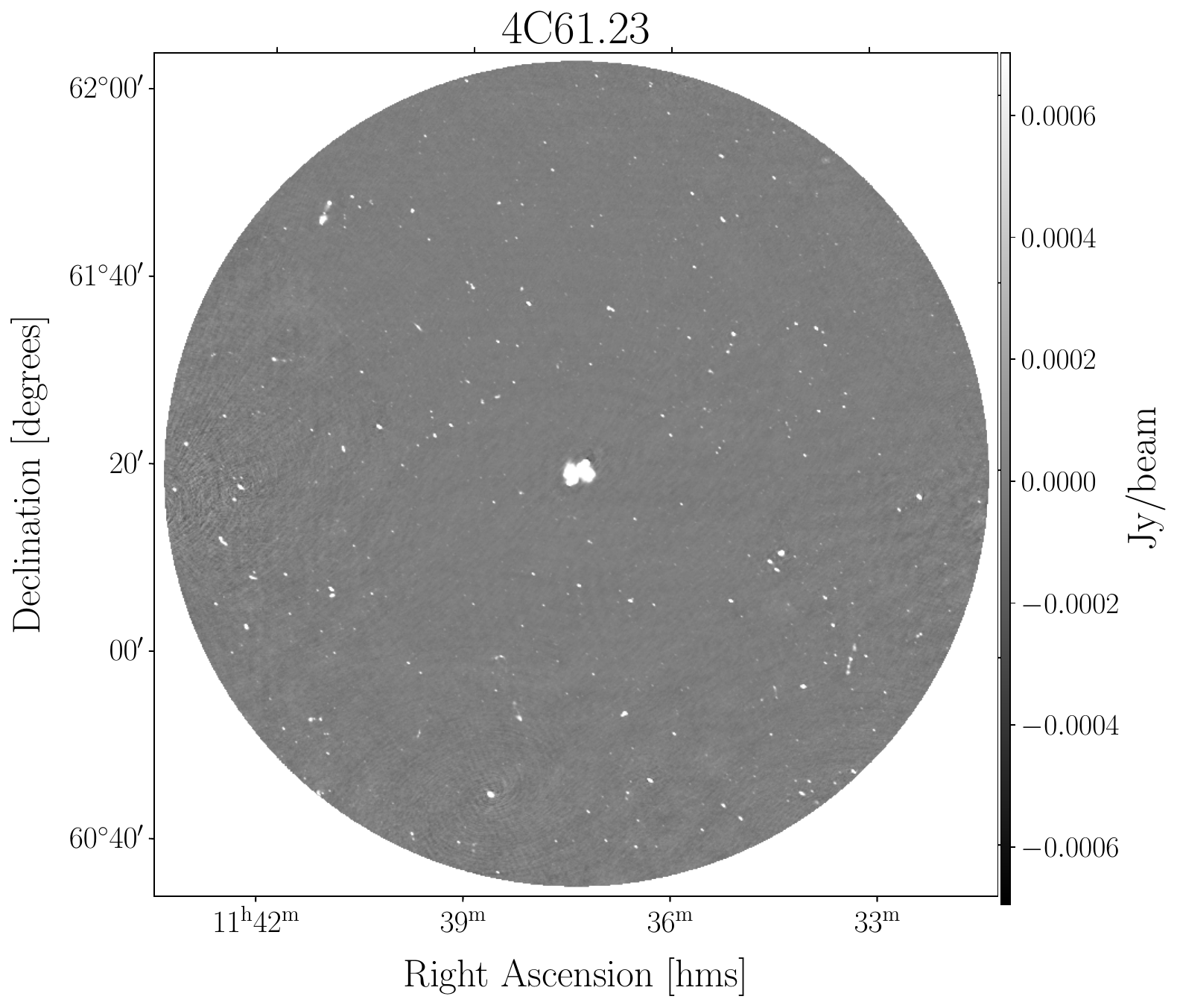}\includegraphics[clip,scale=0.38]{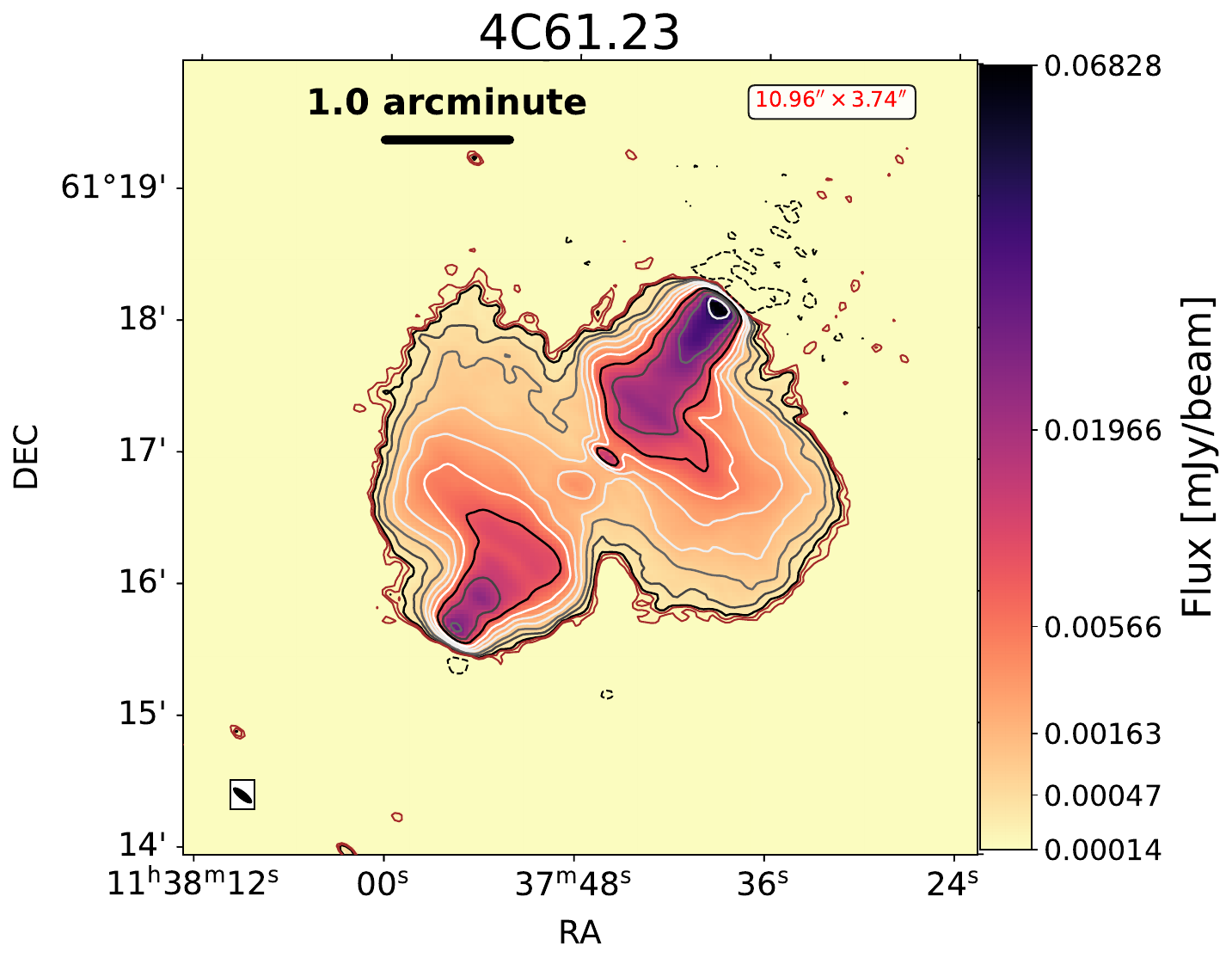}
\par\end{centering}
\centering{}\includegraphics[clip,scale=0.33]{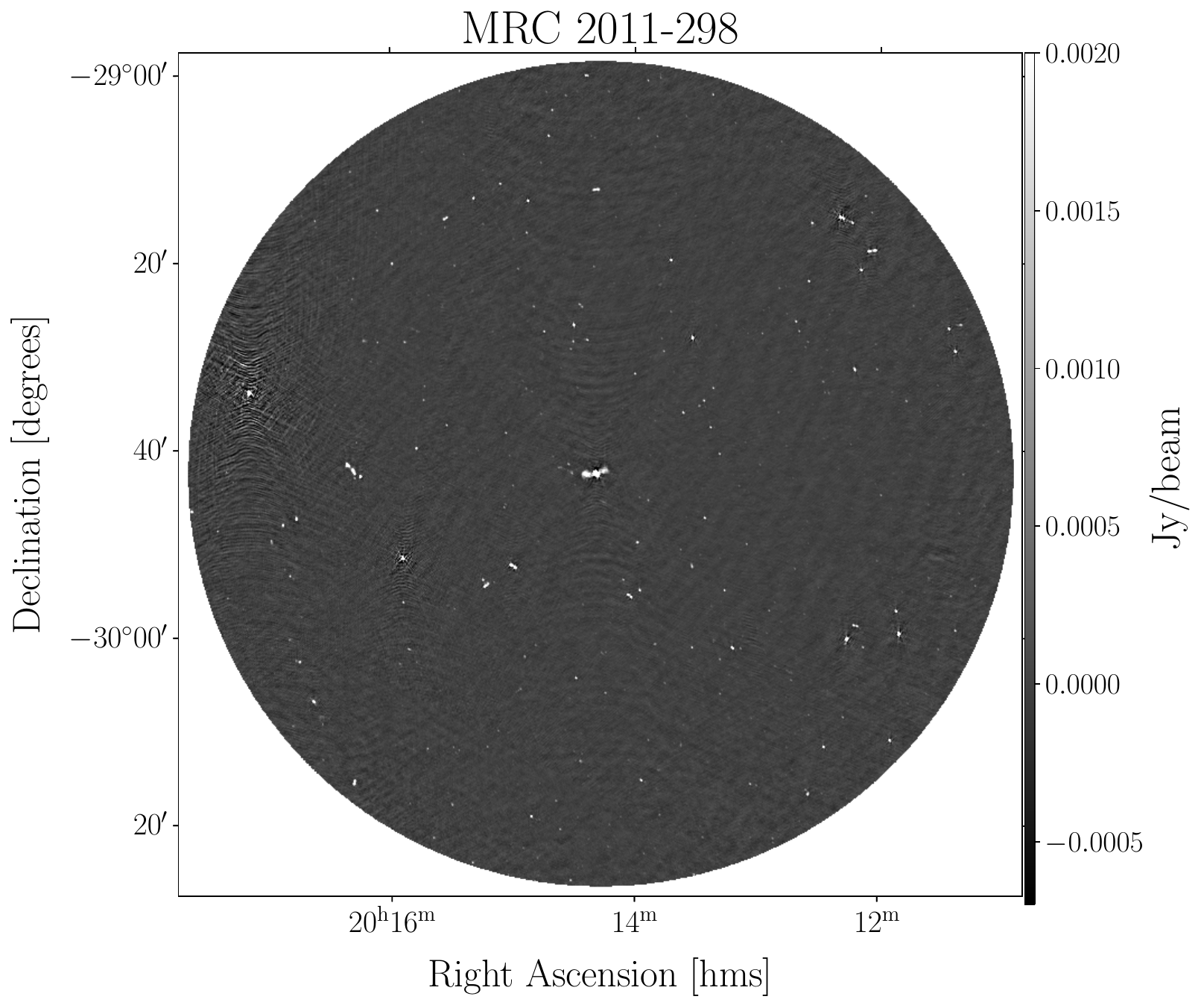}\includegraphics[clip,scale=0.38]{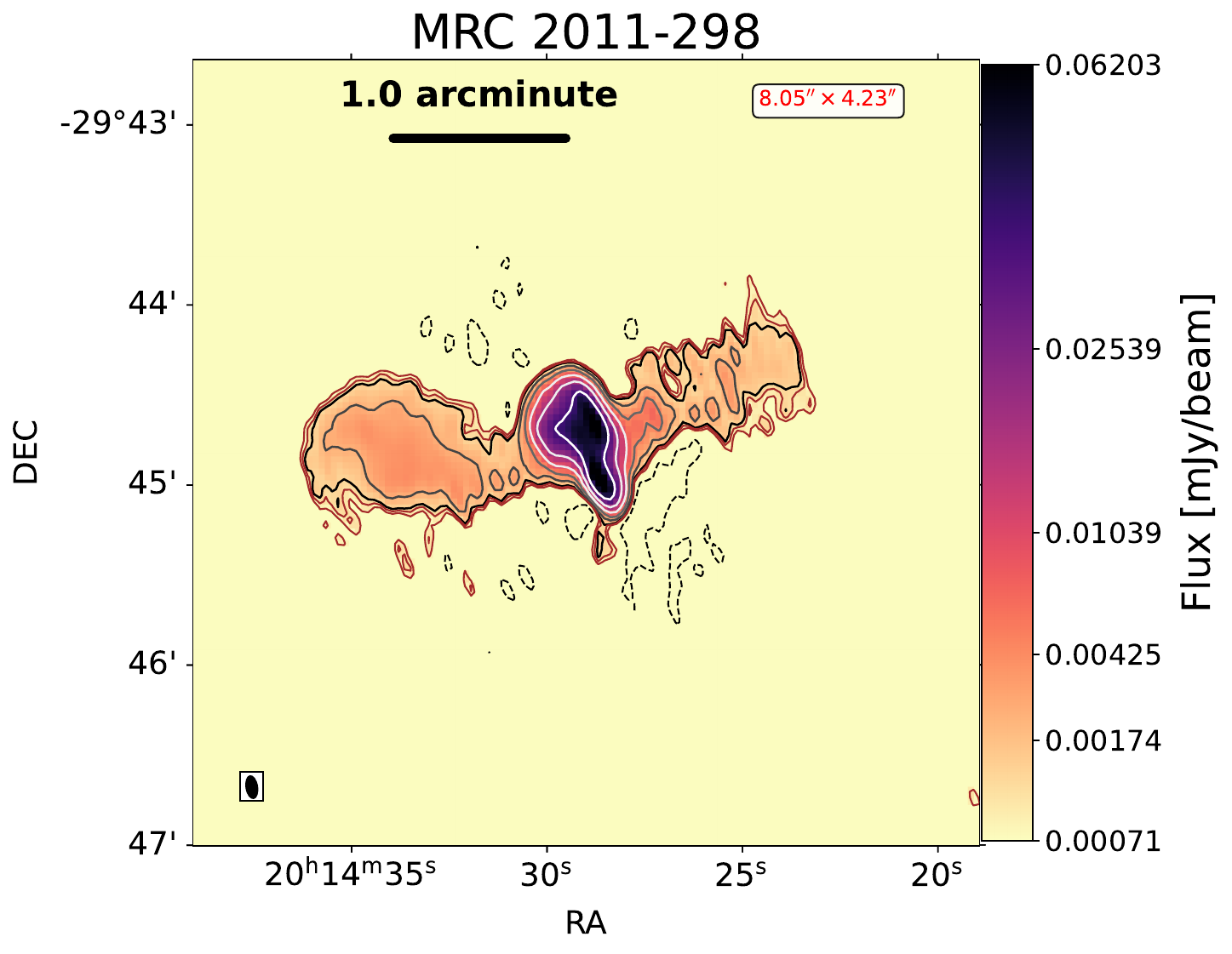}\centering\caption{\label{fig:mosaics}\textbf{ }Right: 400 MHz uGMRT mosaics of the
4C32.25, 4C61.23, and MRC 2011-298 fields. The area of each mosaic
is approximately 1.72 $\textrm{deg}^{2}.$ The details of each field
are summarized in Table \ref{fig:summary_mosaics}. Left: Full-resolution
uGMRT 400 MHz Band-3 images of the three XRGs 4C32.25, 4C61.23, and
MRC 2011-298. The contour levels are $(1,2,4,8,...)\times3\sigma$
mJy/beam, where $\sigma$ is the local noise level. The $-3\sigma$
contours are denoted by dashed lines. The restoring beam is indicated
by an ellipse in the bottom left corner. The black line indicates
the corresponding angular scale.}
\end{figure*}

The final mosaics are shown in Figure \ref{fig:mosaics}. Table \ref{fig:summary_mosaics}
summarizes the central root mean square (RMS) noise, angular resolution,
and position angle for each mosaic. 
\begin{table*}
\noindent \begin{centering}
\caption{Summary of the mosaics for the three fields considered. \label{fig:summary_mosaics}}
\begin{tabular}{cccc}
\hline 
Field & Central RMS & Resolution & Position Angle\tabularnewline
\hline 
 & {[}$\mu\,$Jy{]} & {[}arcseconds{]} & {[}degrees{]}\tabularnewline
\hline 
4C32.25 & $\sim36$ & 13.32$\times$3.6 & 63.40\tabularnewline
4C61.23 & $\sim29$ & 10.8$\times$3.6 & 51.14\tabularnewline
MRC 2011-298 & $\sim115$ & 7.92$\times$3.96 & 7.32\tabularnewline
\hline 
\end{tabular}
\par\end{centering}
\centering{}\centering
\end{table*}

We employ the Python Blob Detection and Source Finder (PyBDSF, \citealt{2015ascl.soft02007M})
package to create initial catalogs for the three fields. We use the
pre-beam-corrected mosaics as the detection images, while the primary-beam
corrected mosaics are the extraction images. The format of the three
catalogs follows the same convections introduced in RM18.

The final catalogs of the fields 4C61.23, 4C32.25, and MRC 2011-298
contain a total of 766, 505, and 375 entries, respectively, detected
above a $5\sigma$ flux density threshold. The astrometry and flux
densities have been corrected using the procedures described in RM18.
Figure \ref{fig:fluxes_comparison} shows the flux density scale comparison
between the uGMRT and the flux densities scaled to 400 MHz from the
reference surveys LoTSS and RACS. Figure \ref{fig:position_offsets}
shows the corrected positional offsets. After we apply the corrections,
the uGMRT astrometry and flux densities are in good agreement with
those of FIRST and LoTSS/RACS, respectively. The catalogs are made
available online\footnote{https://vizier.cds.unistra.fr/}. 

\section{Results \label{sec:results}}

\subsection{Spectral properties \label{sec:spectral_properties}}

\subsubsection{uGMRT images \label{sec:xrgs_images}}

The full-resolution 400 MHz uGMRT images of the XRGs are presented
in Fig. \ref{fig:mosaics}. Our images confirm the X-shape morphology
of the three radio-galaxies, i.e. a pair of bright radio-lobes and
a pair of fainter wings. This is consistent with radio-maps from the
same sources presented previously by other authors \citep{2007MNRAS.374.1085L,2019A&A...631A.173B,2024MNRAS.530.4902S,2024A&A...690A.160B}.
The radio-lobes are the brightest regions, while the core and wings
are fainter. Each of the three XRGs exhibits a different orientation
of its primary lobe axes. Both 4C32.25 and 4C61.23 are FRII sources
due to the presence of hot spots in the jets, while MRC 2011-298 lacks
hot spots in the jets making it a FRI source. The wings in 4C32.25
and MRC 2011-298 are significantly larger than their jets, which is
rare \citep{2009ApJ...695..156S,2019ApJ...887..266J}. The most striking
feature is the large-scale wings of 4C32.25, which extend over $\sim5^{'}$
(293 kpc) in length. The northern 4C32.25 wing is fainter than the
southern wing, although it is approximately $\sim11\%$ longer. MRC\,2011-298
has wings of similar length but displays asymmetrical morphologies
with deflections at the edges, which along with the S shape of the
jets could be related to jet precession \citep{2024A&A...690A.160B}.
The signature of deflections are weaker in 4C32.25, but it has been
proposed that precession could explain its X-shaped morphology \citep{1995A&A...303..427K,2024MNRAS.530.4902S}.
The radio-morphology of 4C61.23 is more symmetric with wings and jets
of similar size and a faint radio core.

Our XRGs radio-maps demonstrate the robustness of our calibration
approach to obtain high-fidelity, high-sensitivity images of diffuse
radio emission. For instance, \citet{2024A&A...690A.160B} analyzed
the same dataset using the SPAM pipeline \citep{2009A&A...501.1185I}.
In comparison, we not only do we improve the DD artifacts around the
XRG, but also we also reach deeper noise levels at higher angular
resolution (see Figs. \ref{fig:dd_selfcal} and \ref{fig:mosaics}).
Compared to \citet{2024A&A...690A.160B}, our image exhibits improvements
of about 45\% in noise levels and 50\% in angular resolution. A detailed
comparison with SPAM is outside the scope of this work. 

\subsubsection{Integrated spectra and spectral aging \label{sec:spectral_index} }

For each XRG, we combine all the flux densities available along with
prior measurements from the literature to construct integrated synchrotron
spectra. Particularly, LOFAR and uGMRT measurements trace the lowest
energy particle population and provide better constraints of the radio
spectra at low-frequencies. These spectra are calculated for the entire
source, and also for the individual components: primary lobes and
wings. Spectral ages of these components are estimated by fitting
synchrotron aging models. 

First, we compile integrated flux density measurements of 4C32.25
and 4C61.23, as shown in Table \ref{fig:total_fluxes_xrgs-1}.1. These
flux densities are taken from the literature, measured in our uGMRT,
VLA and archival images (NVSS, RACS, FIRST, LoTSS, VLASS, NVAS). The
flux density measurements from the literature have been adjusted to
bring them onto the SH12 scale adopted for our uGMRT catalogs. To
account for the uncertainty in the SH12 flux density scale, we add
in quadrature a flux density scale error factor, $\sigma_{F}$, to
the total flux density uncertainties. For frequencies below 1 GHz,
we adopt $\sigma_{F}=0.15$,, and for frequencies above 1 GHz, $\sigma_{F}=0.10$.
We notice that the flux density measurements for 4C32.25 appear to
have been underestimated in the 4C survey \citep{1965MmRAS..69..183P,1967MmRAS..71...49G},
RACS-Low, and RACS-Mid. These flux density values at 178, 887.5, and
1655 MHz have not been considered in the fits to the integrated spectrum
of the source. This flux density underestimation for 4C32.25 could
be caused by sparse UV-coverage, as the source has a large angular
extent. The flux densities can be found in Table \ref{fig:total_fluxes_xrgs-1}.1.
MRC 2011-298 is excluded from this analysis because its synchrotron
spectra was studied in detail by \citet{2024A&A...690A.160B}. For
reference, Table \ref{fig:a3670_fluxes} lists the flux densities
of the different components (primary lobes, wings) of MRC 2011-298
measured in our 400 MHz mosaic.

To determine the spectral age in different parts of the XRGs, the
time elapsed since the radiating particles were last accelerated,
we use the theory of synchrotron aging proposed by \citet{1962SvA.....6..317K}
and \citet{1970ranp.book.....P} (KP model) and later expanded by
\citet{1973A&A....26..423J} (JP model). Both models are based on
the assumption that a single, impulsive injection event produces the
initial power-law distribution of electron energies, $N\left(E\right)\varpropto E^{-p}$,
where $s=2\alpha_{\textrm{inj}}+1$, and $\alpha_{\textrm{inj}}$
is the injection index. Another model, the continuous injection (CI)
model \citep{1962SvA.....6..317K,1970ranp.book.....P} assumes that
relativistic electrons are continuously injected over the source\textquoteright s
lifetime. High-energy electrons lose energy faster than their low-energy
electrons due to various processes (synchrotron, inverse-Compton,
and adiabatic expansion losess). As a consequence, the initial power-law
spectrum develops a break, beyond which the initial $\alpha_{\textrm{int}}$
becomes steeper for frequencies higher than the break frequency, $\nu_{\textrm{b}}$.
The break frequency is related to the spectral age \citep{1987MNRAS.225....1A},

\begin{equation}
t=1.59\times10^{3}\frac{\sqrt{B}}{B^{2}+B_{\textrm{IC}}^{2}}\frac{1}{\sqrt{\nu_{\textrm{B}}\left(1+z\right)}}\left[\textrm{Myr}\right],\label{eq:t_age}
\end{equation}

\noindent where $B$ is the source magnetic field in $\mu\textrm{G}$,
$B_{\textrm{IC}}=3.25\left(1+z\right)^{2}$ is the equivalent field
strength of inverse-Compton scattering with the cosmic background
radiation in $\mu\textrm{G}$, $z$ is the source redshift, and $\nu_{\textrm{b}}$
is the break frequency in GHz. Values of $\alpha_{\textrm{int}}$
and $\nu_{\textrm{b}}$ are found by fitting the observed radio spectra.
The strength of the magnetic field can be estimated using the minimum
energy conditions \citep{1980ARA&A..18..165M,2004IJMPD..13.1549G}.
However, this formula may underestimate the magnetic field strength
because of inadequate assumptions of the frequency integration limits
\citep{1997A&A...325..898B,2005AN....326..414B}. Therefore, we consider
instead the maximum age of the radio-source that occurs when $B_{\textrm{IC}}=\sqrt{3}\,B$
\citep{1969A&A.....3..468V}. Replacing this in eq. \ref{eq:t_age}
yields the maximum spectral age 

\begin{equation}
t_{\textrm{max}}=1.55\times10^{2}\frac{1}{\sqrt{\nu_{\textrm{B}}\left(1+z\right)^{3}}}\left[\textrm{Myr}\right].\label{eq:max_age}
\end{equation}

\noindent 
\begin{figure}[tp]
\begin{centering}
\includegraphics[clip,scale=0.36]{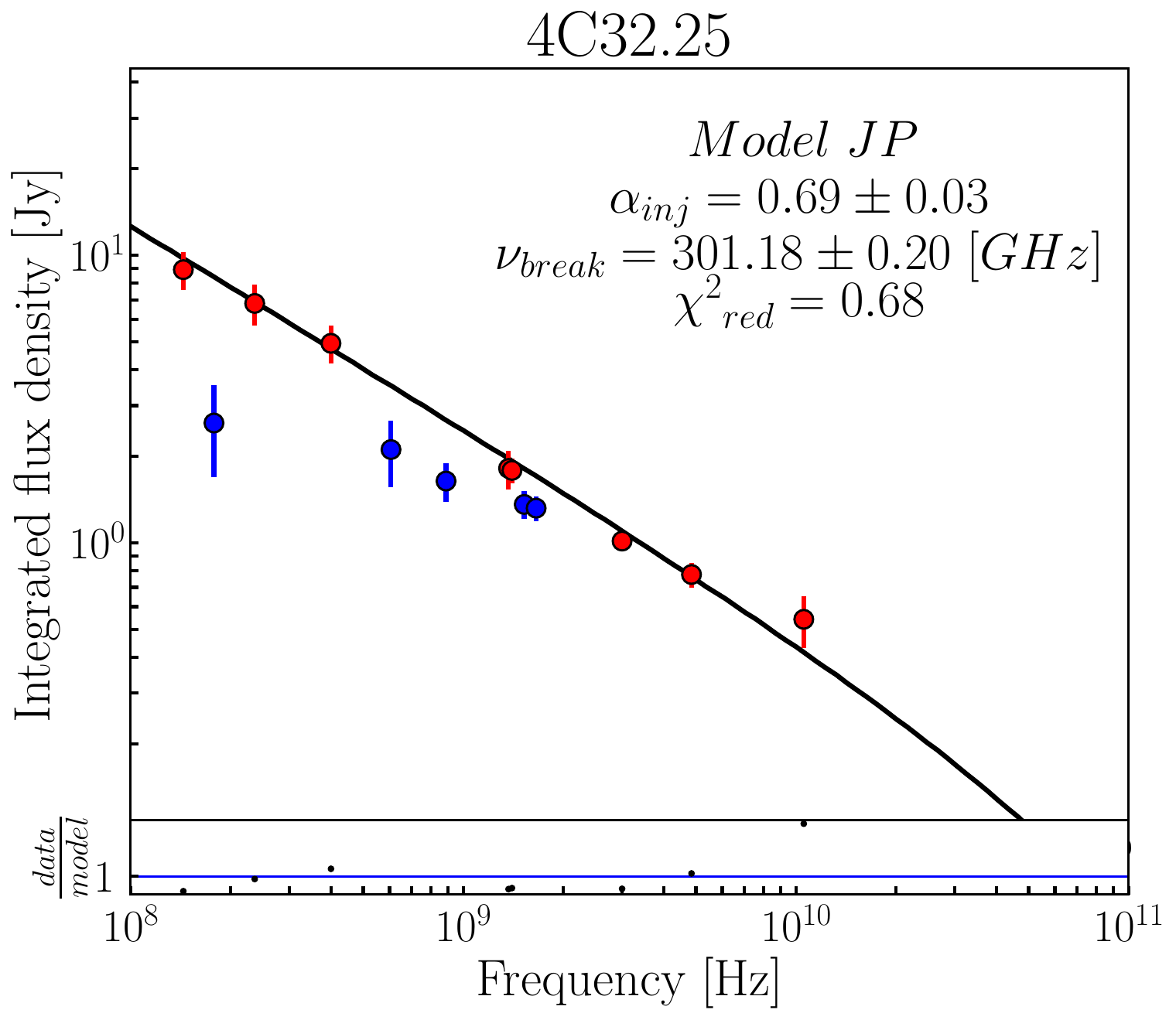}
\par\end{centering}
\begin{centering}
\includegraphics[clip,scale=0.36]{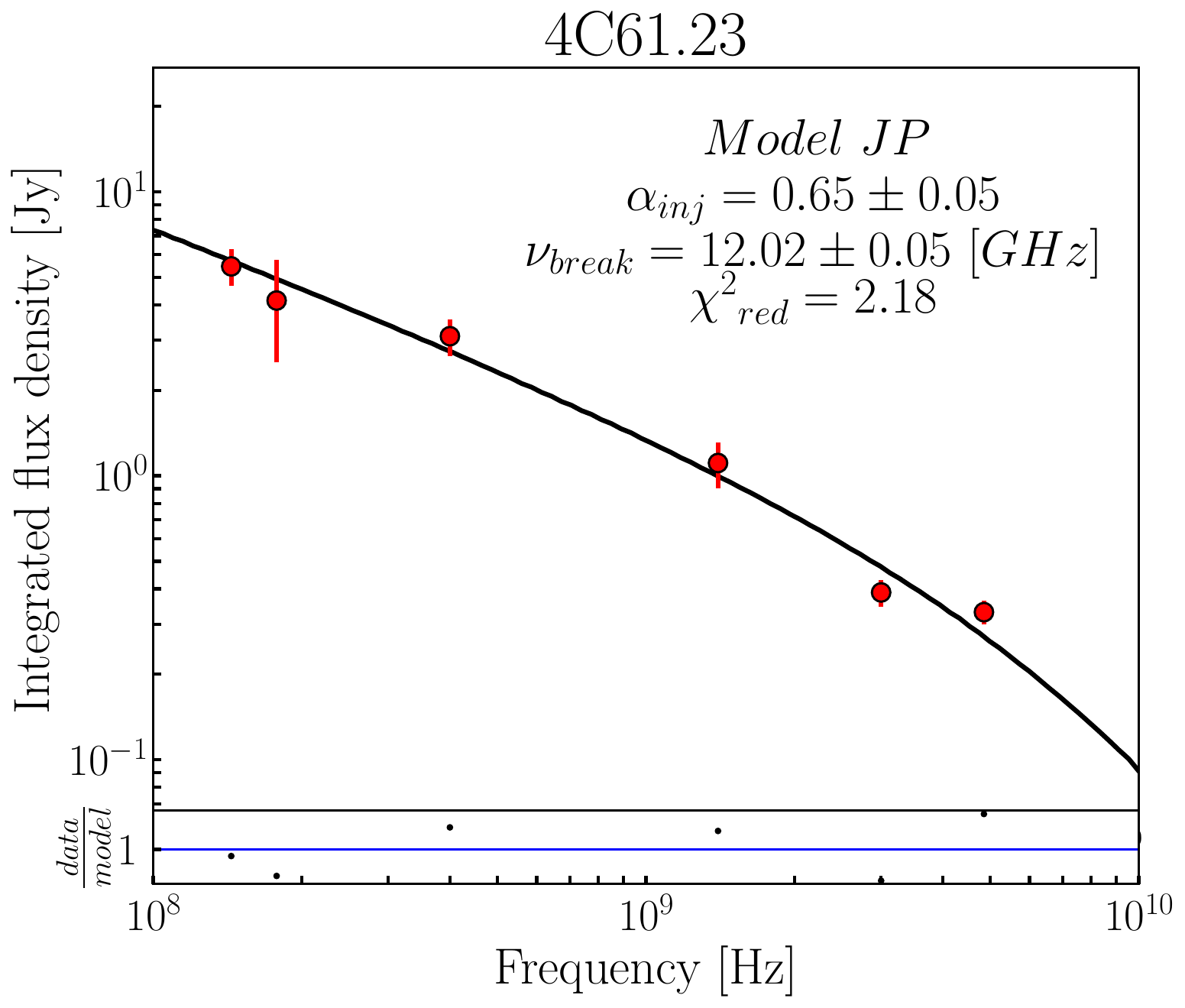}
\par\end{centering}
\begin{centering}
\centering
\par\end{centering}
\centering{}\caption{\label{fig:integrated_spectra} Integrated radio-spectra of 4C32.25
and 4C61.23. The flux densities are taken from the reduced datasets
and the literature. The solid black line is the fitted JP model. Red
points are flux densities used in the fitting, while blue points are
flux densities excluded from the fitting. The fitted parameters and
errors are indicated in the legend.}
\end{figure}

The integrated spectra of 4C32.25 and 4C61.23 and the results from
the best model fitting are presented in Figure \ref{fig:integrated_spectra}.
In general, our 400 MHz uGMRT and 3 GHz VLA flux densities are consistent
with those measured from the other surveys. For frequencies larger
than $1\,\textrm{GHz}$, there are some minor inconsistencies between
the predicted and observed flux densities. We suspect that the reason
for these discrepancies is that there are systematic uncertainties
in the flux density scale of these images. The solid lines represent
the JP model, which provides the best fit for both XRGs among the
three synchrotron models considered. The fitting is done using the
Python implementation of the Powell algorithm in LMFIT\footnote{https://lmfit.github.io/lmfit-py/}.
The 4C61.23 spectrum shows curvature, while the 4C32.25 spectrum is
fitted by a power-law within the frequency ranges considered. 

\begin{figure}[tp]
\begin{centering}
\includegraphics[clip,scale=0.46]{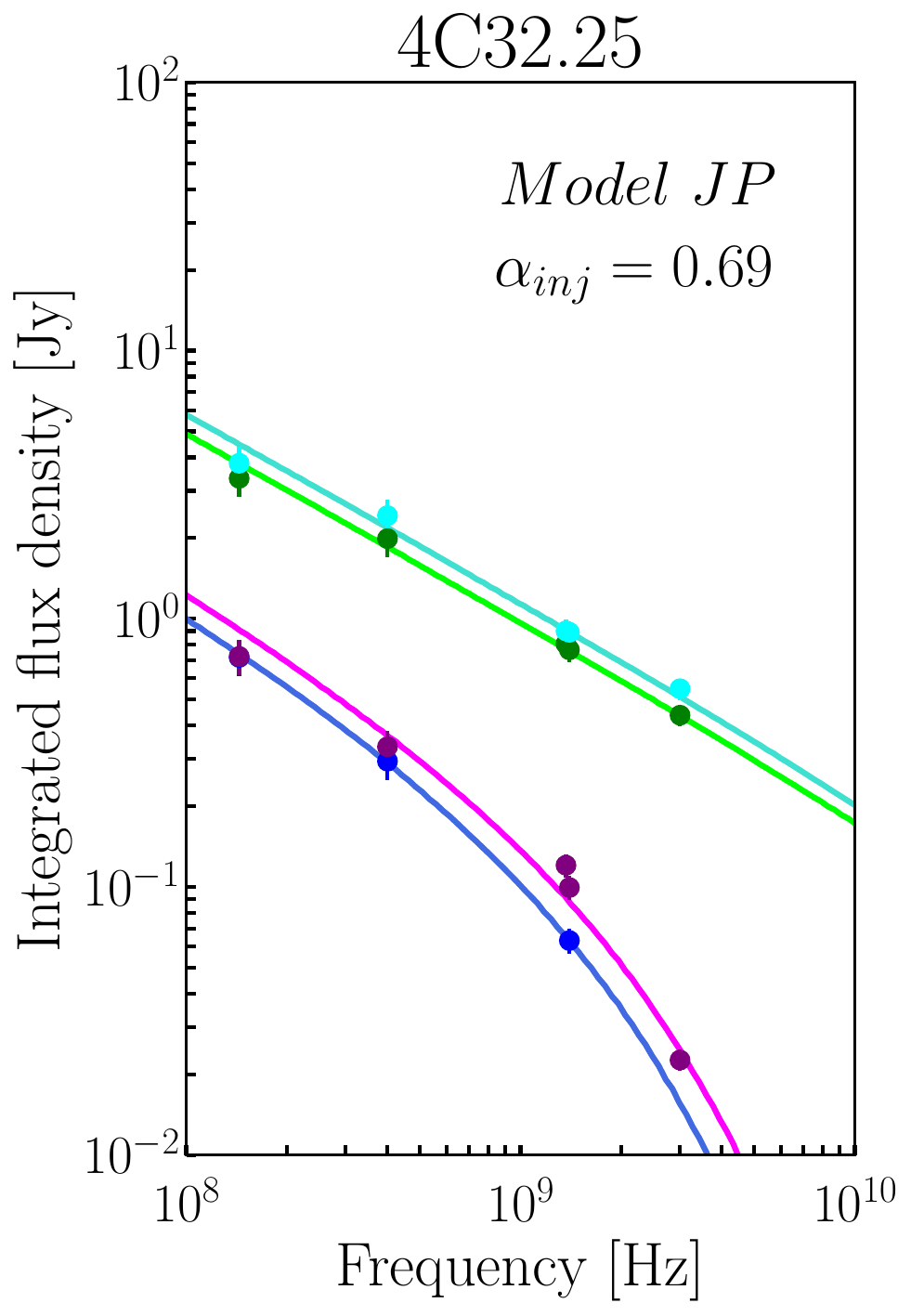}
\par\end{centering}
\begin{centering}
\includegraphics[clip,scale=0.46]{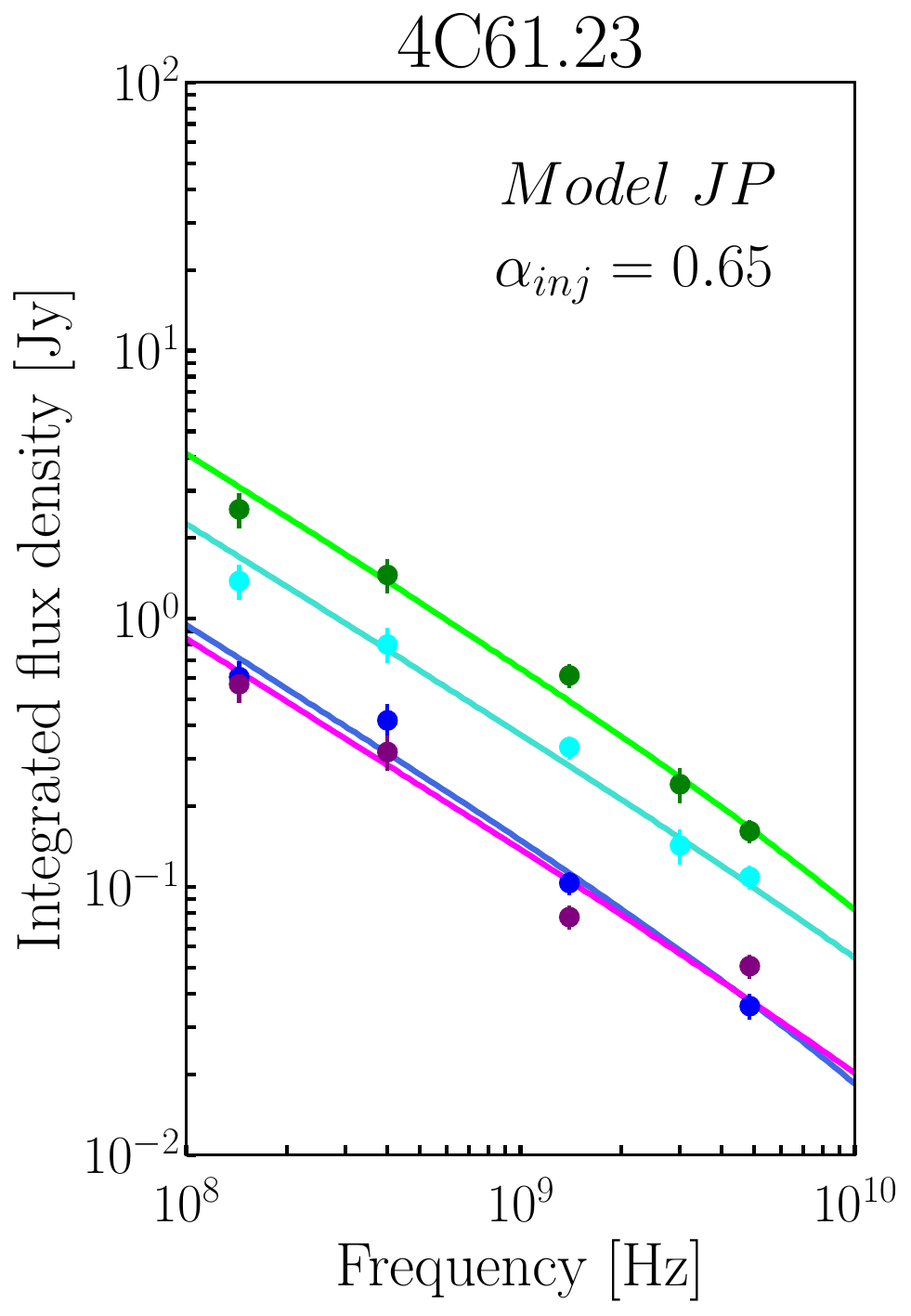}
\par\end{centering}
\begin{centering}
\centering
\par\end{centering}
\centering{}\caption{\label{fig:integrated_spectra-components} Integrated radio-spectra
of the different components (primary lobes and wings) of 4C32.25 and
4C61.23. Colored points indicate the observed flux densities of the
primary lobes (green and cyan), and wings (blue and purple). The solid
green and cyan lines show the best-fit synchrotron model of the primary
lobes, while the blue and purple lines show the best-fit synchrotron
models of the wings, respectively. All the models are calculated keeping
the injection index, $\alpha_{\textrm{int}}$ fixed to the value indicated. }
\end{figure}

Due to the limited number of observed flux density measurements available
to fit the spectra of the components (primary lobes, wings), the injection
indices are fixed to the corresponding values found for the integrated
spectra. The only free parameter of the fits is $\nu_{\textrm{b}}$.
The flux densities for each component are calculated using manually
defined DS9 regions. The flux densities are only extracted from the
images mentioned in Section \ref{sec:data}. The potential impact
of different image resolutions on the flux density measurements is
investigated. First, all images are re-gridded to the largest pixel
scale and smoothed to a common resolution matching that of the NVSS
image, which has the lowest resolution in our dataset. Next, we extract
the component flux densities and repeat the fitting process. We find
that the differences in the fits are negligible, and we conclude that
the variations in image resolution do not significantly affect our
analysis. To maintain consistency with the previous fittings, we use
the JP model to fit the radio spectra of all the components. These
fittings are displayed in Fig. \ref{fig:integrated_spectra-components}.
The radio spectra of the wings of 4C32.25 exhibit spectral curvature,
a key indicator of synchrotron aging, suggesting that these components
are the oldest in 4C32.25. Table \ref{fig:fitting_spectra_summary}.1
presents the derived physical properties and ages of 4C32.25 and 4C61.23.

We use eq. \ref{eq:t_age} and the fitted $\nu_{\textrm{b}}$ values
to determine the spectral ages of the components. For 4C32.25, we
find that the western and eastern primary lobes are the youngest components
with inferred ages of $8.59\pm0.34\:\textrm{Myr}$ and $8.09\pm0.61\,\textrm{Myr}$,
respectively. The southern and northern wings are the oldest components
with spectral ages of $82.16\pm5.32\:\textrm{Myr}$ and $90.77\pm10.35\,\textrm{Myr}$,
respectively. The northern wing could be older due to a more turbulent
magnetic field \citep{1995A&A...303..427K}. For 4C61.23, we find
that southeastern and northwestern jets are the youngest components
with spectral ages of $7.45\pm0.55\:\textrm{Myr}$ and $13.40\pm1.74\,\textrm{Myr}$,
respectively. Finally, the ages of the northeastern and southwestern
wings are $7.66\pm0.71\,\textrm{Myr}$ and $14.11\pm2.11\,\textrm{Myr}$,
respectively. 

In general, the spectral ages we have derived for 4C32.25 and 4C61.23
agree with the expectations of precessional and backflow models \citep{10.1093/mnras/210.4.929,1985A&AS...59..511P,1994A&AS..103..157M},
where the wings should be older than the primary lobes. For instance,
\citet{1995A&A...303..427K} found that the ages of the southern and
northern wings are $69\:\textrm{Myr}$ and $74\,\textrm{Myr}$, respectively.
This implies that our age estimates agree within the uncertainties.
Recently, \citet{2024MNRAS.530.4902S} estimated the spectral ages
of 4C32.25 and 4C61.23 using only uGMRT maps at 400 and 1215 MHz.
These authors calculated that for 4C61.23 (4C32.25) the ages of the
primary lobes and wings are 37 Myr ($51\:\textrm{Myr}$) and 30 Myr
($22\:\textrm{Myr}$), respectively. These findings contradict our
age estimates, as well the predictions of the models. This discrepancy
could be explained by the fact that these authors had only two frequency
measurements and assumed that the highest frequency in their data
corresponded to the break frequency. Finally, our fittings include
low-frequency observations, which are sensitive to diffuse, low-surface-brightness
emission. This allows us to detect the faint wings with higher significance
and to determine the break frequency more precisely, thereby providing
better estimates of the spectral ages.

\subsubsection{Spectral index maps \label{sec:spectral_index_maps}}

The images described in Section \ref{sec:xrgs_images} allow us to
investigate in detail the spectral index distributions of 4C32.25
and 4C61.23. The spectral index maps are determined using the standard
expression for $\alpha=log(S_{1}/S_{2})/\log(\nu_{1}/\nu_{2})$. Firstly,
we find the largest pixel scale and lowest resolution of the image
pair. Secondly, the images are re-gridded to the largest common pixel
scale. Thirdly, the images are convolved with a Gaussian kernel to
produce a circular PSF matching the size found in the first step.
The uncertainties due to the flux density scale were added in quadrature,
as described in Section \ref{sec:spectral_index}. 

\begin{figure*}[tp]
\begin{centering}
\includegraphics[clip,scale=0.62]{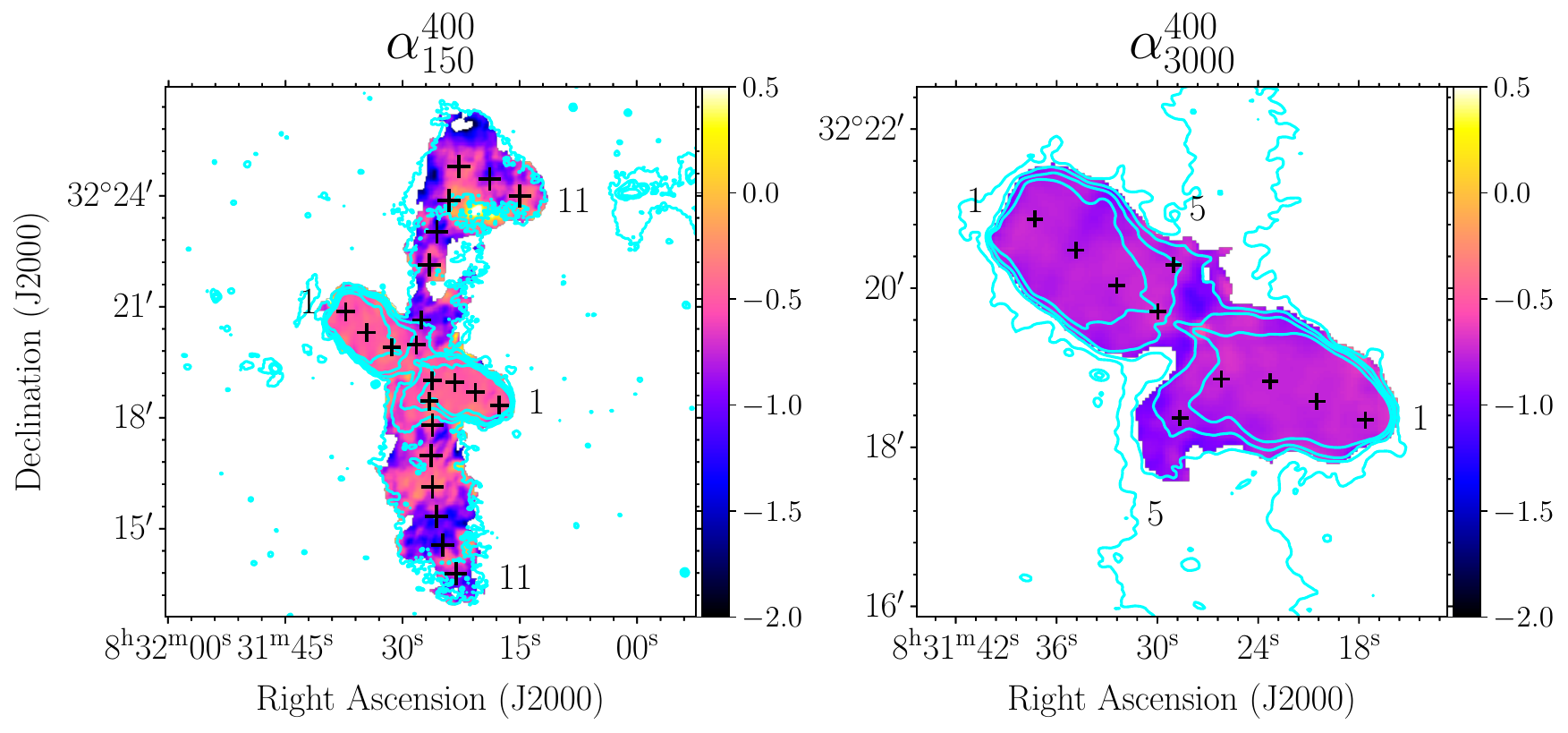}
\par\end{centering}
\begin{centering}
\includegraphics[clip,scale=0.55]{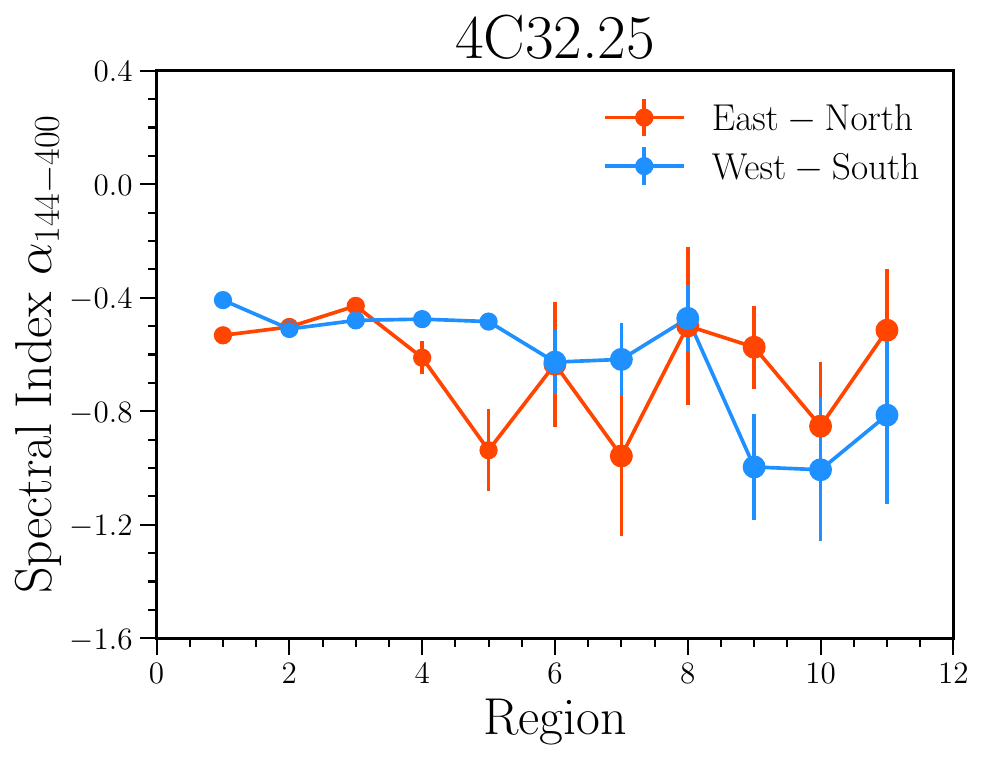}\includegraphics[clip,scale=0.55]{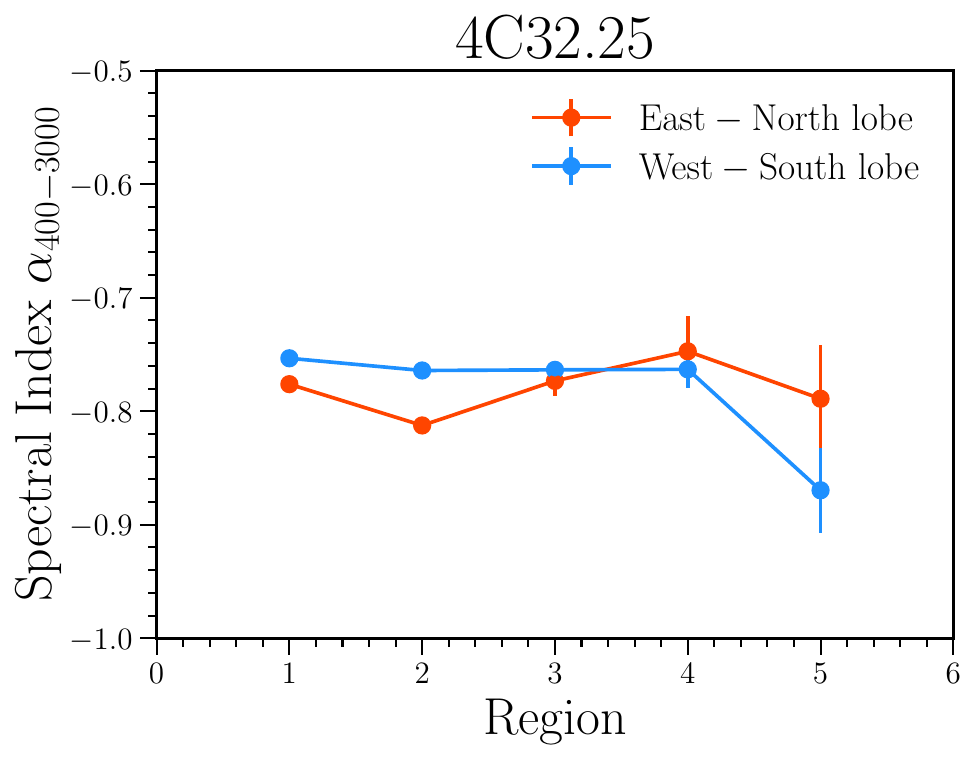}
\par\end{centering}
\begin{centering}
\centering
\par\end{centering}
\centering{}\caption{\label{fig:spectral_index_4c32-1} Top row: Spectral index maps of
4C32.25 between 144-400 MHz (left) and 400-3000 MHz (right). The resolution
of the maps is $10.1^{\prime\prime}\times10.1^{\prime\prime}$. The
spectral index maps are derived using images from LOFAR/uGMRT (left)
and uGMRT/VLA (right). The overlaid contour levels are $[3,12,24,48]\times\sigma_{\textrm{rms}}$,
where $\sigma_{\textrm{rms}}$ is the local noise from the corresponding
lower frequency image. The locations of the centers of the boxes used
to calculate the flux densities and corresponding spectral indices
are indicated by the marked positions. Larger cross symbols correspond
to areas where larger integration boxes were used for flux density
estimation. . Bottom row: Spectral index profiles of 4C32.25 along
the East-North and West-South directions along the primary lobe to
wing transition line computed from the spectral maps. The flux densities
are calculated using $11\times11$ pixel boxes (2.75 times the beam
size) in the primary lobe, and $15\times15$ pixel boxes (3.75 times
the beam size) in the wings. In the spectral maps, the first and last
boxes are indicated by number, with the locations of all boxes also
displayed in the maps. The larger crosses in the 144-400 MHz spectral
profiles denote that larger boxes were used to compute the flux densities.}
\end{figure*}

The upper panels in Figure \ref{fig:spectral_index_4c32-1} presents
the low-frequency and high-frequency spectral index maps of 4C32.25
derived between 150 MHz and 400 MHz, $\alpha_{\textrm{150}}^{\textrm{400}}$,
and between 400 MHz and 3.0 GHz, $\alpha_{\textrm{3000}}^{\textrm{400}}$;
respectively. The maps only include pixels with flux densities above
$3\sigma$. Additionally, the spectral index profiles in both maps
are calculated by integrating the flux density within $11\times11$
pixel boxes (2.75 times the beam size) placed along positions following
the ridge of maximum brightness along the East-North (EN) and West-South
(WS) directions from the primary lobes to the wings. The centers of
the boxes are indicated with crosses in Figure \ref{fig:spectral_index_4c32-1}.
These profiles are displayed in the bottom panels of Figure \ref{fig:spectral_index_4c32-1}.
In the high-frequency spectral index map, the regions in the primary
lobes exhibit values ranging from $-0.82$ to $-0.74$ with a steepening
trend towards the start of the wings with typical values of $-0.83$.
This gradual steepening of the high-frequency spectral index was also
observed by \citet{1995A&A...303..427K} and \citet{2001PhDT.......173R}.
To increase the S/N in the wings, we use larger box sizes of $15\times15$
pixels (3.75 times the beam size) to compute the flux densities in
the 150 MHz and 400 MHz maps. In the $\alpha_{\textrm{150}}^{\textrm{400}}$
map, the WS direction shows flat values of $\sim-0.45$ in the primary
lobe region, with a sudden steepening occurring approximately in the
latter half of the wing with values of $\sim-1.0$. The trend in the
EN direction appears consistent with a steepening from the hot spot
($\sim-0.55$) to the tip of the wing ($\sim-1.1$), although the
large error bars make it difficult to draw a definitive conclusion.

\begin{figure*}[tp]
\begin{centering}
\includegraphics[clip,scale=0.62]{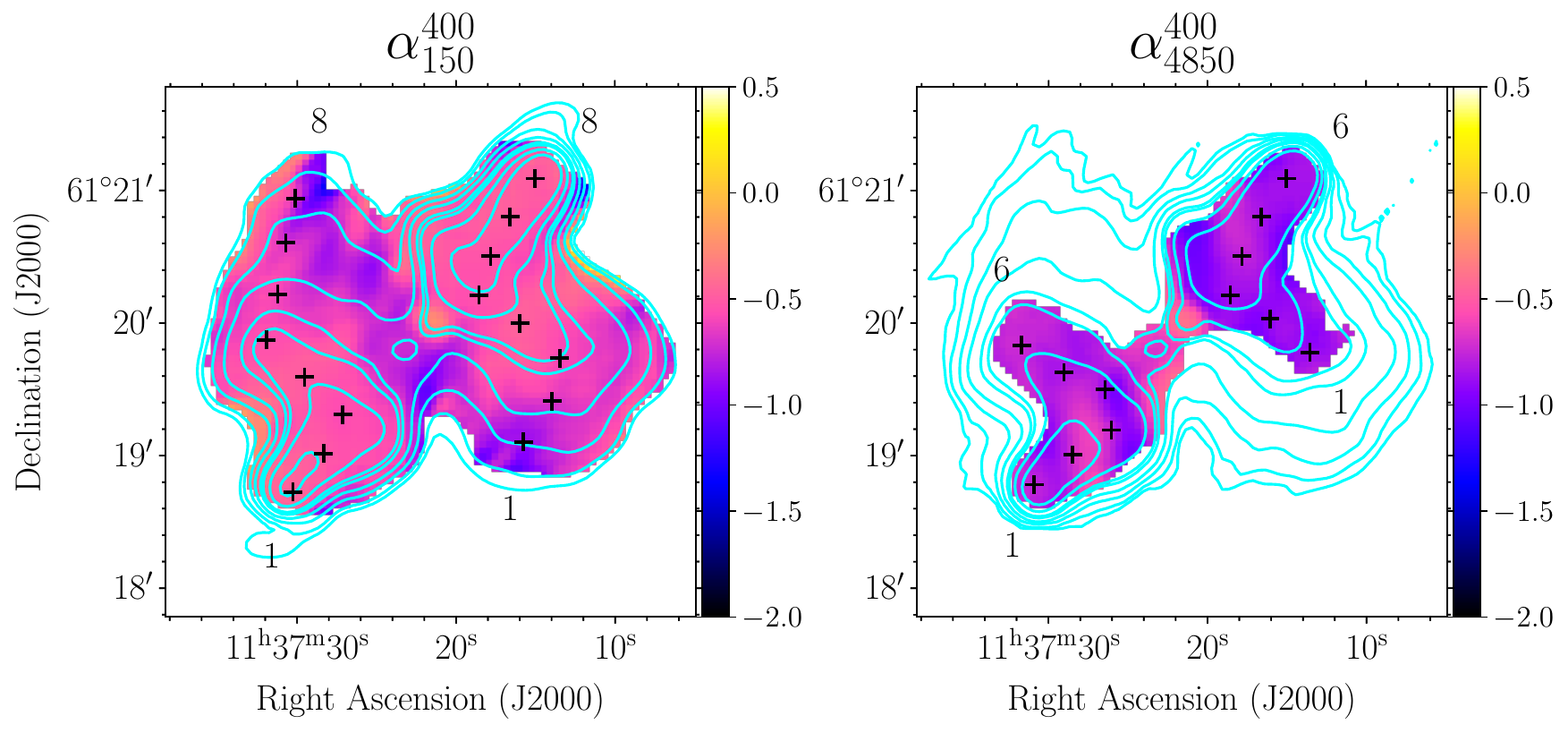}
\par\end{centering}
\begin{centering}
\includegraphics[clip,scale=0.55]{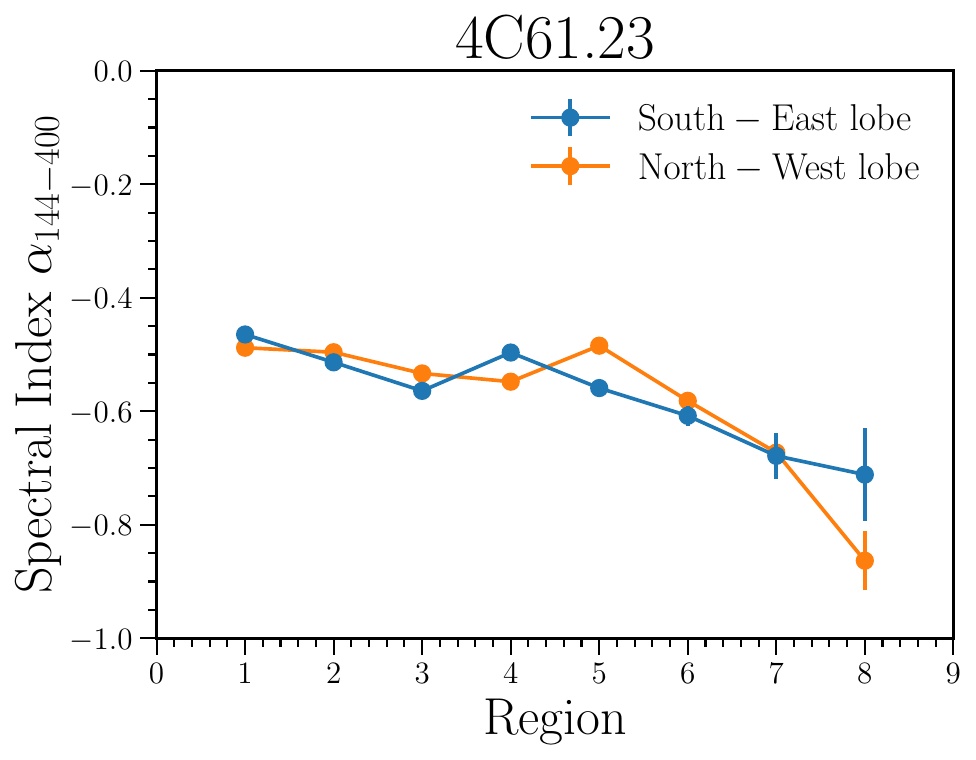}\includegraphics[clip,scale=0.55]{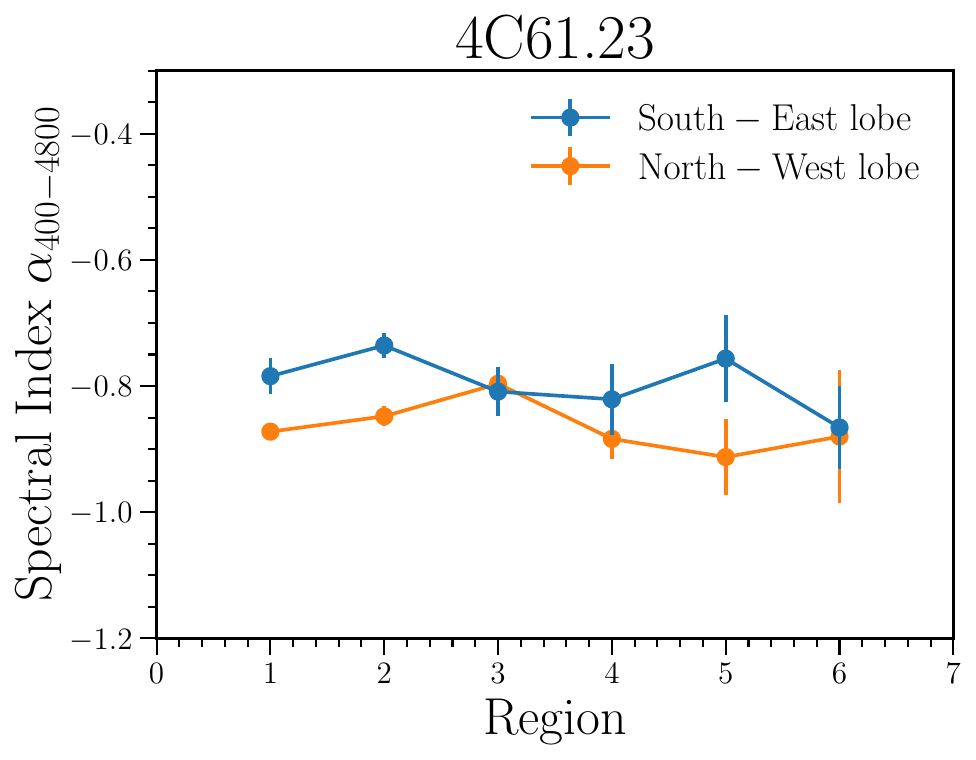}
\par\end{centering}
\begin{centering}
\centering
\par\end{centering}
\centering{}\caption{\label{fig:spectral_index_4c61}\textbf{ }Top row: Same as top row
in Fig. \ref{fig:spectral_index_4c32-1}, but for 4C61.23 using 4850
MHz instead of 3000 MHz as the highest frequency radio-map. The resolution
of the maps is $11.1^{\prime\prime}\times11.1^{\prime\prime}$. The
spectral index between 144-400 MHz map is displayed in the left, while
the 400-4850 MHz map is shown in the right. Bottom row: Same as bottom
row of Fig. \ref{fig:spectral_index_4c32-1}, but for 4C61.23. The
primary lobe to wing transition line is calculated along the South-East
and North-West directions using the spectral index maps. The box size
used is $6\times6$ pixels (1.5 times the beam size) to compute the
flux densities. The first and last boxes are labeled in the spectral
maps, with the positions of all boxes additionally displayed in the
maps.}
\end{figure*}

The low-frequency (150 MHz - 400 MHz) and high-frequency (400 MHz
- 5.85 GHz) spectral index maps of 4C61.23 are shown in the upper
panels of Figure \ref{fig:spectral_index_4c61}. Flux densities are
integrated using $6\times6$ pixel boxes (1.5 times the beam size).
The South-East (SE) and North-West (NW) profiles (bottom panels of
Figure \ref{fig:spectral_index_4c61}) along the maximum brightness
ridges obtained from the low-frequency spectral index map indicate
a moderate and gradual steepening from the hot spots ($\sim-0.47$)
in the primary lobes to the tips of the wings ($\sim-0.55$). In the
high-frequency spectral index map, there is no significant change
in the overall shape of the SE and NW profiles considering the associated
uncertainties. The SE (NW) profile varies from $-0.78$ ($-0.87$)
at the hot spot to $-0.87$ ($-0.88$) at the tip of the wing. 

\noindent 

\noindent 

\subsection{Constraints on models derived from radio-observations \label{sec:explanation_diffuse_emission}}

The spectral ages and gradients found in Sections \ref{sec:spectral_index}
and \ref{sec:spectral_index_maps}, respectively can be useful to
constrain the formation models of 4C32.25 and 4C61.23. Particularly,
in the jet reorientation \citep{1985A&AS...59..511P,1994A&AS..103..157M}
and backflow \citep{10.1093/mnras/210.4.929} models, it is expected
that the wings are relic emission of previously active radio-lobes.
For instance, it has been proposed that jet precession could explain
the radio-morphology of 4C32.25 \citep{1985A&AS...59..511P,1995A&A...303..427K,2024MNRAS.530.4902S}.
On the other hand, the radio morphology of 4C61.23 suggests the wings
are formed by strong backflows from both radio lobes, which are deflected
in opposite directions perpendicular to the jet axis. This ``double-boomerang''
morphology has been explained in some XRGs using the backflow model
(e.g., \citealt{2020MNRAS.495.1271C,2023MNRAS.524.3270P}). Our spectral
analyses of both XRGs yield older spectral ages for the wings and
a gradual steepening from the young lobes to the old wings, as expected
in the precession and backflow models, respectively. Also, \citet{2024MNRAS.530.4902S}
proposed that precession could also explain the radio-morphology of
4C61.23 with a precession period of $4.5\leq P_{\textrm{prec}}\left(\textrm{Myr}\right)\leq70$,
which is compatible with our spectral age estimates. 

Finally, a sudden reorientation of the jet axis, or a spin-flip event,
could plausibly account for the radio-morphology observed in some
XRGs \citep{2002MNRAS.330..609D,2002Sci...297.1310M}. For 4C61.23
and 4C32.25, the detection of double-peaked emission lines \citep{2024MNRAS.530.4902S}
suggests the presence of binary black holes (BBH), whose coalescence
may have caused the sudden spin change \citep{2002Sci...297.1310M}.
VLBI observations support the presence of a BBH in 4C61.23 \citep{2018ApJ...854..169L,2024MNRAS.530.4902S}.
However, for 4C61.23, jet-axis reorientation can be ruled out on radio-morphological
grounds, as the wings appear to originate from backflow processes.
For 4C32.25, the situation is different. First, we cannot reach the
same conclusion as for 4C61.23, since no evidence of backflows is
observed in 4C32.25. Second, the spin-flip scenario cannot be discarded,
as a sudden jet reorientation may have occurred prior to the wings
expanding to their current extent. Therefore, we conclude that more
detailed observational and simulation studies are needed to understand
the role of a spin-flip in shaping the radio-morphology of 4C32.25.

\subsection{Serendipitous radio-sources \label{sec:diffuse_sources} }

\begin{figure*}[tp]
\begin{centering}
\includegraphics[clip,scale=1.05]{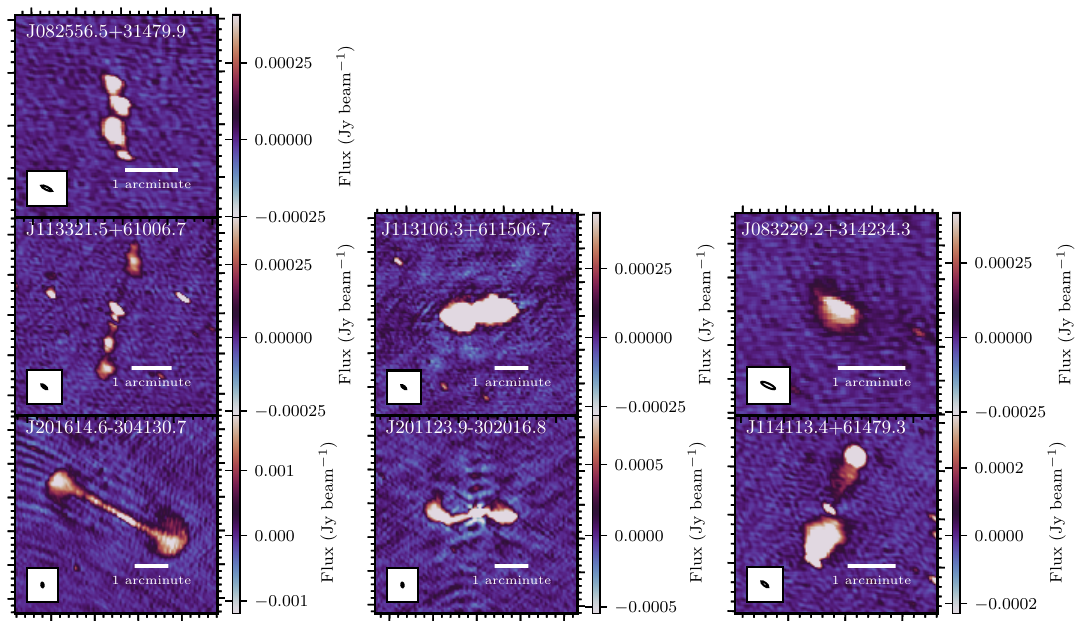}
\par\end{centering}
\centering{}\centering\caption{Images showing the radio-galaxies discussed in Section \ref{sec:diffuse_sources}.
The corresponding J2000 source names are indicated in the top center
of each image. The beam size is shown in the white inset in the bottom
left corner. The white bar in each image denotes indicates a scale
of $1^{'}.$ \label{fig:stamps_1}}
\end{figure*}

The discovery of serendipitous radio-sources is important because
it allows us to uncover previously unknown objects without requiring
new observations (e.g., \citealt{2022A&A...660A...2O,2022MNRAS.513.1300N}).
Furthermore, serendipitous discoveries maximize the scientific output
of radio-surveys by providing information in target fields that were
observed for other purposes. In this section, we discuss some serendipitous
radio-sources located in our XRGs fields and cross-reference them
with the literature. Some of these sources lie beyond the first sidelobe
of our mosaic, where beam attenuation affects the accuracy of their
flux density measurements \citep{2015frao.book.....M}. Nevertheless,
our images still reveal important morphological features that provide
insights into the evolution of these radio-sources. We calculate photometric
redshifts for objects without spectroscopic confirmation. We use optical,
and near and mid infrared photometry from the Pan-STARRS \citep{2016arXiv161205560C},
2MASS \citep{2006AJ....131.1163S}, and unWISE surveys \citep{2019ApJS..240...30S},
respectively. The coverage of the Legacy Survey \citep{2019AJ....157..168D}
in our XRG fields is only partial, with some fields missing two bands.
For this reason, we employ Pan-STARRS optical photometry to estimate
the photometric redshifts. We follow the procedure described in \citet{2020AA}
using the EAZY galaxy templates \citep{2008ApJ...686.1503B} modified
to take into account dust extinction in the host galaxies. Figures
\ref{fig:cluster_4c32} and \ref{fig:stamps_1} display all the serendipitous
sources included in our sample. 

\subsubsection{Galaxy cluster RMJ083056.4+322412.2}

\begin{figure*}[tp]
\begin{centering}
\includegraphics[clip,scale=0.3]{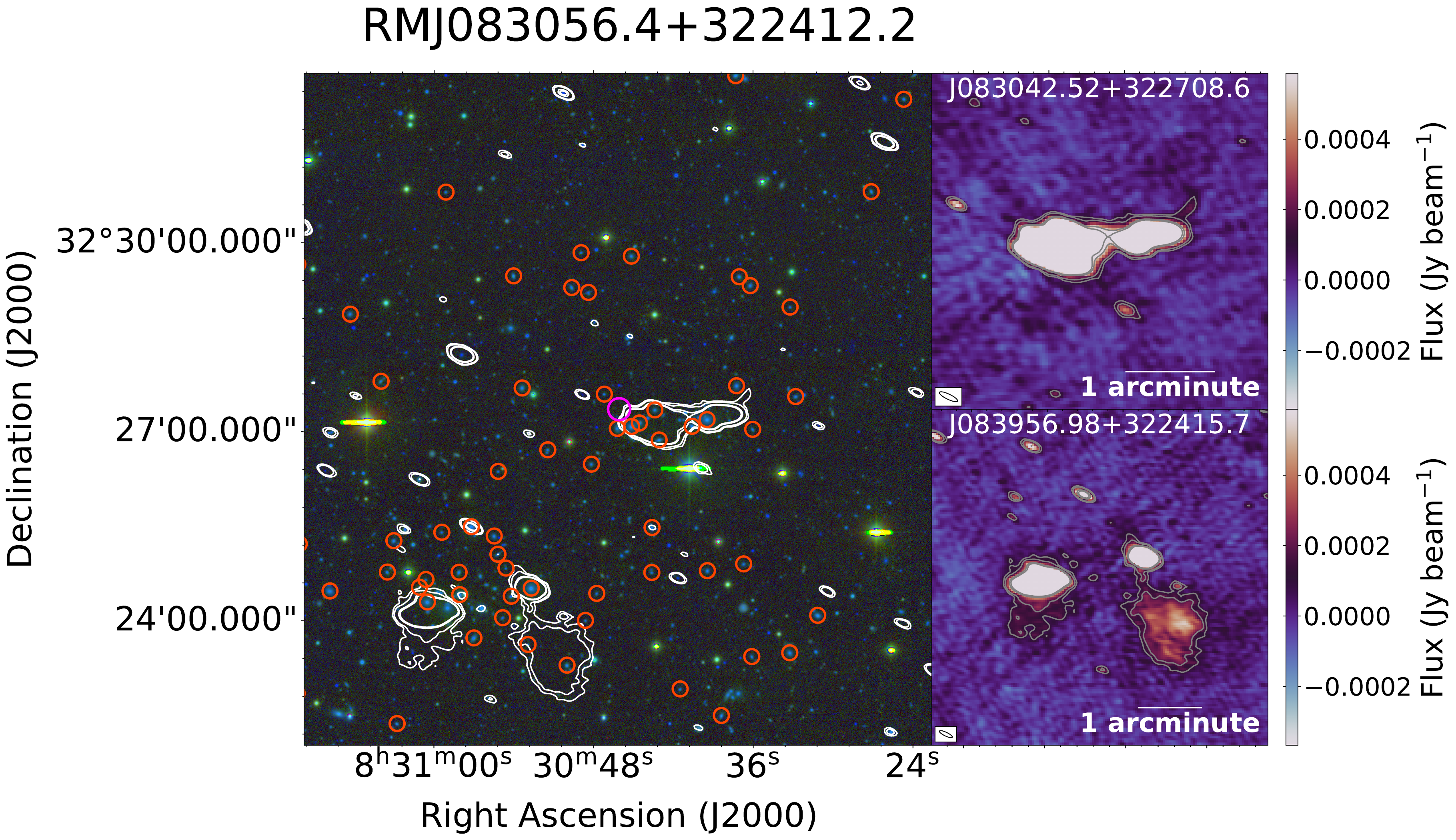}
\par\end{centering}
\centering{}\centering\caption{Left: False color RGB (R\LyXThreePerEmSpace =\LyXThreePerEmSpace g,
G\LyXThreePerEmSpace =\LyXThreePerEmSpace r, B\LyXThreePerEmSpace =\LyXThreePerEmSpace z)
Legacy survey image \citep{2019AJ....157..168D} centered on the galaxy
cluster RMJ083056.4+322412.2 located at $z_{\textrm{spec}}=0.255$.
The image covers $600\lyxmathsym{\textacutedbl}\LyXFourPerEmSpace\times\LyXFourPerEmSpace600\lyxmathsym{\textacutedbl}$.
The contours are $[3,5,15,20]\times\sigma$ times the local noise
level in the 400 MHz uGMRT map. The purple circle indicates the cluster
center, and the red circles denote spectroscopically confirmed cluster
members. Right: 400 MHz uGMRT image of two radio-sources likely associated
with RMJ083056.4+322412.2. The contour levels are the same as in the
left panel. The beam size is shown in the white inset in the bottom
left corner. \label{fig:cluster_4c32}}
\end{figure*}

RMJ083056.4+322412.2 (RMJ0830) was first identified as an optically-selected
galaxy cluster candidate \citep{2010ApJS..191..254H,2014ApJ...785..104R}.
The cluster redshift $z_{\textrm{spec}}=0.255$ was confirmed in the
HeCS-red spectroscopic survey \citep{2018ApJ...862..172R}. The core
of this cluster is $11.51^{\prime}$ from 4C32.25 (see Figure \ref{fig:cluster_4c32}).
Two extended radio sources, J083956.98+322415.7 and J083042.52+322708.6,
appear to be associated with two clumps of spectroscopic cluster members
near the cluster core. J083956.98+322415.7 was detected in previous
observations of 4C32.25 \citep{1985A&AS...59..511P,1994A&AS..103..157M,1995A&A...303..427K},
but it is only in this work that it has been associated with the cluster
RMJ0830. The total flux densities are $S_{\textrm{400MHz}}=38.9\pm5.87\:\textrm{mJy}$
and $S_{\textrm{400MHz}}=210.9\pm31\:\textrm{mJy}$, respectively.
Both sources are also detected in the LoTSS \citep{2022A&A...659A...1S}
and NVSS \citep{1998AJ....115.1693C} surveys. 
 
\subsubsection{Radio-galaxy J201614.6-304130.7}

As shown in Figure \ref{sec:diffuse_sources}, the radio-galaxy J201614.6-304130.7
(also known as PKS 2013-308, \citealt{1974AuJPA..32....1S}) is resolved
at the resolution of the uGMRT Band-3. The optical counterpart has
a magnitude of $i_{\textrm{PS1}}=16$ with a redshift of $z_{\textrm{spec}}=0.088$
\citep{2009MNRAS.399..683J}. This radio-galaxy has been observed
over the course of several decades \citep{1974AuJPA..32....1S,1981MNRAS.194..693L,1991Obs...111...72L,1992ApJS...80..137J}.
As the source lies beyond the first sidelobe, its flux density measurements
in our mosaic are not accurate due to significant beam attenuation
at that distance. The reported NVSS flux density is $S_{\textrm{1.4GHz}}=853.6\pm15.5\:\textrm{mJy}$
\citep{1998AJ....115.1693C}. The source is also detected in RACS
maps \citep{2020PASA...37...48M}. 

\subsubsection{Radio-galaxy J201123.9-302016.8}

The optical counterpart is clearly identified between the radio lobes,
with a redshift of $z_{\textrm{spec}}=0.088$ \citep{2009MNRAS.399..683J}.
No 400 MHz flux density is reported for this source because it lies
beyond the first sidelobe in our mosaic.. However, it has a total
flux density of $S_{\textrm{1.4GHz}}=132.70\pm4.3\:\textrm{mJy}$
in the NVSS catalog \citep{1998AJ....115.1693C}, and it is detected
in RACS maps \citep{2020PASA...37...48M}. This source exhibits asymmetric
jets, with both jets showing signs of bending. Both bends are likely
caused by the galaxy's orbital motion, as well as interactions with
the intergalactic medium, as seen previously in bent radio-galaxies
\citep{2024A&A...691A.193W,2025A&A...695A.178V}. 

\subsubsection{Radio-galaxy J114113.4+61479.3}

The radio-source is associated with a Pan-STARRS source of $i_{\textrm{PS1}}=19.22$
with a photometric redshift of $z_{\textrm{photo}}=0.46$. The source
displays an FRII-type radio morphology, characterized by asymmetric
jets. The total flux densities are $S_{\textrm{400MHz}}=50.8\pm7.6\:\textrm{mJy}$
(this work) and $S_{\textrm{1.4GHz}}=63.4\pm2.24\:\textrm{mJy}$ (NVSS),
respectively.

\subsubsection{Radio-galaxy J113321.5+61006.7}

This radio-source exhibits evidence of recurrent activity, resulting
in a morphology reminiscent of that of double-double radio-galaxies
(DDRGs) with two pair of radio-lobes \citep{2006MNRAS.366.1391S,2007MNRAS.378..581J}.
We associate this source with a faint optical source with $i_{\textrm{PS1}}=21.70$
and with a photometric redshift of $z_{\textrm{photo}}=0.08$ located
between the inner double. We find that the total flux density in our
mosaic is $S_{\textrm{400MHz}}=17.85\pm4.3\:\textrm{mJy}$, while
$S_{\textrm{1.4GHz}}=16\pm0.9\:\textrm{mJy}$ according to the NVSS
catalog. 

\subsubsection{Radio-galaxy J113106.3+611506.7}

The radio-source J113106.3+611506.7 (also known as 4C61.22, \citealt{1965MmRAS..69..183P,1967MmRAS..71...49G})
was first discovered in the 4C survey \citep{1967MmRAS..71...49G}.
The radio structure of the source revealed by our 400 MHz maps consists
of two bright lobes but with undistinguished core. The source is located
beyond the first sidelobe in our mosaic thus its flux density cannot
be determined reliably. The reported NVSS flux density is $S_{\textrm{1.4GHz}}=546.6\pm13.80\:\textrm{mJy}$.
The host galaxy, with $i_{\textrm{PS1}}=19.37$, has been identified
as the BCG of a galaxy cluster at $z_{\textrm{spec}}=0.349$ \citep{2012ApJS..199...34W,2015ApJ...807..178W}. 

\subsubsection{Extended radio-source J083229.2+314234.3}

Our 400 MHz mosaic reveals large-scale radio-emission surrounding
a compact radio-core. There is a clear optical counterpart ($i_{\textrm{PS1}}=21.24$
and $z_{\textrm{photo}}=0.49$) coincident with the radio-peak of
this source. Moreover, the diffuse component's extension is also confirmed
in the LoTSS images, where it appears fuzzier. We find that the total
flux densities are $S_{\textrm{400MHz}}=13.50\pm2.05\:\textrm{mJy}$,
and $S_{\textrm{1.4GHz}}=5.6\pm0.4\:\textrm{mJy}$ as listed in the
NVSS catalog. The extended emission associated with this optical source
could be related to previous episodes of nonthermal nuclear activity
as observed in fossil radio-galaxies \citep{2004rcfg.proc..335K,2023MNRAS.524.6052R}
and other galaxies \citep{2024A&A...683A.175K,2024MNRAS.52711233K}.

\subsection{Radio-galaxy J082556.5+31479.9}

The source J082556.5+31479.9 (also known as B2 0822+31, \citealt{1970A&AS....1..281C})
shows a radio-morphology with signatures of intermittent jet-formation
in the form of two pairs of jets as seen in DDRGs \citep{2000MNRAS.315..371S,2019A&A...622A..13M,2019MNRAS.486.5158N}.
The source location lies beyond the first sidelobe thus we do not
determine its flux density using our mosaic. The listed NVSS flux
density is $S_{\textrm{1.4GHz}}=162.2\pm5.2\:\textrm{mJy}$. We find
a unWISE counterpart ($W_{\textrm{1}}=18.07$) at the expected location
of the host galaxy between the pairs of inner lobes, but no optical
counterpart in the Pan-STARRS images. This could indicate that there
is significant dust obscuration in the host (e.g., \citealp{2021MNRAS.506.3641G}).

\section{Summary and Conclusions \label{subsec:conclusions}}

In this paper, we have presented a modification of the direction-dependent
calibration method based on the extreme peeling technique to produce
wide-field and high-resolution uGMRT Band-3 images. Our calibration
approach assumes that the FOV can be divided into different regions,
or facets, with each facet containing a source or source group that
is bright enough to obtain high S/N calibration solutions. These solutions
are applied, and the visibilities are imaged to obtain a new facet
skymodel. Later, this skymodel is subtracted from the UV data. The
process is iterated over all facets, allowing the correction of DDES
effects across the FOV. We conclude that this calibration method is
robust enough to study not only the central XRG targets but also other
diffuse radio sources located far from the field centers. This method
could be extended to be used with other uGMRT bands. For instance,
Bands-4 and Bands-5 have better angular resolution and reduced ionospheric
effects, making the technique promising for detailed studies of compact
AGN features. On the other hand, applying this approach to Band-2
could help mitigate phase errors caused by stronger ionospheric distortions.
Finally, our method offers a promising tool for calibrating both archival
and future uGMRT observations of XRGs as well as other AGN.

Our 400 MHz uGMRT images are combined with archival radio-data to
perform a spectral analysis of 4C32.25 and 4C61.23 (see Sections \ref{sec:xrgs_images},
\ref{sec:spectral_index}, \ref{sec:spectral_index_maps}, and \ref{sec:diffuse_sources}).
We find that the wings in 4C32.25 and 4C61.23 are the oldest components
of the XRGs. The average spectral ages of the wings of the former
is $79.03\pm10.96\,\textrm{Myr},$ and that of the latter is $9.74\pm1.93\,\textrm{Myr}$.
In the spectral index map, we observe a steepening from the hotspots
in the primary lobes toward the wings, consistent with the derived
spectral ages. These estimates agree with previous results from the
literature and align with the predictions from the precession and
backflow models. Finally, we highlight several serendipitous radio-sources
found in our XRG fields and investigate their associations reported
in the literature (Section \ref{sec:diffuse_sources}). These radio-sources
include the galaxy cluster RMJ0830, as well as various radio-galaxies
showing signatures of intermittent jet activity and fossil radio-emission.

\begin{acknowledgements}

The financial assistance of the South African Radio Astronomy Observatory (SARAO) towards this research is hereby acknowledged. 

We thank the anonymous referee for the helpful comments that improved this work. 

Computations were performed on Hippo at the University of KwaZulu-Natal, and at the Centre for High Performance Computing (project ASTR1534). 

We thank the staff of the GMRT that made these observations possible. GMRT is run by the National Centre for Radio Astrophysics of the Tata Institute of Fundamental Research. 

LOFAR, the Low Frequency Array designed and constructed by ASTRON, has facilities in several countries, that are owned by various parties (each with their own funding sources), and that are collectively operated by the International LOFAR Telescope (ILT) foundation under a joint scientific policy. The Open University is incorporated by Royal Charter (RC 000391), an exempt charity in England \& Wales and a charity registered in Scotland (SC 038302). The Open University is authorized and regulated by the Financial Conduct Authority. 

This scientific work uses data obtained from Inyarrimanha Ilgari Bundara / the Murchison Radio-astronomy Observatory. We acknowledge the Wajarri Yamaji People as the Traditional Owners and native title holders of the Observatory site. CSIRO's ASKAP radio telescope is part of the Australia Telescope National Facility (https://ror.org/05qajvd42). Operation of ASKAP is funded by the Australian Government with support from the National Collaborative Research Infrastructure Strategy. ASKAP uses the resources of the Pawsey Supercomputing Research Centre. Establishment of ASKAP, Inyarrimanha Ilgari Bundara, the CSIRO Murchison Radio-astronomy Observatory and the Pawsey Supercomputing Research Centre are initiatives of the Australian Government, with support from the Government of Western Australia and the Science and Industry Endowment Fund. This paper includes archived data obtained through the CSIRO ASKAP Science Data Archive, CASDA (https://data.csiro.au). 

The National Radio Astronomy Observatory is a facility of the National Science Foundation operated under cooperative agreement by Associated Universities, Inc.

These work uses NVAS images that were produced as part of the NRAO VLA Archive Survey, (c) AUI/NRAO. 

The Pan-STARRS1 Surveys (PS1) have been made possible through contributions of the Institute for Astronomy, the University of Hawaii, the Pan-STARRS Project Office, the Max-Planck Society and its participating institutes, the Max Planck Institute for Astronomy, Heidelberg and the Max Planck Institute for Extraterrestrial Physics, Garching, The Johns Hopkins University, Durham University, the University of Edinburgh, Queen's University Belfast, the Harvard-Smithsonian Center for Astrophysics, the Las Cumbres Observatory Global Telescope Network Incorporated, the National Central University of Taiwan, the Space Telescope Science Institute, the National Aeronautics and Space Administration under Grant No. NNX08AR22G issued through the Planetary Science Division of the NASA Science Mission Directorate, the National Science Foundation under Grant No. AST-1238877, the University of Maryland, and Eotvos Lorand University (ELTE). 

The Legacy Surveys consist of three individual and complementary projects: the Dark Energy Camera Legacy Survey (DECaLS; Proposal ID \#2014B-0404; PIs: David Schlegel and Arjun Dey), the Beijing-Arizona Sky Survey (BASS; NOAO Prop. ID \#2015A-0801; PIs: Zhou Xu and Xiaohui Fan), and the Mayall z-band Legacy Survey (MzLS; Prop. ID \#2016A-0453; PI: Arjun Dey). DECaLS, BASS and MzLS together include data obtained, respectively, at the Blanco telescope, Cerro Tololo Inter-American Observatory, NSF's NOIRLab; the Bok telescope, Steward Observatory, University of Arizona; and the Mayall telescope, Kitt Peak National Observatory, NOIRLab. Pipeline processing and analyses of the data were supported by NOIRLab and the Lawrence Berkeley National Laboratory (LBNL). The Legacy Surveys project is honored to be permitted to conduct astronomical research on Iolkam Duag (Kitt Peak), a mountain with particular significance to the Tohono O'odham Nation.

This publication makes use of data products from the Wide-field Infrared Survey Explorer, which is a joint project of the University of California, Los Angeles, and the Jet Propulsion Laboratory/California Institute of Technology, funded by the National Aeronautics and Space Administration.

\end{acknowledgements}

\bibliographystyle{aa}
\addcontentsline{toc}{section}{\refname}\bibliography{mybib}

\begin{thebibliography}{128}
\expandafter\ifx\csname natexlab\endcsname\relax\def\natexlab#1{#1}\fi

\bibitem[{{Alexander} {et~al.}(2025){Alexander}, {Hickox}, {Aird}, {Combes},
  {Costa}, {Habouzit}, {Harrison}, {Leng}, {Morabito}, {Uckelman}, \&
  {Vickers}}]{2025NewAR.10101733A}
{Alexander}, D.~M., {Hickox}, R.~C., {Aird}, J., {et~al.} 2025, \nar, 101,
  101733

\bibitem[{{Alexander} \& {Leahy}(1987)}]{1987MNRAS.225....1A}
{Alexander}, P. \& {Leahy}, J.~P. 1987, \mnras, 225, 1

\bibitem[{{Andreon} {et~al.}(1992){Andreon}, {Garilli}, {Maccagni},
  {Gregorini}, \& {Vettolani}}]{1992A&A...266..127A}
{Andreon}, S., {Garilli}, B., {Maccagni}, D., {Gregorini}, L., \& {Vettolani},
  G. 1992, \aap, 266, 127

\bibitem[{{Beck} \& {Krause}(2005)}]{2005AN....326..414B}
{Beck}, R. \& {Krause}, M. 2005, Astronomische Nachrichten, 326, 414

\bibitem[{{Becker} {et~al.}(1995){Becker}, {White}, \&
  {Helfand}}]{1995ApJ...450..559B}
{Becker}, R.~H., {White}, R.~L., \& {Helfand}, D.~J. 1995, \apj, 450, 559

\bibitem[{{Bera} {et~al.}(2022){Bera}, {Sasmal}, {Patra}, \&
  {Mondal}}]{2022ApJS..260....7B}
{Bera}, S., {Sasmal}, T.~K., {Patra}, D., \& {Mondal}, S. 2022, \apjs, 260, 7

\bibitem[{{Bhukta} {et~al.}(2022){Bhukta}, {Pal}, \&
  {Mondal}}]{2022MNRAS.512.4308B}
{Bhukta}, N., {Pal}, S., \& {Mondal}, S.~K. 2022, \mnras, 512, 4308

\bibitem[{{Brammer} {et~al.}(2008){Brammer}, {van Dokkum}, \&
  {Coppi}}]{2008ApJ...686.1503B}
{Brammer}, G.~B., {van Dokkum}, P.~G., \& {Coppi}, P. 2008, \apj, 686, 1503

\bibitem[{{Brienza} {et~al.}(2020){Brienza}, {Morganti}, {Harwood}, {Duchet},
  {Rajpurohit}, {Shulevski}, {Hardcastle}, {Mahatma}, {Godfrey}, {Prandoni},
  {Shimwell}, \& {Intema}}]{2020A&A...638A..29B}
{Brienza}, M., {Morganti}, R., {Harwood}, J., {et~al.} 2020, \aap, 638, A29

\bibitem[{{Briggs}(1995)}]{1995PhDT.......238B}
{Briggs}, D.~S. 1995, PhD thesis, New Mexico Institute of Mining and Technology

\bibitem[{{Brunetti} {et~al.}(1997){Brunetti}, {Setti}, \&
  {Comastri}}]{1997A&A...325..898B}
{Brunetti}, G., {Setti}, G., \& {Comastri}, A. 1997, \aap, 325, 898

\bibitem[{{Bruno} {et~al.}(2024){Bruno}, {Brienza}, {Zanichelli}, {Gitti},
  {Ubertosi}, {Rajpurohit}, {Venturi}, \& {Dallacasa}}]{2024A&A...690A.160B}
{Bruno}, L., {Brienza}, M., {Zanichelli}, A., {et~al.} 2024, \aap, 690, A160

\bibitem[{{Bruno} {et~al.}(2019){Bruno}, {Gitti}, {Zanichelli}, \&
  {Gregorini}}]{2019A&A...631A.173B}
{Bruno}, L., {Gitti}, M., {Zanichelli}, A., \& {Gregorini}, L. 2019, \aap, 631,
  A173

\bibitem[{{Capetti} {et~al.}(2002){Capetti}, {Zamfir}, {Rossi}, {Bodo},
  {Zanni}, \& {Massaglia}}]{2002A&A...394...39C}
{Capetti}, A., {Zamfir}, S., {Rossi}, P., {et~al.} 2002, \aap, 394, 39

\bibitem[{{Chambers} {et~al.}(2016){Chambers}, {Magnier}, {Metcalfe},
  {Flewelling}, {Huber}, {Waters}, {Denneau}, {Draper}, {Farrow}, {Finkbeiner},
  {Holmberg}, {Koppenhoefer}, {Price}, {Rest}, {Saglia}, {Schlafly}, {Smartt},
  {Sweeney}, {Wainscoat}, {Burgett}, {Chastel}, {Grav}, {Heasley}, {Hodapp},
  {Jedicke}, {Kaiser}, {Kudritzki}, {Luppino}, {Lupton}, {Monet}, {Morgan},
  {Onaka}, {Shiao}, {Stubbs}, {Tonry}, {White}, {Ba{\~n}ados}, {Bell},
  {Bender}, {Bernard}, {Boegner}, {Boffi}, {Botticella}, {Calamida},
  {Casertano}, {Chen}, {Chen}, {Cole}, {Deacon}, {Frenk}, {Fitzsimmons},
  {Gezari}, {Gibbs}, {Goessl}, {Goggia}, {Gourgue}, {Goldman}, {Grant},
  {Grebel}, {Hambly}, {Hasinger}, {Heavens}, {Heckman}, {Henderson}, {Henning},
  {Holman}, {Hopp}, {Ip}, {Isani}, {Jackson}, {Keyes}, {Koekemoer}, {Kotak},
  {Le}, {Liska}, {Long}, {Lucey}, {Liu}, {Martin}, {Masci}, {McLean}, {Mindel},
  {Misra}, {Morganson}, {Murphy}, {Obaika}, {Narayan}, {Nieto-Santisteban},
  {Norberg}, {Peacock}, {Pier}, {Postman}, {Primak}, {Rae}, {Rai}, {Riess},
  {Riffeser}, {Rix}, {R{\"o}ser}, {Russel}, {Rutz}, {Schilbach}, {Schultz},
  {Scolnic}, {Strolger}, {Szalay}, {Seitz}, {Small}, {Smith}, {Soderblom},
  {Taylor}, {Thomson}, {Taylor}, {Thakar}, {Thiel}, {Thilker}, {Unger},
  {Urata}, {Valenti}, {Wagner}, {Walder}, {Walter}, {Watters}, {Werner},
  {Wood-Vasey}, \& {Wyse}}]{2016arXiv161205560C}
{Chambers}, K.~C., {Magnier}, E.~A., {Metcalfe}, N., {et~al.} 2016, arXiv
  e-prints, arXiv:1612.05560

\bibitem[{{Cheung}(2007)}]{2007AJ....133.2097C}
{Cheung}, C.~C. 2007, \aj, 133, 2097

\bibitem[{{Clews} {et~al.}(2025){Clews}, {Croston}, {Dickinson}, {Mingo},
  {Hardcastle}, {Barkus}, {de Jong}, \& {R{\"o}ttgering}}]{2025MNRAS.541.3452C}
{Clews}, L., {Croston}, J.~H., {Dickinson}, H., {et~al.} 2025, \mnras, 541,
  3452

\bibitem[{{Cohen} {et~al.}(2007){Cohen}, {Lane}, {Cotton}, {Kassim}, {Lazio},
  {Perley}, {Condon}, \& {Erickson}}]{2007AJ....134.1245C}
{Cohen}, A.~S., {Lane}, W.~M., {Cotton}, W.~D., {et~al.} 2007, \aj, 134, 1245

\bibitem[{{Colla} {et~al.}(1970){Colla}, {Fanti}, {Ficarra}, {Formiggini},
  {Gandolfi}, {Grueff}, {Lari}, {Padrielli}, {Roffi}, {Tomasi}, \&
  {Vigotti}}]{1970A&AS....1..281C}
{Colla}, G., {Fanti}, C., {Ficarra}, A., {et~al.} 1970, \aaps, 1, 281

\bibitem[{{Condon} {et~al.}(1998){Condon}, {Cotton}, {Greisen}, {Yin},
  {Perley}, {Taylor}, \& {Broderick}}]{1998AJ....115.1693C}
{Condon}, J.~J., {Cotton}, W.~D., {Greisen}, E.~W., {et~al.} 1998, \aj, 115,
  1693

\bibitem[{{Cotton} {et~al.}(2020){Cotton}, {Thorat}, {Condon}, {Frank},
  {J{\'o}zsa}, {White}, {Deane}, {Oozeer}, {Atemkeng}, {Bester}, {Fanaroff},
  {Kupa}, {Smirnov}, {Mauch}, {Krishnan}, \& {Camilo}}]{2020MNRAS.495.1271C}
{Cotton}, W.~D., {Thorat}, K., {Condon}, J.~J., {et~al.} 2020, \mnras, 495,
  1271

\bibitem[{{Coziol} {et~al.}(2009){Coziol}, {Andernach}, {Caretta},
  {Alamo-Mart{\'\i}nez}, \& {Tago}}]{2009AJ....137.4795C}
{Coziol}, R., {Andernach}, H., {Caretta}, C.~A., {Alamo-Mart{\'\i}nez}, K.~A.,
  \& {Tago}, E. 2009, \aj, 137, 4795

\bibitem[{{Crossley} {et~al.}(2008){Crossley}, {Sjouwerman}, {Fomalont}, \&
  {Radziwill}}]{2008SPIE.7016E..0OC}
{Crossley}, J.~H., {Sjouwerman}, L.~O., {Fomalont}, E.~B., \& {Radziwill},
  N.~M. 2008, in Society of Photo-Optical Instrumentation Engineers (SPIE)
  Conference Series, Vol. 7016, Observatory Operations: Strategies, Processes,
  and Systems II, ed. R.~J. {Brissenden} \& D.~R. {Silva}, 70160O

\bibitem[{{Dennett-Thorpe} {et~al.}(2002){Dennett-Thorpe}, {Scheuer}, {Laing},
  {Bridle}, {Pooley}, \& {Reich}}]{2002MNRAS.330..609D}
{Dennett-Thorpe}, J., {Scheuer}, P.~A.~G., {Laing}, R.~A., {et~al.} 2002,
  \mnras, 330, 609

\bibitem[{{Dey} {et~al.}(2019){Dey}, {Schlegel}, {Lang}, {Blum}, {Burleigh},
  {Fan}, {Findlay}, {Finkbeiner}, {Herrera}, {Juneau}, {Landriau}, {Levi},
  {McGreer}, {Meisner}, {Myers}, {Moustakas}, {Nugent}, {Patej}, {Schlafly},
  {Walker}, {Valdes}, {Weaver}, {Y{\`e}che}, {Zou}, {Zhou}, {Abareshi},
  {Abbott}, {Abolfathi}, {Aguilera}, {Alam}, {Allen}, {Alvarez}, {Annis},
  {Ansarinejad}, {Aubert}, {Beechert}, {Bell}, {BenZvi}, {Beutler}, {Bielby},
  {Bolton}, {Brice{\~n}o}, {Buckley-Geer}, {Butler}, {Calamida}, {Carlberg},
  {Carter}, {Casas}, {Castander}, {Choi}, {Comparat}, {Cukanovaite}, {Delubac},
  {DeVries}, {Dey}, {Dhungana}, {Dickinson}, {Ding}, {Donaldson}, {Duan},
  {Duckworth}, {Eftekharzadeh}, {Eisenstein}, {Etourneau}, {Fagrelius},
  {Farihi}, {Fitzpatrick}, {Font-Ribera}, {Fulmer}, {G{\"a}nsicke},
  {Gaztanaga}, {George}, {Gerdes}, {Gontcho}, {Gorgoni}, {Green}, {Guy},
  {Harmer}, {Hernandez}, {Honscheid}, {Huang}, {James}, {Jannuzi}, {Jiang},
  {Joyce}, {Karcher}, {Karkar}, {Kehoe}, {Kneib}, {Kueter-Young}, {Lan},
  {Lauer}, {Le Guillou}, {Le Van Suu}, {Lee}, {Lesser}, {Perreault Levasseur},
  {Li}, {Mann}, {Marshall}, {Mart{\'\i}nez-V{\'a}zquez}, {Martini}, {du Mas des
  Bourboux}, {McManus}, {Meier}, {M{\'e}nard}, {Metcalfe},
  {Mu{\~n}oz-Guti{\'e}rrez}, {Najita}, {Napier}, {Narayan}, {Newman}, {Nie},
  {Nord}, {Norman}, {Olsen}, {Paat}, {Palanque-Delabrouille}, {Peng},
  {Poppett}, {Poremba}, {Prakash}, {Rabinowitz}, {Raichoor}, {Rezaie},
  {Robertson}, {Roe}, {Ross}, {Ross}, {Rudnick}, {Safonova}, {Saha},
  {S{\'a}nchez}, {Savary}, {Schweiker}, {Scott}, {Seo}, {Shan}, {Silva},
  {Slepian}, {Soto}, {Sprayberry}, {Staten}, {Stillman}, {Stupak}, {Summers},
  {Sien Tie}, {Tirado}, {Vargas-Maga{\~n}a}, {Vivas}, {Wechsler}, {Williams},
  {Yang}, {Yang}, {Yapici}, {Zaritsky}, {Zenteno}, {Zhang}, {Zhang}, {Zhou}, \&
  {Zhou}}]{2019AJ....157..168D}
{Dey}, A., {Schlegel}, D.~J., {Lang}, D., {et~al.} 2019, \aj, 157, 168

\bibitem[{{Duchesne} {et~al.}(2025){Duchesne}, {Ross}, {Thomson}, {Lenc},
  {Murphy}, {Galvin}, {Hotan}, {Moss}, \& {Whiting}}]{2025PASA...42...38D}
{Duchesne}, S., {Ross}, K., {Thomson}, A.~J.~M., {et~al.} 2025, \pasa, 42, 38

\bibitem[{{Duchesne} {et~al.}(2023){Duchesne}, {Thomson}, {Pritchard}, {Lenc},
  {Moss}, {McConnell}, {Wieringa}, {Whiting}, {Wang}, {Wang}, {Rose}, {Raja},
  {Murphy}, {Leung}, {Huynh}, {Hotan}, {Hodgson}, \&
  {Heald}}]{2023PASA...40...34D}
{Duchesne}, S.~W., {Thomson}, A.~J.~M., {Pritchard}, J., {et~al.} 2023, \pasa,
  40, e034

\bibitem[{{Fanaroff} \& {Riley}(1974)}]{1974MNRAS.167P..31F}
{Fanaroff}, B.~L. \& {Riley}, J.~M. 1974, \mnras, 167, 31P

\bibitem[{{Ferrarese} \& {Merritt}(2000)}]{2000ApJ...539L...9F}
{Ferrarese}, L. \& {Merritt}, D. 2000, \apjl, 539, L9

\bibitem[{{Gab{\'a}nyi} {et~al.}(2021){Gab{\'a}nyi}, {Frey}, \&
  {Perger}}]{2021MNRAS.506.3641G}
{Gab{\'a}nyi}, K.~{\'E}., {Frey}, S., \& {Perger}, K. 2021, \mnras, 506, 3641

\bibitem[{{Giri} {et~al.}(2024){Giri}, {Fendt}, {Thorat}, {Bodo}, \&
  {Rossi}}]{2024FrASS..1171101G}
{Giri}, G., {Fendt}, C., {Thorat}, K., {Bodo}, G., \& {Rossi}, P. 2024,
  Frontiers in Astronomy and Space Sciences, 11, 1371101

\bibitem[{{Gopal-Krishna} {et~al.}(2012){Gopal-Krishna}, {Biermann}, {Gergely},
  \& {Wiita}}]{2012RAA....12..127G}
{Gopal-Krishna}, {Biermann}, P.~L., {Gergely}, L.~{\'A}., \& {Wiita}, P.~J.
  2012, Research in Astronomy and Astrophysics, 12, 127

\bibitem[{{Govoni} \& {Feretti}(2004)}]{2004IJMPD..13.1549G}
{Govoni}, F. \& {Feretti}, L. 2004, International Journal of Modern Physics D,
  13, 1549

\bibitem[{{Gower} {et~al.}(1967){Gower}, {Scott}, \&
  {Wills}}]{1967MmRAS..71...49G}
{Gower}, J.~F.~R., {Scott}, P.~F., \& {Wills}, D. 1967, \memras, 71, 49

\bibitem[{{Gregorini} {et~al.}(1994){Gregorini}, {de Ruiter}, {Parma},
  {Sadler}, {Vettolani}, \& {Ekers}}]{1994A&AS..106....1G}
{Gregorini}, L., {de Ruiter}, H.~R., {Parma}, P., {et~al.} 1994, \aaps, 106, 1

\bibitem[{{Gregorini} {et~al.}(1992{\natexlab{a}}){Gregorini}, {Klein},
  {Parma}, {Schlickeiser}, \& {Wielebinski}}]{1992A&AS...94...13G}
{Gregorini}, L., {Klein}, U., {Parma}, P., {Schlickeiser}, R., \&
  {Wielebinski}, R. 1992{\natexlab{a}}, \aaps, 94, 13

\bibitem[{{Gregorini} {et~al.}(1992{\natexlab{b}}){Gregorini}, {Vettolani}, {de
  Ruiter}, \& {Parma}}]{1992A&AS...95....1G}
{Gregorini}, L., {Vettolani}, G., {de Ruiter}, H.~R., \& {Parma}, P.
  1992{\natexlab{b}}, \aaps, 95, 1

\bibitem[{{Gregory} \& {Condon}(1991)}]{1991ApJS...75.1011G}
{Gregory}, P.~C. \& {Condon}, J.~J. 1991, \apjs, 75, 1011

\bibitem[{{Greisen}(2003)}]{2003ASSL..285..109G}
{Greisen}, E.~W. 2003, in Astrophysics and Space Science Library, Vol. 285,
  Information Handling in Astronomy - Historical Vistas, ed. A.~{Heck}, 109

\bibitem[{{Gupta} {et~al.}(2017){Gupta}, {Ajithkumar}, {Kale}, {Nayak},
  {Sabhapathy}, {Sureshkumar}, {Swami}, {Chengalur}, {Ghosh},
  {Ishwara-Chandra}, {Joshi}, {Kanekar}, {Lal}, \& {Roy}}]{2017CSci..113..707G}
{Gupta}, Y., {Ajithkumar}, B., {Kale}, H.~S., {et~al.} 2017, Current Science,
  113, 707

\bibitem[{{Hale} {et~al.}(2021){Hale}, {McConnell}, {Thomson}, {Lenc}, {Heald},
  {Hotan}, {Leung}, {Moss}, {Murphy}, {Pritchard}, {Sadler}, {Stewart}, \&
  {Whiting}}]{2021PASA...38...58H}
{Hale}, C.~L., {McConnell}, D., {Thomson}, A.~J.~M., {et~al.} 2021, \pasa, 38,
  e058

\bibitem[{{Hao} {et~al.}(2010){Hao}, {McKay}, {Koester}, {Rykoff}, {Rozo},
  {Annis}, {Wechsler}, {Evrard}, {Siegel}, {Becker}, {Busha}, {Gerdes},
  {Johnston}, \& {Sheldon}}]{2010ApJS..191..254H}
{Hao}, J., {McKay}, T.~A., {Koester}, B.~P., {et~al.} 2010, \apjs, 191, 254

\bibitem[{{Intema} {et~al.}(2017){Intema}, {Jagannathan}, {Mooley}, \&
  {Frail}}]{2017A&A...598A..78I}
{Intema}, H.~T., {Jagannathan}, P., {Mooley}, K.~P., \& {Frail}, D.~A. 2017,
  \aap, 598, A78

\bibitem[{{Intema} {et~al.}(2009){Intema}, {van der Tol}, {Cotton}, {Cohen},
  {van Bemmel}, \& {R{\"o}ttgering}}]{2009A&A...501.1185I}
{Intema}, H.~T., {van der Tol}, S., {Cotton}, W.~D., {et~al.} 2009, \aap, 501,
  1185

\bibitem[{{Jaffe} \& {Perola}(1973)}]{1973A&A....26..423J}
{Jaffe}, W.~J. \& {Perola}, G.~C. 1973, \aap, 26, 423

\bibitem[{{Jamrozy} {et~al.}(2007){Jamrozy}, {Konar}, {Saikia}, {Stawarz},
  {Mack}, \& {Siemiginowska}}]{2007MNRAS.378..581J}
{Jamrozy}, M., {Konar}, C., {Saikia}, D.~J., {et~al.} 2007, \mnras, 378, 581

\bibitem[{{Jones} {et~al.}(2009){Jones}, {Read}, {Saunders}, {Colless},
  {Jarrett}, {Parker}, {Fairall}, {Mauch}, {Sadler}, {Watson}, {Burton},
  {Campbell}, {Cass}, {Croom}, {Dawe}, {Fiegert}, {Frankcombe}, {Hartley},
  {Huchra}, {James}, {Kirby}, {Lahav}, {Lucey}, {Mamon}, {Moore}, {Peterson},
  {Prior}, {Proust}, {Russell}, {Safouris}, {Wakamatsu}, {Westra}, \&
  {Williams}}]{2009MNRAS.399..683J}
{Jones}, D.~H., {Read}, M.~A., {Saunders}, W., {et~al.} 2009, \mnras, 399, 683

\bibitem[{{Jones} \& {McAdam}(1992)}]{1992ApJS...80..137J}
{Jones}, P.~A. \& {McAdam}, W.~B. 1992, \apjs, 80, 137

\bibitem[{{Joshi} {et~al.}(2019){Joshi}, {Krishna}, {Yang}, {Shi}, {Yu},
  {Wiita}, {Ho}, {Wu}, {An}, {Wang}, {Subramanian}, \&
  {Yesuf}}]{2019ApJ...887..266J}
{Joshi}, R., {Krishna}, G., {Yang}, X., {et~al.} 2019, \apj, 887, 266

\bibitem[{{Kardashev}(1962)}]{1962SvA.....6..317K}
{Kardashev}, N.~S. 1962, \sovast, 6, 317

\bibitem[{{Kempner} {et~al.}(2004){Kempner}, {Blanton}, {Clarke}, {En{\ss}lin},
  {Johnston-Hollitt}, \& {Rudnick}}]{2004rcfg.proc..335K}
{Kempner}, J.~C., {Blanton}, E.~L., {Clarke}, T.~E., {et~al.} 2004, in The
  Riddle of Cooling Flows in Galaxies and Clusters of galaxies, ed.
  T.~{Reiprich}, J.~{Kempner}, \& N.~{Soker}, 335

\bibitem[{{Klein} {et~al.}(1995){Klein}, {Mack}, {Gregorini}, \&
  {Parma}}]{1995A&A...303..427K}
{Klein}, U., {Mack}, K.~H., {Gregorini}, L., \& {Parma}, P. 1995, \aap, 303,
  427

\bibitem[{{Kormendy} \& {Ho}(2013)}]{2013ARA&A..51..511K}
{Kormendy}, J. \& {Ho}, L.~C. 2013, \araa, 51, 511

\bibitem[{{Kumari} \& {Pal}(2024{\natexlab{a}})}]{2024A&A...683A.175K}
{Kumari}, S. \& {Pal}, S. 2024{\natexlab{a}}, \aap, 683, A175

\bibitem[{{Kumari} \& {Pal}(2024{\natexlab{b}})}]{2024MNRAS.52711233K}
{Kumari}, S. \& {Pal}, S. 2024{\natexlab{b}}, \mnras, 527, 11233

\bibitem[{{Lacy} {et~al.}(2020){Lacy}, {Baum}, {Chandler}, {Chatterjee},
  {Clarke}, {Deustua}, {English}, {Farnes}, {Gaensler}, {Gugliucci},
  {Hallinan}, {Kent}, {Kimball}, {Law}, {Lazio}, {Marvil}, {Mao}, {Medlin},
  {Mooley}, {Murphy}, {Myers}, {Osten}, {Richards}, {Rosolowsky}, {Rudnick},
  {Schinzel}, {Sivakoff}, {Sjouwerman}, {Taylor}, {White}, {Wrobel},
  {Andernach}, {Beasley}, {Berger}, {Bhatnager}, {Birkinshaw}, {Bower},
  {Brandt}, {Brown}, {Burke-Spolaor}, {Butler}, {Comerford}, {Demorest}, {Fu},
  {Giacintucci}, {Golap}, {G{\"u}th}, {Hales}, {Hiriart}, {Hodge}, {Horesh},
  {Ivezi{\'c}}, {Jarvis}, {Kamble}, {Kassim}, {Liu}, {Loinard}, {Lyons},
  {Masters}, {Mezcua}, {Moellenbrock}, {Mroczkowski}, {Nyland},
  {O{\textquoteright}Dea}, {O{\textquoteright}Sullivan}, {Peters}, {Radford},
  {Rao}, {Robnett}, {Salcido}, {Shen}, {Sobotka}, {Witz}, {Vaccari}, {van
  Weeren}, {Vargas}, {Williams}, \& {Yoon}}]{2020PASP..132c5001L}
{Lacy}, M., {Baum}, S.~A., {Chandler}, C.~J., {et~al.} 2020, \pasp, 132, 035001

\bibitem[{{Lal} \& {Rao}(2007)}]{2007MNRAS.374.1085L}
{Lal}, D.~V. \& {Rao}, A.~P. 2007, \mnras, 374, 1085

\bibitem[{{Lal} {et~al.}(2019){Lal}, {Sebastian}, {Cheung}, \& {Pramesh
  Rao}}]{2019AJ....157..195L}
{Lal}, D.~V., {Sebastian}, B., {Cheung}, C.~C., \& {Pramesh Rao}, A. 2019, \aj,
  157, 195

\bibitem[{{Landt} {et~al.}(2010){Landt}, {Cheung}, \&
  {Healey}}]{2010MNRAS.408.1103L}
{Landt}, H., {Cheung}, C.~C., \& {Healey}, S.~E. 2010, \mnras, 408, 1103

\bibitem[{{Lara} {et~al.}(2001{\natexlab{a}}){Lara}, {Cotton}, {Feretti},
  {Giovannini}, {Marcaide}, {M{\'a}rquez}, \& {Venturi}}]{2001A&A...370..409L}
{Lara}, L., {Cotton}, W.~D., {Feretti}, L., {et~al.} 2001{\natexlab{a}}, \aap,
  370, 409

\bibitem[{{Lara} {et~al.}(2001{\natexlab{b}}){Lara}, {M{\'a}rquez}, {Cotton},
  {Feretti}, {Giovannini}, {Marcaide}, \& {Venturi}}]{2001A&A...378..826L}
{Lara}, L., {M{\'a}rquez}, I., {Cotton}, W.~D., {et~al.} 2001{\natexlab{b}},
  \aap, 378, 826

\bibitem[{{Large} {et~al.}(1991){Large}, {Cram}, \&
  {Burgess}}]{1991Obs...111...72L}
{Large}, M.~I., {Cram}, L.~E., \& {Burgess}, A.~M. 1991, The Observatory, 111,
  72

\bibitem[{{Large} {et~al.}(1981){Large}, {Mills}, {Little}, {Crawford}, \&
  {Sutton}}]{1981MNRAS.194..693L}
{Large}, M.~I., {Mills}, B.~Y., {Little}, A.~G., {Crawford}, D.~F., \&
  {Sutton}, J.~M. 1981, \mnras, 194, 693

\bibitem[{{Leahy} \& {Parma}(1992)}]{1992ersf.meet..307L}
{Leahy}, J.~P. \& {Parma}, P. 1992, in Extragalactic Radio Sources. From Beams
  to Jets, ed. J.~{Roland}, H.~{Sol}, \& G.~{Pelletier}, 307--308

\bibitem[{Leahy \& Williams(1984)}]{10.1093/mnras/210.4.929}
Leahy, J.~P. \& Williams, A.~G. 1984, Monthly Notices of the Royal Astronomical
  Society, 210, 929

\bibitem[{{Liu} {et~al.}(2018){Liu}, {Lazio}, {Shen}, \&
  {Strauss}}]{2018ApJ...854..169L}
{Liu}, X., {Lazio}, T. J.~W., {Shen}, Y., \& {Strauss}, M.~A. 2018, \apj, 854,
  169

\bibitem[{{Mack} {et~al.}(1994){Mack}, {Gregorini}, {Parma}, \&
  {Klein}}]{1994A&AS..103..157M}
{Mack}, K.~H., {Gregorini}, L., {Parma}, P., \& {Klein}, U. 1994, \aaps, 103,
  157

\bibitem[{{Magorrian} {et~al.}(1998){Magorrian}, {Tremaine}, {Richstone},
  {Bender}, {Bower}, {Dressler}, {Faber}, {Gebhardt}, {Green}, {Grillmair},
  {Kormendy}, \& {Lauer}}]{1998AJ....115.2285M}
{Magorrian}, J., {Tremaine}, S., {Richstone}, D., {et~al.} 1998, \aj, 115, 2285

\bibitem[{{Mahatma} {et~al.}(2019){Mahatma}, {Hardcastle}, {Williams}, {Best},
  {Croston}, {Duncan}, {Mingo}, {Morganti}, {Brienza}, {Cochrane},
  {G{\"u}rkan}, {Harwood}, {Jarvis}, {Jamrozy}, {Jurlin}, {Morabito},
  {R{\"o}ttgering}, {Sabater}, {Shimwell}, {Smith}, {Shulevski}, \&
  {Tasse}}]{2019A&A...622A..13M}
{Mahatma}, V.~H., {Hardcastle}, M.~J., {Williams}, W.~L., {et~al.} 2019, \aap,
  622, A13

\bibitem[{{Makarov} {et~al.}(2014){Makarov}, {Prugniel}, {Terekhova},
  {Courtois}, \& {Vauglin}}]{2014A&A...570A..13M}
{Makarov}, D., {Prugniel}, P., {Terekhova}, N., {Courtois}, H., \& {Vauglin},
  I. 2014, \aap, 570, A13

\bibitem[{{Marr} {et~al.}(2015){Marr}, {Snell}, \&
  {Kurtz}}]{2015frao.book.....M}
{Marr}, J.~M., {Snell}, R.~L., \& {Kurtz}, S.~E. 2015, {Fundamentals of Radio
  Astronomy: Observational Methods (Series in Astronomy and Astrophysics)}

\bibitem[{{McConnell} {et~al.}(2020){McConnell}, {Hale}, {Lenc}, {Banfield},
  {Heald}, {Hotan}, {Leung}, {Moss}, {Murphy}, {O'Brien}, {Pritchard}, {Raja},
  {Sadler}, {Stewart}, {Thomson}, {Whiting}, {Allison}, {Amy}, {Anderson},
  {Ball}, {Bannister}, {Bell}, {Bock}, {Bolton}, {Bunton}, {Chippendale},
  {Collier}, {Cooray}, {Cornwell}, {Diamond}, {Edwards}, {Gupta}, {Hayman},
  {Heywood}, {Jackson}, {Koribalski}, {Lee-Waddell}, {McClure-Griffiths}, {Ng},
  {Norris}, {Phillips}, {Reynolds}, {Roxby}, {Schinckel}, {Shields},
  {Tremblay}, {Tzioumis}, {Voronkov}, \& {Westmeier}}]{2020PASA...37...48M}
{McConnell}, D., {Hale}, C.~L., {Lenc}, E., {et~al.} 2020, \pasa, 37, e048

\bibitem[{{McMullin} {et~al.}(2007){McMullin}, {Waters}, {Schiebel}, {Young},
  \& {Golap}}]{2007ASPC..376..127M}
{McMullin}, J.~P., {Waters}, B., {Schiebel}, D., {Young}, W., \& {Golap}, K.
  2007, in Astronomical Society of the Pacific Conference Series, Vol. 376,
  Astronomical Data Analysis Software and Systems XVI, ed. R.~A. {Shaw},
  F.~{Hill}, \& D.~J. {Bell}, 127

\bibitem[{{Merritt} \& {Ekers}(2002)}]{2002Sci...297.1310M}
{Merritt}, D. \& {Ekers}, R.~D. 2002, Science, 297, 1310

\bibitem[{{Miley}(1980)}]{1980ARA&A..18..165M}
{Miley}, G. 1980, \araa, 18, 165

\bibitem[{{Mohan} \& {Rafferty}(2015)}]{2015ascl.soft02007M}
{Mohan}, N. \& {Rafferty}, D. 2015, {PyBDSF: Python Blob Detection and Source
  Finder}, Astrophysics Source Code Library, record ascl:1502.007

\bibitem[{{Morris} {et~al.}(2022){Morris}, {Wilcots}, {Hooper}, \&
  {Heinz}}]{2022AJ....163..280M}
{Morris}, M.~E., {Wilcots}, E., {Hooper}, E., \& {Heinz}, S. 2022, \aj, 163,
  280

\bibitem[{{Nandi} {et~al.}(2019){Nandi}, {Saikia}, {Roy}, {Dabhade},
  {Wadadekar}, {Larsson}, {Baes}, {Chandola}, \& {Singh}}]{2019MNRAS.486.5158N}
{Nandi}, S., {Saikia}, D.~J., {Roy}, R., {et~al.} 2019, \mnras, 486, 5158

\bibitem[{{Norris} {et~al.}(2022){Norris}, {Collier}, {Crocker}, {Heywood},
  {Macgregor}, {Rudnick}, {Shabala}, {Andernach}, {da Cunha}, {English},
  {Filipovi{\'c}}, {Koribalski}, {Luken}, {Robotham}, {Sekhar}, {Thorne}, \&
  {Vernstrom}}]{2022MNRAS.513.1300N}
{Norris}, R.~P., {Collier}, J.~D., {Crocker}, R.~M., {et~al.} 2022, \mnras,
  513, 1300

\bibitem[{{Oei} {et~al.}(2022){Oei}, {van Weeren}, {Hardcastle}, {Botteon},
  {Shimwell}, {Dabhade}, {Gast}, {R{\"o}ttgering}, {Br{\"u}ggen}, {Tasse},
  {Williams}, \& {Shulevski}}]{2022A&A...660A...2O}
{Oei}, M. S.~S.~L., {van Weeren}, R.~J., {Hardcastle}, M.~J., {et~al.} 2022,
  \aap, 660, A2

\bibitem[{{Offringa} {et~al.}(2010){Offringa}, {de Bruyn}, {Biehl}, {Zaroubi},
  {Bernardi}, \& {Pandey}}]{2010MNRAS.405..155O}
{Offringa}, A.~R., {de Bruyn}, A.~G., {Biehl}, M., {et~al.} 2010, \mnras, 405,
  155

\bibitem[{{Offringa} {et~al.}(2013){Offringa}, {de Bruyn}, {Zaroubi}, {van
  Diepen}, {Martinez-Ruby}, {Labropoulos}, {Brentjens}, {Ciardi}, {Daiboo},
  {Harker}, {Jeli{\'c}}, {Kazemi}, {Koopmans}, {Mellema}, {Pandey}, {Pizzo},
  {Schaye}, {Vedantham}, {Veligatla}, {Wijnholds}, {Yatawatta}, {Zarka},
  {Alexov}, {Anderson}, {Asgekar}, {Avruch}, {Beck}, {Bell}, {Bell}, {Bentum},
  {Bernardi}, {Best}, {Birzan}, {Bonafede}, {Breitling}, {Broderick},
  {Br{\"u}ggen}, {Butcher}, {Conway}, {de Vos}, {Dettmar}, {Eisloeffel},
  {Falcke}, {Fender}, {Frieswijk}, {Gerbers}, {Griessmeier}, {Gunst},
  {Hassall}, {Heald}, {Hessels}, {Hoeft}, {Horneffer}, {Karastergiou},
  {Kondratiev}, {Koopman}, {Kuniyoshi}, {Kuper}, {Maat}, {Mann}, {McKean},
  {Meulman}, {Mevius}, {Mol}, {Nijboer}, {Noordam}, {Norden}, {Paas}, {Pandey},
  {Pizzo}, {Polatidis}, {Rafferty}, {Rawlings}, {Reich}, {R{\"o}ttgering},
  {Schoenmakers}, {Sluman}, {Smirnov}, {Sobey}, {Stappers}, {Steinmetz},
  {Swinbank}, {Tagger}, {Tang}, {Tasse}, {van Ardenne}, {van Cappellen}, {van
  Duin}, {van Haarlem}, {van Leeuwen}, {van Weeren}, {Vermeulen}, {Vocks},
  {Wijers}, {Wise}, \& {Wucknitz}}]{2013A&A...549A..11O}
{Offringa}, A.~R., {de Bruyn}, A.~G., {Zaroubi}, S., {et~al.} 2013, \aap, 549,
  A11

\bibitem[{{Offringa} {et~al.}(2014){Offringa}, {McKinley}, {Hurley-Walker},
  {Briggs}, {Wayth}, {Kaplan}, {Bell}, {Feng}, {Neben}, {Hughes}, {Rhee},
  {Murphy}, {Bhat}, {Bernardi}, {Bowman}, {Cappallo}, {Corey}, {Deshpande},
  {Emrich}, {Ewall-Wice}, {Gaensler}, {Goeke}, {Greenhill}, {Hazelton},
  {Hindson}, {Johnston-Hollitt}, {Jacobs}, {Kasper}, {Kratzenberg}, {Lenc},
  {Lonsdale}, {Lynch}, {McWhirter}, {Mitchell}, {Morales}, {Morgan},
  {Kudryavtseva}, {Oberoi}, {Ord}, {Pindor}, {Procopio}, {Prabu}, {Riding},
  {Roshi}, {Shankar}, {Srivani}, {Subrahmanyan}, {Tingay}, {Waterson},
  {Webster}, {Whitney}, {Williams}, \& {Williams}}]{2014MNRAS.444..606O}
{Offringa}, A.~R., {McKinley}, B., {Hurley-Walker}, N., {et~al.} 2014, \mnras,
  444, 606

\bibitem[{{Offringa} {et~al.}(2012){Offringa}, {van de Gronde}, \&
  {Roerdink}}]{2012A&A...539A..95O}
{Offringa}, A.~R., {van de Gronde}, J.~J., \& {Roerdink}, J.~B.~T.~M. 2012,
  \aap, 539, A95

\bibitem[{{Oke} \& {Gunn}(1983)}]{1983ApJ...266..713O}
{Oke}, J.~B. \& {Gunn}, J.~E. 1983, \apj, 266, 713

\bibitem[{{Pacholczyk}(1970)}]{1970ranp.book.....P}
{Pacholczyk}, A.~G. 1970, {Radio astrophysics. Nonthermal processes in galactic
  and extragalactic sources}

\bibitem[{{Parma} {et~al.}(1985){Parma}, {Ekers}, \&
  {Fanti}}]{1985A&AS...59..511P}
{Parma}, P., {Ekers}, R.~D., \& {Fanti}, R. 1985, \aaps, 59, 511

\bibitem[{{Patra} {et~al.}(2023){Patra}, {Joshi}, \&
  {Gopal-Krishna}}]{2023MNRAS.524.3270P}
{Patra}, D., {Joshi}, R., \& {Gopal-Krishna}. 2023, \mnras, 524, 3270

\bibitem[{{Pilkington} \& {Scott}(1965)}]{1965MmRAS..69..183P}
{Pilkington}, J.~D.~H. \& {Scott}, J.~F. 1965, \memras, 69, 183

\bibitem[{{Proctor}(2011)}]{2011ApJS..194...31P}
{Proctor}, D.~D. 2011, \apjs, 194, 31

\bibitem[{{Rees}(1978)}]{1978Natur.275..516R}
{Rees}, M.~J. 1978, \nat, 275, 516

\bibitem[{{Rengelink} {et~al.}(1997){Rengelink}, {Tang}, {de Bruyn}, {Miley},
  {Bremer}, {Roettgering}, \& {Bremer}}]{1997A&AS..124..259R}
{Rengelink}, R.~B., {Tang}, Y., {de Bruyn}, A.~G., {et~al.} 1997, \aaps, 124,
  259

\bibitem[{{Retana-Montenegro}(2022)}]{2022A&A...663A.153R}
{Retana-Montenegro}, E. 2022, \aap, 663, A153

\bibitem[{{Retana-Montenegro} \& {R\"ottgering}(2020)}]{2020AA}
{Retana-Montenegro}, E. \& {R\"ottgering}, H. 2020, \aap

\bibitem[{{Retana-Montenegro} {et~al.}(2018){Retana-Montenegro},
  {R{\"o}ttgering}, {Shimwell}, {van Weeren}, {Prandoni}, {Brunetti}, {Best},
  \& {Br{\"u}ggen}}]{2018A&A...620A..74R}
{Retana-Montenegro}, E., {R{\"o}ttgering}, H.~J.~A., {Shimwell}, T.~W.,
  {et~al.} 2018, \aap, 620, A74

\bibitem[{Reynolds(1994)}]{Reynolds1994}
Reynolds, J. 1994, A revised flux scale for the AT Compact Array, Tech. Rep.
  AT/39.3/040, ATNF

\bibitem[{{Rines} {et~al.}(2018){Rines}, {Geller}, {Diaferio}, {Hwang}, \&
  {Sohn}}]{2018ApJ...862..172R}
{Rines}, K.~J., {Geller}, M.~J., {Diaferio}, A., {Hwang}, H.~S., \& {Sohn}, J.
  2018, \apj, 862, 172

\bibitem[{{Riseley} {et~al.}(2023){Riseley}, {Biava}, {Lusetti}, {Bonafede},
  {Bonnassieux}, {Botteon}, {Loi}, {Brunetti}, {Cassano}, {Osinga},
  {Rajpurohit}, {R{\"o}ttgering}, {Shimwell}, {Timmerman}, \& {van
  Weeren}}]{2023MNRAS.524.6052R}
{Riseley}, C.~J., {Biava}, N., {Lusetti}, G., {et~al.} 2023, \mnras, 524, 6052

\bibitem[{{Rottmann}(2001)}]{2001PhDT.......173R}
{Rottmann}, H. 2001, PhD thesis, -

\bibitem[{{Rykoff} {et~al.}(2014){Rykoff}, {Rozo}, {Busha}, {Cunha},
  {Finoguenov}, {Evrard}, {Hao}, {Koester}, {Leauthaud}, {Nord}, {Pierre},
  {Reddick}, {Sadibekova}, {Sheldon}, \& {Wechsler}}]{2014ApJ...785..104R}
{Rykoff}, E.~S., {Rozo}, E., {Busha}, M.~T., {et~al.} 2014, \apj, 785, 104

\bibitem[{{Sabater} {et~al.}(2019){Sabater}, {Best}, {Hardcastle}, {Shimwell},
  {Tasse}, {Williams}, {Br{\"u}ggen}, {Cochrane}, {Croston}, {de Gasperin},
  {Duncan}, {G{\"u}rkan}, {Mechev}, {Morabito}, {Prandoni}, {R{\"o}ttgering},
  {Smith}, {Harwood}, {Mingo}, {Mooney}, \& {Saxena}}]{2019A&A...622A..17S}
{Sabater}, J., {Best}, P.~N., {Hardcastle}, M.~J., {et~al.} 2019, \aap, 622,
  A17

\bibitem[{{Saikia} {et~al.}(2006){Saikia}, {Konar}, \&
  {Kulkarni}}]{2006MNRAS.366.1391S}
{Saikia}, D.~J., {Konar}, C., \& {Kulkarni}, V.~K. 2006, \mnras, 366, 1391

\bibitem[{{Saripalli} \& {Subrahmanyan}(2009)}]{2009ApJ...695..156S}
{Saripalli}, L. \& {Subrahmanyan}, R. 2009, \apj, 695, 156

\bibitem[{{Scaife} \& {Heald}(2012)}]{2012MNRAS.423L..30S}
{Scaife}, A.~M.~M. \& {Heald}, G.~H. 2012, \mnras, 423, L30

\bibitem[{{Schlafly} {et~al.}(2019){Schlafly}, {Meisner}, \&
  {Green}}]{2019ApJS..240...30S}
{Schlafly}, E.~F., {Meisner}, A.~M., \& {Green}, G.~M. 2019, \apjs, 240, 30

\bibitem[{{Schoenmakers} {et~al.}(2000){Schoenmakers}, {de Bruyn},
  {R{\"o}ttgering}, {van der Laan}, \& {Kaiser}}]{2000MNRAS.315..371S}
{Schoenmakers}, A.~P., {de Bruyn}, A.~G., {R{\"o}ttgering}, H.~J.~A., {van der
  Laan}, H., \& {Kaiser}, C.~R. 2000, \mnras, 315, 371

\bibitem[{{Schwab}(1984)}]{1984AJ.....89.1076S}
{Schwab}, F.~R. 1984, \aj, 89, 1076

\bibitem[{{Sebastian} {et~al.}(2024){Sebastian}, {Caproni}, {Kharb}, {Nayana},
  {Ali}, {Rubinur}, {O'Dea}, {Baum}, \& {Nandi}}]{2024MNRAS.530.4902S}
{Sebastian}, B., {Caproni}, A., {Kharb}, P., {et~al.} 2024, \mnras, 530, 4902

\bibitem[{{Sekhar} \& {Athreya}(2018)}]{2018AJ....156....9S}
{Sekhar}, S. \& {Athreya}, R. 2018, \aj, 156, 9

\bibitem[{{Shimmins} \& {Bolton}(1974)}]{1974AuJPA..32....1S}
{Shimmins}, A.~J. \& {Bolton}, J.~G. 1974, Australian Journal of Physics
  Astrophysical Supplement, 32, 1

\bibitem[{{Shimwell} {et~al.}(2022){Shimwell}, {Hardcastle}, {Tasse}, {Best},
  {R{\"o}ttgering}, {Williams}, {Botteon}, {Drabent}, {Mechev}, {Shulevski},
  {van Weeren}, {Bester}, {Br{\"u}ggen}, {Brunetti}, {Callingham}, {Chy{\.z}y},
  {Conway}, {Dijkema}, {Duncan}, {de Gasperin}, {Hale}, {Haverkorn}, {Hugo},
  {Jackson}, {Mevius}, {Miley}, {Morabito}, {Morganti}, {Offringa}, {Oonk},
  {Rafferty}, {Sabater}, {Smith}, {Schwarz}, {Smirnov}, {O'Sullivan},
  {Vedantham}, {White}, {Albert}, {Alegre}, {Asabere}, {Bacon}, {Bonafede},
  {Bonnassieux}, {Brienza}, {Bilicki}, {Bonato}, {Calistro Rivera}, {Cassano},
  {Cochrane}, {Croston}, {Cuciti}, {Dallacasa}, {Danezi}, {Dettmar}, {Di
  Gennaro}, {Edler}, {En{\ss}lin}, {Emig}, {Franzen}, {Garc{\'\i}a-Vergara},
  {Grange}, {G{\"u}rkan}, {Hajduk}, {Heald}, {Heesen}, {Hoang}, {Hoeft},
  {Horellou}, {Iacobelli}, {Jamrozy}, {Jeli{\'c}}, {Kondapally}, {Kukreti},
  {Kunert-Bajraszewska}, {Magliocchetti}, {Mahatma}, {Ma{\l}ek}, {Mandal},
  {Massaro}, {Meyer-Zhao}, {Mingo}, {Mostert}, {Nair}, {Nakoneczny},
  {Nikiel-Wroczy{\'n}ski}, {Orr{\'u}}, {Pajdosz-{\'S}mierciak}, {Pasini},
  {Prandoni}, {van Piggelen}, {Rajpurohit}, {Retana-Montenegro}, {Riseley},
  {Rowlinson}, {Saxena}, {Schrijvers}, {Sweijen}, {Siewert}, {Timmerman},
  {Vaccari}, {Vink}, {West}, {Wo{\l}owska}, {Zhang}, \&
  {Zheng}}]{2022A&A...659A...1S}
{Shimwell}, T.~W., {Hardcastle}, M.~J., {Tasse}, C., {et~al.} 2022, \aap, 659,
  A1

\bibitem[{{Shimwell} {et~al.}(2017){Shimwell}, {R{\"o}ttgering}, {Best},
  {Williams}, {Dijkema}, {de Gasperin}, {Hardcastle}, {Heald}, {Hoang},
  {Horneffer}, {Intema}, {Mahony}, {Mandal}, {Mechev}, {Morabito}, {Oonk},
  {Rafferty}, {Retana-Montenegro}, {Sabater}, {Tasse}, {van Weeren},
  {Br{\"u}ggen}, {Brunetti}, {Chy{\.z}y}, {Conway}, {Haverkorn}, {Jackson},
  {Jarvis}, {McKean}, {Miley}, {Morganti}, {White}, {Wise}, {van Bemmel},
  {Beck}, {Brienza}, {Bonafede}, {Calistro Rivera}, {Cassano}, {Clarke},
  {Cseh}, {Deller}, {Drabent}, {van Driel}, {Engels}, {Falcke}, {Ferrari},
  {Fr{\"o}hlich}, {Garrett}, {Harwood}, {Heesen}, {Hoeft}, {Horellou},
  {Israel}, {Kapi{\'n}ska}, {Kunert-Bajraszewska}, {McKay}, {Mohan},
  {Orr{\'u}}, {Pizzo}, {Prandoni}, {Schwarz}, {Shulevski}, {Sipior}, {Smith},
  {Sridhar}, {Steinmetz}, {Stroe}, {Varenius}, {van der Werf}, {Zensus}, \&
  {Zwart}}]{2017A&A...598A.104S}
{Shimwell}, T.~W., {R{\"o}ttgering}, H.~J.~A., {Best}, P.~N., {et~al.} 2017,
  \aap, 598, A104

\bibitem[{{Skrutskie} {et~al.}(2006){Skrutskie}, {Cutri}, {Stiening},
  {Weinberg}, {Schneider}, {Carpenter}, {Beichman}, {Capps}, {Chester},
  {Elias}, {Huchra}, {Liebert}, {Lonsdale}, {Monet}, {Price}, {Seitzer},
  {Jarrett}, {Kirkpatrick}, {Gizis}, {Howard}, {Evans}, {Fowler}, {Fullmer},
  {Hurt}, {Light}, {Kopan}, {Marsh}, {McCallon}, {Tam}, {Van Dyk}, \&
  {Wheelock}}]{2006AJ....131.1163S}
{Skrutskie}, M.~F., {Cutri}, R.~M., {Stiening}, R., {et~al.} 2006, \aj, 131,
  1163

\bibitem[{{Tasse}(2014)}]{2014arXiv1410.8706T}
{Tasse}, C. 2014, arXiv e-prints, arXiv:1410.8706

\bibitem[{{Tasse} {et~al.}(2018){Tasse}, {Hugo}, {Mirmont}, {Smirnov},
  {Atemkeng}, {Bester}, {Hardcastle}, {Lakhoo}, {Perkins}, \&
  {Shimwell}}]{2018A&A...611A..87T}
{Tasse}, C., {Hugo}, B., {Mirmont}, M., {et~al.} 2018, \aap, 611, A87

\bibitem[{{Tasse} {et~al.}(2021){Tasse}, {Shimwell}, {Hardcastle},
  {O'Sullivan}, {van Weeren}, {Best}, {Bester}, {Hugo}, {Smirnov}, {Sabater},
  {Calistro-Rivera}, {de Gasperin}, {Morabito}, {R{\"o}ttgering}, {Williams},
  {Bonato}, {Bondi}, {Botteon}, {Br{\"u}ggen}, {Brunetti}, {Chy{\.z}y},
  {Garrett}, {G{\"u}rkan}, {Jarvis}, {Kondapally}, {Mandal}, {Prandoni},
  {Repetti}, {Retana-Montenegro}, {Schwarz}, {Shulevski}, \&
  {Wiaux}}]{2021A&A...648A...1T}
{Tasse}, C., {Shimwell}, T., {Hardcastle}, M.~J., {et~al.} 2021, \aap, 648, A1

\bibitem[{{Ulrich} \& {Roennback}(1996)}]{1996A&A...313..750U}
{Ulrich}, M.~H. \& {Roennback}, J. 1996, \aap, 313, 750

\bibitem[{{van der Laan} \& {Perola}(1969)}]{1969A&A.....3..468V}
{van der Laan}, H. \& {Perola}, G.~C. 1969, \aap, 3, 468

\bibitem[{{van Diepen} {et~al.}(2018){van Diepen}, {Dijkema}, \&
  {Offringa}}]{2018ascl.soft04003V}
{van Diepen}, G., {Dijkema}, T.~J., \& {Offringa}, A. 2018, {DPPP: Default
  Pre-Processing Pipeline}, Astrophysics Source Code Library, record
  ascl:1804.003

\bibitem[{{van Weeren} {et~al.}(2016){van Weeren}, {Williams}, {Hardcastle},
  {Shimwell}, {Rafferty}, {Sabater}, {Heald}, {Sridhar}, {Dijkema}, {Brunetti},
  {Br{\"u}ggen}, {Andrade-Santos}, {Ogrean}, {R{\"o}ttgering}, {Dawson},
  {Forman}, {de Gasperin}, {Jones}, {Miley}, {Rudnick}, {Sarazin}, {Bonafede},
  {Best}, {B{\^\i}rzan}, {Cassano}, {Chy{\.z}y}, {Croston}, {Ensslin},
  {Ferrari}, {Hoeft}, {Horellou}, {Jarvis}, {Kraft}, {Mevius}, {Intema},
  {Murray}, {Orr{\'u}}, {Pizzo}, {Simionescu}, {Stroe}, {van der Tol}, \&
  {White}}]{2016ApJS..223....2V}
{van Weeren}, R.~J., {Williams}, W.~L., {Hardcastle}, M.~J., {et~al.} 2016,
  \apjs, 223, 2

\bibitem[{{Vardoulaki} {et~al.}(2025){Vardoulaki}, {Back{\"o}fer},
  {Finoguenov}, {Vazza}, {Comparat}, {Gozaliasl}, {Whittam}, {Hale}, {Weaver},
  {Koekemoer}, {Collier}, {Frank}, {Heywood}, {Sekhar}, {Taylor}, {Pinjarkar},
  {Hardcastle}, {Shimwell}, {Hoeft}, {White}, {An}, {Tabatabaei},
  {Randriamanakoto}, \& {Filipovic}}]{2025A&A...695A.178V}
{Vardoulaki}, E., {Back{\"o}fer}, V., {Finoguenov}, A., {et~al.} 2025, \aap,
  695, A178

\bibitem[{{Wen} \& {Han}(2015)}]{2015ApJ...807..178W}
{Wen}, Z.~L. \& {Han}, J.~L. 2015, \apj, 807, 178

\bibitem[{{Wen} {et~al.}(2012){Wen}, {Han}, \& {Liu}}]{2012ApJS..199...34W}
{Wen}, Z.~L., {Han}, J.~L., \& {Liu}, F.~S. 2012, \apjs, 199, 34

\bibitem[{{We{\.z}gowiec} {et~al.}(2024){We{\.z}gowiec}, {Jamrozy},
  {Chy{\.z}y}, {Hardcastle}, {Ku{\'z}micz}, {Heald}, \&
  {Shimwell}}]{2024A&A...691A.193W}
{We{\.z}gowiec}, M., {Jamrozy}, M., {Chy{\.z}y}, K.~T., {et~al.} 2024, \aap,
  691, A193

\bibitem[{{Williams} {et~al.}(2016){Williams}, {van Weeren}, {R{\"o}ttgering},
  {Best}, {Dijkema}, {de Gasperin}, {Hardcastle}, {Heald}, {Prandoni},
  {Sabater}, {Shimwell}, {Tasse}, {van Bemmel}, {Br{\"u}ggen}, {Brunetti},
  {Conway}, {En{\ss}lin}, {Engels}, {Falcke}, {Ferrari}, {Haverkorn},
  {Jackson}, {Jarvis}, {Kapi{\'n}ska}, {Mahony}, {Miley}, {Morabito},
  {Morganti}, {Orr{\'u}}, {Retana-Montenegro}, {Sridhar}, {Toribio}, {White},
  {Wise}, \& {Zwart}}]{2016MNRAS.460.2385W}
{Williams}, W.~L., {van Weeren}, R.~J., {R{\"o}ttgering}, H.~J.~A., {et~al.}
  2016, \mnras, 460, 2385

\bibitem[{{Xie} {et~al.}(2024){Xie}, {Ba{\~n}ados}, {Belladitta},
  {Mazzucchelli}, {Schindler}, {Davies}, \& {Venemans}}]{2024ApJ...964...98X}
{Xie}, Z.-L., {Ba{\~n}ados}, E., {Belladitta}, S., {et~al.} 2024, \apj, 964, 98

\bibitem[{{Yang} {et~al.}(2022){Yang}, {Ji}, {Joshi}, {Yang}, {An}, {Wang},
  {Ho}, {Roberts}, \& {Saripalli}}]{2022ApJ...933...98Y}
{Yang}, X., {Ji}, J., {Joshi}, R., {et~al.} 2022, \apj, 933, 98

\bibitem[{{Yang} {et~al.}(2019){Yang}, {Joshi}, {Gopal-Krishna}, {An}, {Ho},
  {Wiita}, {Liu}, {Yang}, {Wang}, {Wu}, \& {Yang}}]{2019ApJS..245...17Y}
{Yang}, X., {Joshi}, R., {Gopal-Krishna}, {et~al.} 2019, \apjs, 245, 17

\end{thebibliography}

\begin{appendix} \section{Corrected positional offsets and total flux comparison of the three X-shaped radio-galaxies fields} 

 \label{sec:appendix_A}

\begin{figure}[H]
\centering{}\includegraphics[clip,scale=0.45]{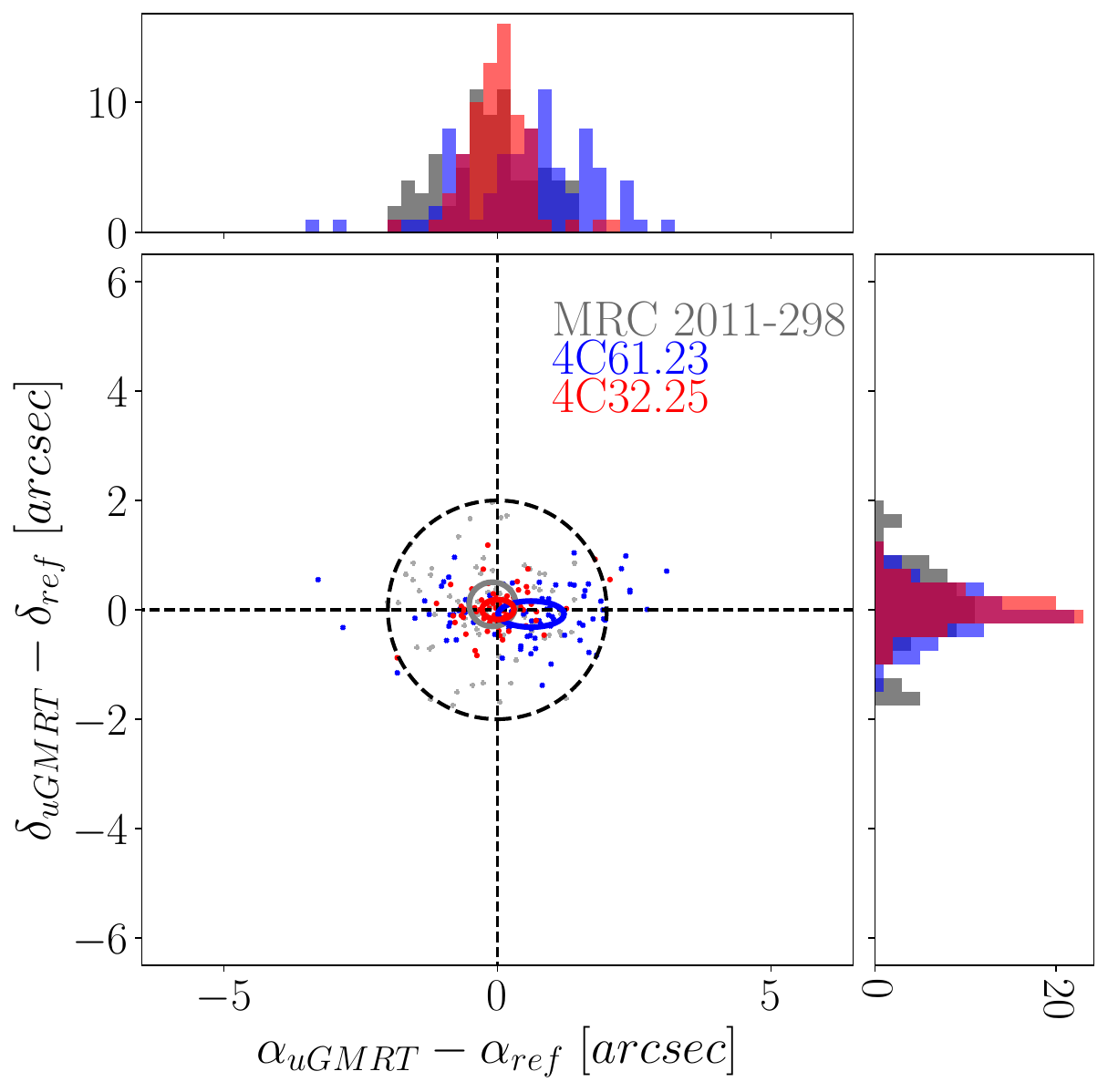}\centering\caption{\label{fig:position_offsets} The corrected positional offsets between
high S/N and compact uGMRT sources and their FIRST/RACS \citep{1995ApJ...450..559B,2020PASA...37...48M}
counterparts. The dashed lines denote a circle with radius $r=2^{\prime\prime}$,
which is the mosaic pixel scale. The colored ellipses centered on
the right ascension and declination mean offsets indicate the standard
deviation for both directions.}
\end{figure}

\begin{figure}[H]
\begin{centering}
\includegraphics[clip,scale=0.56]{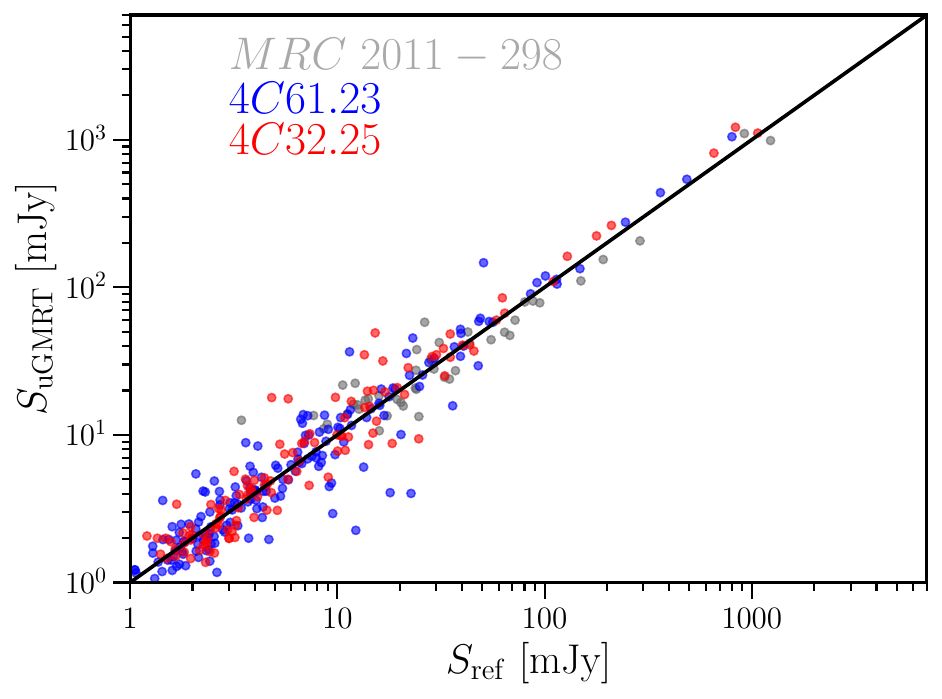}
\par\end{centering}
\centering{}\centering\caption{\label{fig:fluxes_comparison} Total flux densities of matches between
our 400 MHz uGMRT catalogs and the reference surveys LoTSS and RACS
\citep{2022A&A...659A...1S,2020PASA...37...48M} The dashed line indicates
a flux density ratio equal to unity. A correction factor is applied
to the RACS flux densities to convert them from the \citet{Reynolds1994}
scale to the SH12 flux density scale adopted in this work.}
\end{figure}

In Figure \ref{fig:position_offsets}, we show the corrected positional
offsets between high S/N and compact uGMRT sources and their counterparts
in the reference high-resolution surveys FIRST/RACS \citep{1995ApJ...450..559B,2020PASA...37...48M}.
Figure \ref{fig:fluxes_comparison} shows the flux density scale comparison
between the scaled uGMRT and the reference flux densities at 400 MHz.
The methods to correct the astrometry and flux densities are described
in RM18.

 \section{Flux densities of 4C32.25, 4C61.23, and MRC 2011-298.}
 \label{sec:appendix_B}

Table \ref{fig:total_fluxes_xrgs-1}.1 reports the total flux densities
of 4C32.25, and 4C61.23 compiled from the literature and this work
to fit the integrated spectra of 4C32.25 and 4C61.23 in Section \ref{sec:spectral_index}.
Table \ref{fig:a3670_fluxes} presents the 400 MHz flux densities
of the different components of MRC 2011-298 estimated using the radio-map
from Section \ref{sec:images_catalogs}.
\begin{table}[H]
\noindent \begin{centering}
\caption{Total flux densities of 4C32.25 and 4C61.23}
\begin{tabular}{cccc}
\hline 
Source & Frequency & Flux \textbf{Density} & Reference\tabularnewline
\hline 
 & {[}$ $MHz$ ${]} & {[}$ $Jy$ ${]} & \tabularnewline
\hline 
4C32.25 &  &  & \tabularnewline
 & 144 & $8.91\pm1.33$ & (1)\tabularnewline
 & 178 & $2.61\pm0.92^{\dagger}$ & (2)\tabularnewline
 & 236 & $6.80\pm1.10$ & (6)\tabularnewline
 & 400 & $4.94\pm0.74$ & P\tabularnewline
 & 606 & $2.11\pm0.55^{\dagger}$ & (6)\tabularnewline
 & 887 & $1.64\pm0.25^{\dagger}$ & (4)\tabularnewline
 & 1367 & $1.81\pm0.28$ & (4)\tabularnewline
 & 1400 & $1.79\pm0.18$ & (3)\tabularnewline
 & 1525 & $1.36\pm0.15^{\dagger}$ & P\tabularnewline
 & 1655 & $1.32\pm0.13^{\dagger}$ & (4)\tabularnewline
 & 3000 & $1.01\pm0.06$ & P\tabularnewline
 & 4850 & $0.78\pm0.08$ & (8)\tabularnewline
 & 10550 & $0.54\pm0.11$ & (7)\tabularnewline
4C61.23 &  &  & \tabularnewline
 & 144 & $5.46\pm0.81$ & (1)\tabularnewline
 & 178 & $4.14\pm1.63$ & (2)\tabularnewline
 & 400 & $3.10\pm0.46$ & P\tabularnewline
 & 1400 & $1.11\pm0.20$ & (3)\tabularnewline
 & 3000 & $0.39\pm0.04$ & (5)\tabularnewline
 & 4850 & $0.33\pm0.03$ & P\tabularnewline
\hline 
\end{tabular}
\par\end{centering}
\begin{centering}
\centering
\par\end{centering}
$\;$

Notes: Total flux densities of 4C32.25 and 4C61.23. Notes: P: present
paper. 1: \citet{2022A&A...659A...1S}; 2: \citep{1965MmRAS..69..183P,1967MmRAS..71...49G}
; 3: \citet{1998AJ....115.1693C}; 4: \citet{2020PASA...37...48M,2023PASA...40...34D,2025PASA...42...38D};
5: \citet{2020PASP..132c5001L}; 6: \citet{2007MNRAS.374.1085L};
7: \citet{1992A&AS...94...13G}; 8: \citet{1991ApJS...75.1011G};
$\dagger$: flux density not used in fitting. \label{fig:total_fluxes_xrgs-1}
\end{table}

\begin{table}[H]
\noindent \begin{centering}
\caption{Flux densities of the various components of MRC 2011-298 measured
at 400 MHz.\textbf{ }\label{fig:a3670_fluxes}}
\begin{tabular}{cc}
\hline 
Source & Flux\tabularnewline
\hline 
 & {[}$ $Jy$ ${]}\tabularnewline
\hline 
MRC 2011-298 & \tabularnewline
Northern Primary Lobe & $0.712\pm0.11$\tabularnewline
Southern Primary Lobe & $0.393\pm0.06$\tabularnewline
Western Wing & $0.191\pm0.03$\tabularnewline
Eastern Wing & $0.175\pm0.03$\tabularnewline
\hline 
\end{tabular}
\par\end{centering}
\centering{}\centering
\end{table}

\section{Physical properties and spectral ages of 4C32.25 and 4C61.23} 
\label{sec:appendix_C}

Table \ref{fig:fitting_spectra_summary}.1 reports the physical properties
and spectral ages derived in Section \ref{sec:spectral_index}. 

\begin{table}[H]
\noindent \begin{centering}
\caption{Physical properties and ages of 4C32.25 and 4C61.23.}
\par\end{centering}
\begin{centering}
\begin{tabular}{cccc}
\hline 
Source/Comp & Luminosity & $\nu_{\textrm{b}}$ & $t_{\textrm{max}}$\tabularnewline
\hline 
 & {[}$1\times10^{40}]$ &  & \tabularnewline
(1) & (2) & (3) & (4)\tabularnewline
\hline 
4C32.25 &  &  & \tabularnewline
Western P. L. & $3.42$ & $282.5\pm22.3$ & \textbf{$8.6\pm0.3$}\tabularnewline
Eastern P. L. & $3.01$ & $316.2\pm47.5$ & \textbf{$8.1\pm0.6$}\tabularnewline
Southern Wing & $0.662$ & $3.06\pm0.4$ & \textbf{$82.2\pm5.3$}\tabularnewline
Northern Wing & $0.660$ & $1.02\pm0.04$ & \textbf{$90.8\pm10.2$}\tabularnewline
4C61.23 &  &  & \tabularnewline
SouthEastern P. L. & $6.15$ & $316.3\pm46.6$ & \textbf{$7.4\pm0.5$}\tabularnewline
NorthWestern P. L. & $14.1$ & $97.3\pm25.3$ & \textbf{$13.4\pm1.7$}\tabularnewline
NorthEastern Wing & $2.55$ & $308.1\pm37.8$ & \textbf{$7.7\pm0.7$}\tabularnewline
SouthWestern Wing & $2.65$ & $88.3\pm26.4$ & \textbf{$14.1\pm2.1$}\tabularnewline
\hline 
\end{tabular}
\par\end{centering}
\begin{centering}
\centering
\par\end{centering}
$\;$

Notes: (1) Source/Component; (2) radio-luminosity at 150 MHz in erg/s;
(3) break frequency in GHz; (4) maximum spectral age in Myr. \label{fig:fitting_spectra_summary}
\end{table}
` 

\end{appendix}

\end{document}